\documentclass[11pt,a4paper]{article}
\usepackage{jheppub}
\usepackage{xcolor}
\usepackage{amsmath,amssymb,amsfonts}
\usepackage{physics}
\usepackage{bbm}
\usepackage{graphicx}
\usepackage{svg}
\graphicspath{{./images}}
\usepackage{subcaption}
\usepackage{array}
\usepackage{tikz}
\usetikzlibrary{arrows.meta, positioning}
\usepackage{url}
\usepackage{multirow}
\usepackage{tabularray}
\usepackage{adjustbox}
\usepackage{comment}
\usepackage{tabularx}
\usepackage{slashed}
\newcolumntype{Y}{>{\centering\arraybackslash}X}
\usepackage{makecell}
\newcommand{\eq}[1]{eq.~(\ref{#1})}

\newcommand{\bea}{\begin{eqnarray}}
\newcommand{\eea}{\end{eqnarray}}

\newcommand{\E}{E}
\newcommand{\nn}{\nonumber}

\def\CaI{\textbf{a}_{1,1}}
\def\CaII{\textbf{a}_{1,2}}
\def\CaIII{\delta\textbf{a}_{1,3}}
\def\CaIV{\delta\textbf{a}_{1,4}}
\def\CaV{\delta\textbf{a}_{1,5}}
\def\CaVI{\delta\textbf{a}_{1,6}}
\def\rr{\varepsilon_2}
\def\KaI{\delta^3\textbf{a}_{2,1}}
\def\KaII{\delta^3\textbf{a}_{2,2}}

\def\BaIsX{\delta\textbf{a}_{1,7}^0}
\def\BaIsI{\delta\textbf{a}_{1,7}^1 + \alpha_0\alpha_1\,\delta\textbf{a}_{1,7}^{1,\,3p}}

\def\BaIIsX{\delta\textbf{a}_{1,8}^0}
\def\BaIIsI{\delta\textbf{a}_{1,8}^1 + \alpha_0\alpha_1\,\delta\textbf{a}_{1,8}^{1,\,3p}}

\def\BaIIIsX{\delta^2\textbf{a}_{1,9}^{0} + \alpha_0\,\delta^2\textbf{a}_{1,9}^{0,\,3p}}

\def\BaIVsX{\delta^2\textbf{a}_{1,10}^{0} + \alpha_0\,\delta^2\textbf{a}_{1,10}^{0,\,3p}}

\def\BaVsX{\alpha_0(\alpha_1 - \varepsilon_1)\,\delta^2\textbf{a}_{1,11}^{0,\,3p}}

\def\BaVIsX{\alpha_0(\alpha_1 - \varepsilon_1)\,\delta^2\textbf{a}_{1,12}^{0,\,3p}}

\def\Qr{Q^Z_{f_R}}
\def\Ql{Q^Z_{f_L}}
\def\mom{A}
\def\smPL{G_{+-0}}
\def\smML{G_{-+0}}
\def\smPT{G_{+-}}
\def\smMT{G_{-+}}
\def\qr{q_{LR}}

\usepackage{hyperref}

\usepackage{hyperref}

\def\hc{h.c.}
\def\stw{s_{\theta_W}}
\def\ctw{c_{\theta_W}}

\def\hata{\hat{a}}
\def\mom{A}

\begin{document}

\definecolor{lime}{HTML}{A6CE39}
\DeclareRobustCommand{\orcidicon}{\hspace{-1mm}
	\begin{tikzpicture}
	\draw[lime, fill=lime] (0,0) 
	circle [radius=0.12] 
	node[white] {{\fontfamily{qag}\selectfont \tiny \,ID}};
	\draw[white, fill=white] (-0.0525,0.095) 
	circle [radius=0.007];
	\end{tikzpicture}
	\hspace{-3mm}
}

\foreach \x in {A, ..., Z}{\expandafter\xdef\csname orcid\x\endcsname{\noexpand\href{https://orcid.org/\csname orcidauthor\x\endcsname}
		{\noexpand\orcidicon}}
}
\newcommand{\orcidauthorD}{0009-0009-3206-9970}
\newcommand{\orcidauthorA}{0009-0000-2346-2273}
\newcommand{\orcidauthorB}{0000-0001-5815-4182}
\newcommand{\orcidauthorC}{0000-0003-4398-4698}

\title{Differential observables for the Higgs-strahlung process to all orders in EFT}
\preprint{TIFR/TH/26-3}

\author[a]{Sourav Bera\orcidD{}}
\author[a]{Debsubhra Chakraborty\orcidB{}}
\author[a]{Susobhan Chattopadhyay\orcidA{}}
\author[a]{Rick S. Gupta\orcidC{}}
\affiliation[a]{Department of Theoretical Physics, Tata Institute of Fundamental Research, Homi Bhabha Road, Colaba, Mumbai 400005, India}

\emailAdd{sourav.bera@tifr.res.in}
\emailAdd{debsubhra.chakraborty@tifr.res.in}
\emailAdd{susobhan.chattopadhyay@tifr.res.in}
\emailAdd{rsgupta@theory.tifr.res.in}

\begin{abstract}
{   We develop methods to obtain the fully differential cross-section for the   $f \bar{f} \to Z(\ell\ell)\,h$ process  to any desired order in  effective field theory (EFT). To achieve this, we first derive a mapping between the partial wave expansion and the EFT expansion  to all orders. We find that at lower orders,  EFT predicts correlations between the different partial wave coefficients. This allows us to construct linear combinations of partial wave coefficients that get their leading contributions from a higher dimension EFT operator. We then introduce experimental observables, the so called angular moments---that   probe these  linear combinations of partial wave coefficients---and can be determined from a fully differential analysis of the angular distribution of the leptons arising from the $Z$  decay. We show that  analysing the dependence of these angular moments on the $Zh$ invariant mass allows us to systematically probe all higher dimension EFT operators contributing to this process. While we take  the Higgs-strahlung process as an example, the methods developed here are completely general and can be applied to other 2-to-2 collider processes.
}

\end{abstract}
\maketitle

\setcounter{page}{0}
\pagenumbering{arabic}

\section{Introduction}

Effective Field theories (EFT) have become the standard language to parametrise indirect deviations beyond the Standard Model (SM) in  Large Hadron Collider (LHC) measurements. In particular, the Standard Model Effective Field Theory (SMEFT)~\cite{bw,warsaw}  provides a framework for such a parametrisation for a wide variety of processes such as $W$ and $Z$ decays, production and decay of the Higgs boson, diboson production, vector boson scattering, top quark production and decay, flavour physics etc. Another  parameterisation that has gained popularity in recent times is the Higgs Effective Field Theory (HEFT)~\cite{heft1,heft2,heft3,georgi,Feruglio:1992wf,Bagger:1993zf,Koulovassilopoulos:1993pw,Burgess:1999ha,Grinstein:2007iv,Alonso:2012px,Espriu:2013fia,Buchalla:2013rka,Brivio:2013pma,Alonso:2015fsp,Alonso:2016oah,Buchalla:2017jlu,Alonso:2017tdy,deBlas:2018tjm}  which is even more general than the SMEFT~\cite{ajm1,ajm2,rf,craig,Alonso:2025ksv}.

Given the current level of experimental sensitivity, most studies using EFT parametrisations include only the lowest order corrections in the EFT expansion, i.e. dimension-6 operators in the SMEFT and next to leading order (NLO) operators in the HEFT. We are now entering  the era of higher luminosities and  higher experimental sensitivities. The computation of higher order QCD and electroweak effects---including    loop effects with one EFT insertion---has also shown impressive progress (see for instance Ref.~\cite{grober} and references therein).  To keep pace with these developments, it is important to go beyond the lowest order EFT effects.  This has led to many recent studies on higher order EFT effects, in particular the effect of dimension-8 SMEFT operators in Higgs and electroweak physics~\cite{Bellm,Martin1, Li:2020gnx, Murphy:2020rsh, Ellis, Chala:2021pll,Martin3,Banerjee:2022kfu,Degrande, Arzate2024, Martin5, Martin4, Briviogrober}.  In this work we will present a systematic approach to study the effect of higher dimensional operators on the differential observables for the  Higgs-strahlung process. While we take this process as an example, our methods can be  generalised to other 2-to-2 collider processes like diboson production.

As higher order EFT effects become important, it is worth highlighting a  defining property of EFTs. Although an EFT   lagrangian includes all operators respecting the symmetries of a theory, the observables predicted  at lower orders of the expansion are not the most general ones allowed by the symmetries. One of  the most well-known examples of this is   the fact that although SM symmetries do not forbid $B$ and $L$  violation, these possibilities are not realised perturbatively at the renormalizable level. One needs to include higher dimensional  SMEFT operators to allow such effects.  In the context of the present study  this gives rise to the following question: are there features of the differential distributions of the process $f\bar{f} \to Z(\ell\ell) h$ that arise only at higher   EFT orders? 

One way to answer the question posed above is to begin with the partial wave expansion that provides the most general  parametrisation of the amplitude. As we will see, using general principles like unitarity and CPT invariance one can then derive---at a given center of mass energy---the most general angular distribution  for the final state particles in the $f \bar{f} \to Z(\ell\ell) h$ process. The most general angular distribution thus derived, however, contains an infinite number of free parameters, namely the infinite number of partial wave coefficients. Unlike the EFT parametrisation, there is \emph{a priori} no way to organise these parameters in varying degrees of relevance as in the EFT expansion. In order to incorporate  EFT dimensional analysis into the partial wave expansion, we  derive a mapping between the partial wave coefficients and the EFT Wilson coefficients to all orders in the EFT. This then allows us to find how many  partial wave coefficients are non-zero at a given order in the EFT. 

We  find that at lower  orders, EFT predicts that certain linear combinations of  partial wave coefficients must vanish---i.e.  correlations exist between these  partial wave coefficients. We construct such linear combinations of partial wave coefficients and map them to higher dimensional EFT operators.

The partial wave  expansion is connected in a straightforward way to differential observables. In particular, using our results we can write the amplitude as a sum over the finite number of partial waves that are non-zero at a given EFT order. Squaring this amplitude  we then express the angular distribution for the  $f\bar{f} \to Z(\ell\ell) h$ process, at that order, as a sum of a finite number angular functions. The coefficients of these angular functions, the so-called angular moments, can be  determined from a fully differential analysis of the process. In fact, a similar approach is already being employed by the CMS collaboration to study the  $pp \to Z(\ell\ell) h (bb)$ process differentially at the dimension-6 level in the SMEFT~\cite{cmseft}. We show how our all-orders mapping between partial wave coefficients and EFT  Wilson coefficients can be used to compute these angular moments to any given EFT order. 


The plan of this paper is as follows. In Sec.~\ref{sec2} we review the formalism introduced in Ref.~\cite{Luty2023} that presents   the EFT expansion to all orders, and will serve as our starting point. In Sec.~\ref{sec3} we introduce our own EFT basis and present the mapping between the partial wave expansion and the EFT expansion to all orders. In Sec.~\ref{sec4} we discuss how our results can be interpreted in the SMEFT at dimension-8 level. Finally, in Sec.~\ref{sec5} we show how a  mapping between differential observables and EFT Wilson coefficients can be obtained up to any given EFT order.


\section{Principals and descendants} 
\label{sec2}

 In Ref.~\cite{Chang2023, Arzate2024, Luty2023}, the EFT expansion  for standard  Higgs/electroweak processes was organised in a    way that will be particularly useful for us.\footnote{See also Ref.\cite{Shadmi:2018xan,Durieux:2019eor}.   where a similar analysis has been performed in the massive spinor helicity formalism.} In these works it was shown that the  tree-level EFT amplitude for a given process can be  written in terms of  a finite number of \textit{principal} amplitudes and higher order \textit{descendants} obtained by  multiplying the principal amplitudes  by powers of $s/\Lambda^2$ and $t/\Lambda^2$; here $s$ and $t$ are the usual Mandelstam invariants.\footnote{We will take the liberty to  use the term  `principals', instead of the term `primaries' originally used in  Ref.~\cite{Chang2023, Arzate2024, Luty2023}, as the latter term  has been used in other contexts in the EFT literature~\cite{Gupta:2014rxa,hepp}. The principal operators were called `stripped contact terms' in  Ref.\cite{Shadmi:2018xan,Durieux:2019eor}. }

 Let us discuss this in more detail for the  process, $f\,\bar{f} \longrightarrow Z\,h$. The  tree-level amplitude for this process can be written as, 
 \begin{equation}
     \mathcal{M}_{f_1\,\bar{f}_2 \to Z\,h}=\mathcal{M}^{3-point}_{f_1\,\bar{f}_2 \to Z\,h}+\delta \mathcal{M}_{f_1\,\bar{f}_2 \to Z\,h}.
     \label{eq:amp}
 \end{equation}
The  first term above  is the contribution due to 3-point  corrections to the SM amplitude---i.e. EFT corrections to the $Zff$ and $hZZ$ vertices---and the second term contains the contributions due to contact interactions in the EFT. There are a finite number of 3-point vertices that contribute to this amplitude; these have been shown in Table~\ref{tab:3-point_ops}. The second term above can be organised  in terms of 4-point principal  and descendant amplitudes  as follows,
\begin{align}
    \delta \mathcal{M}(f_1\,\bar{f}_2 \to Z\,h) &= \frac{(\bar{v}_2 \slashed{\epsilon}^* u_1)}{\Lambda} \Bigg[C_{(0,0)}^1 + C_{(1,0)}^1\frac{s}{\Lambda^2} + C_{(0,1)}^1\frac{(t-u)}{\Lambda^2} + \mathcal{O}(\frac{E^4}{\Lambda^4})\Bigg] \nonumber\\
    + \frac{i}{\Lambda} \epsilon_{\mu\nu\rho\sigma}&(\bar{v}_2 \gamma^{\mu}u_1) (p_1 - p_2)^{\nu} p_{3}^{\rho}\,{\epsilon}_3^{*\sigma}\left[C_{(0,0)}^2 + C_{(1,0)}^2\frac{s}{\Lambda^2} + C_{(0,1)}^2\frac{(t-u)}{\Lambda^2} + \mathcal{O}(E^4/\Lambda^4)\right] \nonumber \\
    + \sum_{k = 3}^{N_p=12} &\sum_{m,n=0}^\infty C_{(m,n)}^k\frac{P_{k}}{\Lambda^{d_k-4}}   \frac{s^m (t-u)^n}{(\Lambda^2)^{m+n}}.
    \label{eq:des}
\end{align}
 We see that the expression for $\delta \mathcal{M}_{f_1\,\bar{f}_2 \to Z\,h}$, to all orders in the EFT expansion, can be written as a sum of $N_p=12$ contributions. Each of these 12 contributions is an infinite series in powers of Mandelstam variables suppressed  by the cutoff. The first term in each  series defines the  principal  amplitude and is denoted by $P_k$. The principal amplitude is followed by an  infinite series of descendants, i.e. terms obtained by multiplying factors of,   $s^m (t-u)^n/\Lambda^{2(m+n)}$, to the principal amplitudes. The coefficients $C^k_{(m,n)}$  are  EFT Wilson coefficients. In Table~\ref{tab:4-pt_principals} we present the operators corresponding to the 12 principal amplitudes as listed in Ref.~\cite{Luty2023}.\footnote{Note that Ref.~\cite{Luty2023} considers both the 3-point operators in Table~\ref{tab:3-point_ops} and the four-point operators  to be principal operators. In this work we will refer to only the 4-point operators as principal operators.}

\begin{table}[t]
\renewcommand{\arraystretch}{1.5}
   \begin{subtable}{.5\linewidth}
      \centering
        \begin{tabular}{|c|c|c|c|}
    \hline
    $i$ & $\mathcal{O}^{3-point}_{i}$ & CP & $d_{\mathcal{O}}$  \\[0.1cm]
    \hline
     1 & $\dfrac{m_Z}{2}hZ_{\mu}Z^{\mu}$ & + & 3\\[0.1cm] 
     \hline
     2 & $Q^Z_{f_R} Z_{\mu}\bar{\psi}_R\gamma^{\mu} \psi_R$ & + & \multirow{2}{*}{4} \\[0.1cm]
     3 & $Q^Z_{f_L} Z_{\mu}\bar{\psi}_L\gamma^{\mu} \psi_L$ & + & \\[0.1cm]
     \hline
     4 & $hZ_{\mu\nu}Z^{\mu\nu}$ & + & \multirow{4}{*}{5} \\
     5 & $hZ_{\mu\nu}\tilde{Z}^{\mu\nu}$ & $-$ & \\
    
     6 & $Z_{\mu\nu}\bar{\psi}_L \sigma^{\mu\nu} \psi_R$ + h.c. & + &  \\
     7 & $iZ_{\mu\nu}\bar{\psi}_L \sigma^{\mu\nu} \psi_R$ + h.c. & $-$ & \\
     \hline
\end{tabular}
        \caption{}
        \label{tab:3-point_ops}
    \end{subtable} 
    \begin{subtable}{.5\linewidth}
      \centering
        
        \begin{tabular}{|c|c|c|c|}
        \hline
            $k$ & $\mathcal{O}^{k}_{(0,0)}$ & CP & $d_{\mathcal{O}}$  \\[0.15em]
             \hline
            \multicolumn{4}{|c|}{$J = 1$} \\
            \hline
            1 & $hZ^{\mu}\bar{\psi}_R\gamma_{\mu}\psi_R$ & + & \multirow{2}{*}{5} \\[0.15em]
            2 & $hZ^{\mu}\bar{\psi}_L\gamma_{\mu}\psi_L$ & + &   \\[0.15em]
            \cline{1-4}
            
            3 & $ih\widetilde{Z}_{\mu\nu}(\bar{\psi}_R\gamma^{\mu}\overset{\leftrightarrow}{\partial^{\nu}}\psi_R)$ & + & \multirow{4}{*}{7}\\[0.15em]
            4 & $ih\widetilde{Z}_{\mu\nu}(\bar{\psi}_L\gamma^{\mu}\overset{\leftrightarrow}{\partial^{\nu}}\psi_L)$ & $+$ & \\[0.15em]
            5 & $h\widetilde{Z}_{\mu\nu}\partial^{\mu}(\bar{\psi}_R \gamma^{\nu}\psi_R)$ & $-$ &  \\[0.15em]
            6 & $h\widetilde{Z}_{\mu\nu}\partial^{\mu}(\bar{\psi}_L \gamma^{\nu}\psi_L)$ & $-$ &  \\[0.15em]
            \hline
            7 & $hZ^{\mu\nu}\bar{\psi}_L\sigma_{\mu\nu}\psi_R$ + h.c & $+$ & \multirow{4}{*}{6}  \\[0.15em]
            8 & $i h Z^{\mu\nu}\bar{\psi}_L\sigma_{\mu\nu}\psi_R$ + h.c & $-$ &  \\[0.15em]
            
            9 & $ihZ^{\mu}(\bar{\psi}_L\overset{\leftrightarrow}{\partial}_{\mu}\psi_R)$ + h.c. & + &  \\[0.15em] 
            
            10 & $hZ^{\mu}(\bar{\psi}_L\overset{\leftrightarrow}{\partial}_{\mu}\psi_R)$ + h.c. & $-$ &   \\[0.15em]
            \cline{1-4}
            \hline
            \hline
             \multicolumn{4}{|c|}{$J = 0$} \\
             \hline
            11 & $hZ^{\mu}\partial_{\mu}(\bar{\psi}_L\psi_R)$ + h.c. & $-$ & \multirow{2}{*}{6} \\[0.15em]
            12 & $ihZ^{\mu}\partial_{\mu}(\bar{\psi}_L\psi_R)$ + h.c. & $+$ & \\[0.15em] 
            \hline
        \end{tabular}
        \caption{}
        \label{tab:4-pt_principals}
    \end{subtable}%
     \caption{Three point functions and principal operators for the $f\bar{f} \to Zh$ process. Here $Q^Z_{f_L}= (T^3_f - Q_f\textrm{ sin}^2\theta_W)$ and $Q^Z_{f_R}= - Q_f\textrm{ sin}^2\theta_W$. }
     \label{tab:ffZh_ops}
\end{table}


Note that in Ref.~\cite{Luty2023} the  Mandelstam descendants were written as a product of the principal amplitude times a series in $s^m t^n$ and not $s^m (t-u)^n$ as in \eq{eq:des}. These two choices are of course entirely equivalent as only two of the three Mandelstam variables---$s,t$ and $u$---are independent. Any two linear combinations of these variables can thus be used for defining the descendants. Our specific choice is motivated by the fact that multiplying the principal amplitudes by powers of $s^m (t-u)^n$ results in  descendant operators that are either hermitian or anti-hermitian with  Wilson coefficients that are, respectively,   purely real or imaginary (see App.~\ref{app:cpt}).  We will call the amplitudes $s^m { P}_k$ as   $s$-descendants of the amplitude  ${P}_k$ and the amplitudes,  $(t-u)^m {P}_k$, as its $(t-u)$-descendants.

Let us now see why it is possible to write the amplitude for any process  in this way. In the case of the Higgs-strahlung process, for instance, the most general form for the contact term amplitude is given by,
\begin{equation}
   \delta\mathcal{M}_{f\bar{f}\to Zh} = \sum_i C_i\, \bar{v}(p_2)\, \Gamma_i^{\mu}\, u(p_1)\,\epsilon^*_{\mu}(p_3)  
\end{equation}
where $u(p_1)$ and $\bar{v}(p_2)$ are the spinors corresponding to the incoming fermion and anti-fermion respectively. Here, $\Gamma_i^{\mu}$ is a four vector formed by combining the external momenta $p_1^{\mu}$, $p_2^{\mu}$, $p_3^{\mu}$, the Dirac matrices $\gamma^{\mu}$, $\gamma^5$, $\sigma^{\mu\nu} = i[\gamma^{\mu}, \gamma^{\nu}]/4$ and the Levi-Civita tensor.   To derive the principal amplitudes, we   systematically enumerate all possible $\Gamma_i^\mu$  that do not contain powers of $p_i.p_j$, as these can be expressed in terms of $s, t$ and $u$. It is clear that that this leaves only a finite number of possibilities; for the $f \bar{f} \to Zh$ process we obtain 32 terms (see  Ref.~\cite{Luty2023} where all these have been listed explicitly), 
\begin{align}
   \sum_i C_i\, \Gamma_i ^{\mu} &= C_1\,\gamma^{\mu} + C_2\, \gamma^{\mu}\gamma^5+ C_3 \,p_1^{\mu} + C_4\, p_2^{\mu} + + C_5\, p_1^{\mu}\gamma^5 \nonumber\\
    & + C_6 \,p_2^{\mu}\gamma^5 + C_7 \,\sigma^{\mu\nu} p_{3\nu} + \epsilon^{\mu\nu\alpha\beta} \sigma_{\nu\alpha} (C_8\, p_{1\beta} + C_9\, p_{2\beta} + C_{10}\,p_{3\beta}) \nonumber\\
    &+ \dots \;\; \dots  \;\; + \epsilon^{\alpha\beta\rho\delta} \gamma_{\alpha}\, p_{1\beta}\,p_{2\rho}\,p_{3\delta} (C_{29}\,p_1^{\mu} + C_{30}\,p_2^{\mu} + C_{31}\,p_1^{\mu}\gamma^5 + C_{32}\,p_2^{\mu}\gamma^5)    \label{eq:list_of_Gmu}.
\end{align}
There are, however, still redundancies in the above list that must be removed in order to arrive at a minimal set of independent principal amplitudes. For the Higgs-stahlung case this process finally leads to the 12 principal operators listed in Table~\ref{tab:ffZh_ops}.  

From our point of view, the most remarkable thing about \eq{eq:amp} and \eq{eq:des} is that these equations  present the $f\bar{f} \to Zh$ amplitude to all orders in the EFT expansion. In fact, as shown in Ref.~\cite{Luty2023}, the number of independent Wilson coefficients  at a given order obtained this way matches the number obtained using Hilbert series~\cite{Henning1, Henning2} methods.
The effective Lagrangian corresponding to the EFT expansion in \eq{eq:des} can be written as follows, 
\begin{eqnarray}\label{ccll}
    \mathcal{L}_{\rm CCLL} &=&  {\cal L}_{SM}+\sum_i \frac{g}{\ctw}\frac{C^{3p}_{i}}{\Lambda^{d_i-4}} O^{3-point}_i +\sum_{k=1}^{N_p}\sum_{m,n=0}^{\infty}\frac{g^2}{\ctw^2}\frac{C^k_{(m,n)}}{\Lambda^{d_k-4+2(m+n)}} O^k_{(m,n)}
\end{eqnarray}
where the first term is the SM lagrangian; the second term contains all the 3-point vertices;  and $O^k_{(m,n)}$ is the operator  corresponding to the  amplitude $s^m (t-u)^n {P}_k$. While we have provided the explicit form of only the principal operators, i.e. only for the $O^k_{(0,0)}$, it is straightforward to write all the  descendant operators in \eq{ccll}  in terms of the principal operators explicitly. For instance the operators corresponding to descendants of the  first principal amplitude, $P_1$, are given by,
\begin{align}
    {O}^3_{m,0} &= hZ_{\mu} \Box_f^m (\bar{\psi}_R\gamma^{\mu}\psi_R) \\
 {O}^3_{0,n} &= (-i)^n\bar{\psi}_R \gamma^{\mu}\,\Box^n(\psi_R\,h) Z_{\mu} + h.c. \\
    {O}^3_{m,n} &=  (-i)^n\Box_f^m [\bar{\psi}_R \gamma^{\mu} \Box^n(\psi_R\,h)] Z_{\mu}+h.c.
\end{align}
where $\Box_f$ refers to the d'Alembertian operator acting only on the fermionic fields.  It can be explicitly checked that  ${O}^3_{m,n}$ gives the Feynman rule, $s^m (t-u)^n P_1$. 

Note that while the lagrangian in \eq{ccll} will be our starting point, it is still not the most convenient EFT basis  to map  partial wave coefficients to Wilson coefficients. Thus we will finally use a different EFT basis to be introduced in the next section. We will call the basis in \eq{ccll} the CCLL basis in order to distinguish it from the final basis.

\section{Mapping  partial waves to operators}

\label{sec3}

The amplitude for the scattering process, $ab \to cd$, can be decomposed into partial waves as follows~\cite{jacob}, 
\begin{equation}\label{par_wave}
\mathcal{M}_{\lambda_a\lambda_b;\lambda_c\lambda_d}(s,\Theta) = 4\pi \sum_{J=J_{min}(\sigma, \lambda)}^\infty(2J + 1) d^{J}_{\sigma,\, \lambda}(\Theta) \,a^J_{\lambda_a\lambda_b;\lambda_c\lambda_d}(s).
\end{equation}
Here, $\lambda_i$ denotes the helicity of each particle and $J$ is the total angular momentum for a given partial wave; $d^{J}_{\sigma\lambda}(\Theta)$ corresponds to the Wigner-$d$ matrix with $\sigma = \lambda_a - \lambda_b$, $\lambda = \lambda_c - \lambda_d$; $a^J_{\lambda_a\lambda_b;\lambda_c\lambda_d}(s)$ is the partial wave amplitude; $s$ and $\Theta$ are, respectively, the  total energy squared and the scattering angle in the centre of mass frame. Our analysis will be focussed on the process $f \bar{f} \to Zh$, where all the initial and final particles, $a,b,c,$ and $d$, are distinct.   As the helicity of the Higgs boson is fixed to zero, we will write the amplitude as, $\mathcal{M}_{\lambda_f \lambda_{\bar{f}}; \lambda_Z}$ and the partial wave coefficients as, $a^J_{\lambda_f \lambda_{\bar{f}}; \lambda_Z}$.  Our methodology can   be easily extended to other processes involving the scattering of distinct particles such as, $f \bar{f} \to  W^\pm Z,\; W^+W^-$.
 
For a given helicity channel, $\lambda_a\, \lambda_b \to \lambda_c\, \lambda_d$,  all the possible $J$ are not allowed  as angular momentum conservation requires,
  \bea\label{jmin}
  J \geq J_{min}\equiv \max (|\sigma|, |\lambda|).
  \eea
The number of  helicity channels is given by the number of  possible choices,  $\{\lambda_a, \lambda_b, \lambda_c, \lambda_d\}$, i.e.,
\bea
N_\lambda=\prod_i (2 J_i +1),
\eea
$J_i$, being the spin of the $i$-th particle. For the $f\bar{f} \to Zh$ process the number of different helicity channels is $N_\lambda=12$, of which 2 (10) helicity channels have $J_{min}=0$ ($J_{min}=1$). 

In general, the partial wave coefficients, $a^J_{\lambda_a\lambda_b;\lambda_c\lambda_d}$, corresponding to the $N_\lambda$ helicity channels are independent complex quantities representing 2$N_\lambda$ variables. As we are considering the tree-level scattering amplitude at energies, $m_{h,W,Z}^2 < s\ll \Lambda^2$, the amplitude has no absorptive part. We show in App.~\ref{app:cpt} that one can then use unitarity and CPT invariance to derive   the following relationship  between partial wave coefficients in different helicity channels,
\begin{equation}
    a^J_{\lambda_1\;\lambda_2;\;\lambda_{z}}(s)
    = (-1)^{J-m}
    \left(a^J_{-\lambda_2~-\lambda_1;-\lambda_{z}}(s)\right)^{\ast},
    \label{cpt}
    \end{equation}
where  $m = \lambda_1 - \lambda_{2}$. This relation implies that there can be, at most, $N_\lambda$ independent partial wave coefficients for tree-level EFT amplitudes.



Our primary goal in this section is to  derive  the mapping between   the partial wave expansion and the EFT expansion. Before going into the details, let us give a broad outline of our strategy and list the   steps we  take to arrive at this mapping:
\begin{enumerate}
    \item First, we  introduce fixed-$J$ operators that only contribute to partial wave coefficients, $a^J_{\lambda_a\lambda_b;\lambda_c\lambda_d}$, with a single  $J$. We show that for a  processes where $a,b,c$ and $d$, are  distinct, the contributing operators can be written as a sum of fixed $J$ operators, 
     \bea
O_{i}= \sum_{J=0}^{J_{max} }\alpha_{(i|J)} O_{(i|J)}
\label{Jdec2}
\eea
where the $O_{(i|J)}$ are fixed $J$ operators, the $\alpha_{(i|J)}$ are numerical coefficients  and $J_{max}$ is the maximum value of $J$  the operator, $O_{i}$, contributes to. 
    

    \item We   then define our EFT basis. We  start from the principal and descendant operators in \eq{ccll} and then expand them as sum of fixed-$J$ operators as in \eq{Jdec2}. Our EFT basis is defined by keeping   only the largest $J$ component for each operator, i.e. by trading $O_{i} \to O_{(i|J_{max})}$.  We show that in our EFT  basis  only the three point vertices, the principal operators and their $s$-descendants contribute to  partial wave coefficients with $J=J_{min}$ (see \eq{jmin}). 
    


    \item Next we present the  contribution of principal operators and their $s$-descendants  to the $N_\lambda$ independent partial wave coefficients with $J=J_{min}$. We  show that in the high energy limit, $s\gg m^2_{h,W,Z}$ the principal operators---along with their $s$-descendants---become independent directions in the  vector space spanned by the $N_\lambda$ independent partial wave coefficients. We  also show that in this limit, the number of   principal operators must equal the number of helicity channels, i.e. $N_p=N_\lambda$.

    
\item We find that for a given $J$,  the EFT at 
 lower orders predicts that certain linear combinations of the $N_\lambda$ independent partial wave coefficients must vanish, i.e. it predicts correlations between partial wave coefficients in addition to those implied by  \eq{cpt}. We call these linear combinations of partial wave coefficients, \textit{helicity directions}. We map each helicity direction   to the higher dimensional operators that break the particular EFT-predicted correlation. For $J=J_{min}$, we call these  linear combinations \textit{principal directions} and show that they  can be mapped to    principal operators  followed by their $s$-descendants.

 \item We show that these helicity directions can be expressed in terms of the Wilson coefficients to all orders of the EFT expansion in a straightforward way. The helicity directions  then act as the bridge between  the Wilson coefficients and the angular observables described in Sec.~\ref{sec5}. Thus they allow us to  map the latter  to the former.


\end{enumerate}  
We now discuss the  above steps  in more detail.

\subsection{Fixed $J$-operators and choice of EFT basis}
\label{sec:proj}
\paragraph{Fixed $J$ decomposition} For a scattering process $a b \to c d$ where $a,b,c,d$ are all distinct, it is possible to write any contact term operator contributing to the amplitude as a product of two tensors, 
\bea
O_i=\left[ ab\right]_{\mu \nu \rho...} \left[ cd\right]^{\mu \nu \rho...}
\label{eg}
\eea
where the first block, $\left[ ab\right]_{\mu \nu \rho...}$
contains the fields of the initial state particles and  the second block contains  the fields of the final state particles. For instance, for the operator $h Z_\mu \bar{f} \gamma^\mu f$ contributing to the process, $f\bar{f} \to hZ$, these blocks are given by,
\bea
\left[ab\right]^\mu&=& \bar{f} \gamma^\mu f \nonumber\\
\left[ cd\right]_\mu &=& h Z_\mu.
\eea

We want to show that any such operator can be written as a sum of operators, each contributing  to a fixed $J$. To this end we introduce projection operators that act on  a rank $n$ tensor, $T_{\mu_1 \mu_2..\mu_n}$, 
\bea
{\cal T}^J_{\nu_1 \nu_2..\nu_n} =({\cal P}^{i,J})^{\mu_1 \mu_2..\mu_n}_{\nu_1 \nu_2..\nu_n} T_{\mu_1 \mu_2..\mu_n}
\label{eq:PT}
\eea
such that the resulting tensor ${\cal T}^J_{\nu_1 \nu_2..\nu_n}$ transforms under the Lorentz group with a fixed $J$. The index $i$ has been introduced as there may be more than one projection operator for a given $J$.  These projection operators have the usual properties,
\bea
\sum_k {\cal P}^k=\mathbb{I}~~~~~~~{\cal P}^{l}{\cal P}^{m}=\delta^{lm} {\cal P}^l
\eea
where we have suppressed the Lorentz indices above.  We show in App.~\ref{app:proj} that for any given $n$, such projection operators always exist and  discuss their explicit construction. For  rank 2 tensors, the projection operators are given by \eq{Eq.App.B4} to \eq{Eq.App.B9} in  App.~\ref{app:proj}. For our work we will, in fact, need only the projection operator that projects a rank $n$ tensor to its highest $J$ mode, i.e. to the $J=n$ mode; we present in App.~\ref{app:proj} a systematic procedure to explicitly construct the $J=n$ projection operators for any $n$.

Let us go back to the operator in \eq{eg} and insert the factor $\sum_{l,J} ({\cal P}^{l,J})^2= \mathbf{I}$ between the two blocks; this gives,
\bea
O_i&=&\left[ ab\right]_{\mu \nu \rho...}\left(\sum_{l,J}({\cal P}^{l,J})^{\mu \nu \rho...}_{\alpha \beta \gamma...} ({\cal P}^{l,J})^{\alpha \beta \gamma...}_{\mu' \nu' \rho'...}\right)   \left[ cd\right]^{\mu' \nu' \rho'...}\nonumber\\
&=& \sum _J \alpha_{(i|J)}\, O_{(i|J)},
\label{Jdec}
\eea
where the projection operators acting on the right and left blocks  allow us to write $O_i$ as a sum of fixed $J$ operators in the second line,  $\alpha_{(i|J)}$ being  numerical coefficients. 




\begin{table}[t]
    \centering
    \renewcommand{\arraystretch}{1.5}
    \begin{tabular}{|c|c|c|c|}
    \hline
    $i$ & $\mathcal{O}^{k}_{(0,1)}$ & CP & $d_{\mathcal{O}}$  \\[0.25em]
    \hline
    \multicolumn{4}{|c|}{$J = 2$} \\
    \hline
     1 & $i(\partial^{\mu}hZ^{\nu}+\partial^{\nu}hZ^{\mu})(\overline{\psi}_R\gamma_{\mu}\overset{\leftrightarrow}{\partial}_{\nu}\psi_R)/2$ & $-$ & \multirow{2}{*}{$7$}\\
     2 & $i(\partial^{\mu}hZ^{\nu}+\partial^{\nu}hZ^{\mu})(\overline{\psi}_L\gamma_{\mu}\overset{\leftrightarrow}{\partial}_{\nu}\psi_L)/2$ & $-$ & \\
     \hline
     3 & $\partial^{\rho}h\tilde{Z}^{\mu\nu}(\overline{\psi}_R\gamma_{\mu}\overset{\leftrightarrow}{\partial}_{\nu}\overset{\leftrightarrow}{\partial}_{\rho}\psi_R \;+(\partial_{\rho}\partial_{\mu}+\eta_{\rho\mu}\Box)(\overline{\psi}_R\gamma_{\nu}\psi_R)/2 ) $ & $-$ & \multirow{4}{*}{$9$} \\[0.15em] 
     4 & $\partial^{\rho}h\tilde{Z}^{\mu\nu}(\overline{\psi}_L\gamma_{\mu}\overset{\leftrightarrow}{\partial}_{\nu}\overset{\leftrightarrow}{\partial}_{\rho}\psi_L \;+(\partial_{\rho}\partial_{\mu}+\eta_{\rho\mu}\Box)(\overline{\psi}_L\gamma_{\nu}\psi_L)/2 ) $ & $-$ &  \\[0.15em] 
     5 & $i(\partial^{\rho}h\tilde{Z}^{\nu\mu}+\partial^{\nu}h\tilde{Z}^{\rho\mu})\partial_{\mu}(\overline{\psi}_R\gamma_{\nu}\overset{\leftrightarrow}{\partial}_{\rho}\psi_R)/2$ & $+$ &  \\ [0.15em]
     6 & $i(\partial^{\rho}h\tilde{Z}^{\nu\mu}+\partial^{\nu}h\tilde{Z}^{\rho\mu})\partial_{\mu}(\overline{\psi}_L\gamma_{\nu}\overset{\leftrightarrow}{\partial}_{\rho}\psi_L)/2$  & $+$ & \\[0.15em] 
     \hline
     7 & $i\partial^{\rho}hZ^{\mu\nu}(\overline{\psi}_L\sigma_{\mu\nu}\overset{\leftrightarrow}{\partial}_{\rho}\psi_R-2\eta_{\rho\mu}(\overline{\psi}_L\sigma^{\alpha}{}_{\nu}\overset{\leftrightarrow}{\partial}_{\alpha}\psi_R)/3)$ + h.c. & $-$ & \multirow{4}{*}{8} \\[0.15em]
     8 & $\partial^{\rho}hZ^{\mu\nu}(\overline{\psi}_L\sigma_{\mu\nu}\overset{\leftrightarrow}{\partial}_{\rho}\psi_R-2\eta_{\rho\mu}(\overline{\psi}_L\sigma^{\alpha}{}_{\nu}\overset{\leftrightarrow}{\partial}_{\alpha}\psi_R)/3)$ + h.c. & $+$ & \\[0.15em]
    
     9 & $\partial^{\nu}hZ^{\mu}(\overline{\psi}_L\overset{\leftrightarrow}{\partial}_{\mu}\overset{\leftrightarrow}{\partial}_{\nu}\psi_R-(\partial_{\nu}\partial_{\mu}+\eta_{\nu\mu}\Box)(\overline{\psi}_L\psi_R)/3)$ + h.c. & $-$ &  \\[0.15em] 
     10 & $i\partial^{\nu}hZ^{\mu}(\overline{\psi}_L\overset{\leftrightarrow}{\partial}_{\mu}\overset{\leftrightarrow}{\partial}_{\nu}\psi_R-(\partial_{\nu}\partial_{\mu}+\eta_{\nu\mu}\Box)(\overline{\psi}_L\psi_R)/3)$ + h.c. & $+$ & \\[0.15em] 
     \hline
     \hline
     \multicolumn{4}{|c|}{$J = 1$} \\[0.15em]
     \hline
     11 & $i\partial_{\mu}h\partial^{\nu}Z^{\mu}(\overline{\psi}_L\overset{\leftrightarrow}{\partial}_{\nu}\psi_R)$ + h.c. & $+$ & \multirow{2}{*}{$8$} \\[0.15em] 
     12 & $\partial_{\mu}h\partial^{\nu}Z^{\mu}(\overline{\psi}_L\overset{\leftrightarrow}{\partial}_{\nu}\psi_R)$ + h.c. & $-$ &  \\[0.15em] 
     \hline
     \end{tabular}
     \caption{The first $J$-descendant of the principal operators for the $f\bar{f} \to Zh$ process presented in Table~\ref{tab:ffZh_ops}.}
     \label{tab:ffZh_Jdesc_ops}
\end{table}
\paragraph{Definition of EFT basis}

We now introduce the EFT basis that we will use. Our starting point will be   the EFT expansion in terms of principals and descendants  in \eq{ccll}. We want to go to a fixed $J$ basis in order to simplify the mapping of EFT Wilson coefficients to partial waves. First of all the three point corrections in the second term of \eq{ccll} already contribute to a fixed value of $J$ given by the spin of the particle being exchanged. As far as the principals and descendants in the last term of \eq{ccll} are concerned, we expand each of them as a sum of fixed $J$ operators by  inserting projection operators as in \eq{Jdec},
\bea
O^k_{(m,n)}= \sum_{J=0}^{J_{max} }\alpha^k_{(m~n|J)}\, O^k_{(m~n|J)}
\label{Jdec3}
\eea
where  $J_{max}$ is the maximum value of $J$ the operator contributes to. Our EFT basis is defined by trading all operators in \eq{ccll} for their maximum $J$ component in \eq{Jdec3}, i.e., $O^k_{(m,n)} \to { O}^k_{(m~n|J_{max} (n,k))}$. This gives,
\begin{eqnarray}\label{basisdef}
    \mathcal{L} &=&  {\cal L}_{SM}+\sum_i\frac{g}{\ctw}\frac{c^{3p}_{i}}{\Lambda^{d_i-4}} {\cal O}^{3-point}_i +\sum_{k=1}^{N_p}\sum_{m,n=0}^{\infty}\frac{g^2}{\ctw^2}\frac{c^k_{(m,n)}}{\Lambda^{d_k-4+2(m+n)}} {\cal O}^k_{(m, n)}
\end{eqnarray}
where, ${\cal O}^k_{(m, n)}={ O}^k_{(m~n| J_{max}(n,k))}$, is the maximum $J$ component of the operator ${O}^k_{(m, n)}$ from the CCLL basis in \eq{ccll}.  The operator ${\cal O}^k_{(m, n)}$ contributes only to partial wave coefficients with, $J=J_k+n$. 

\begin{figure}
    \centering
    \includegraphics[width=0.8\linewidth]{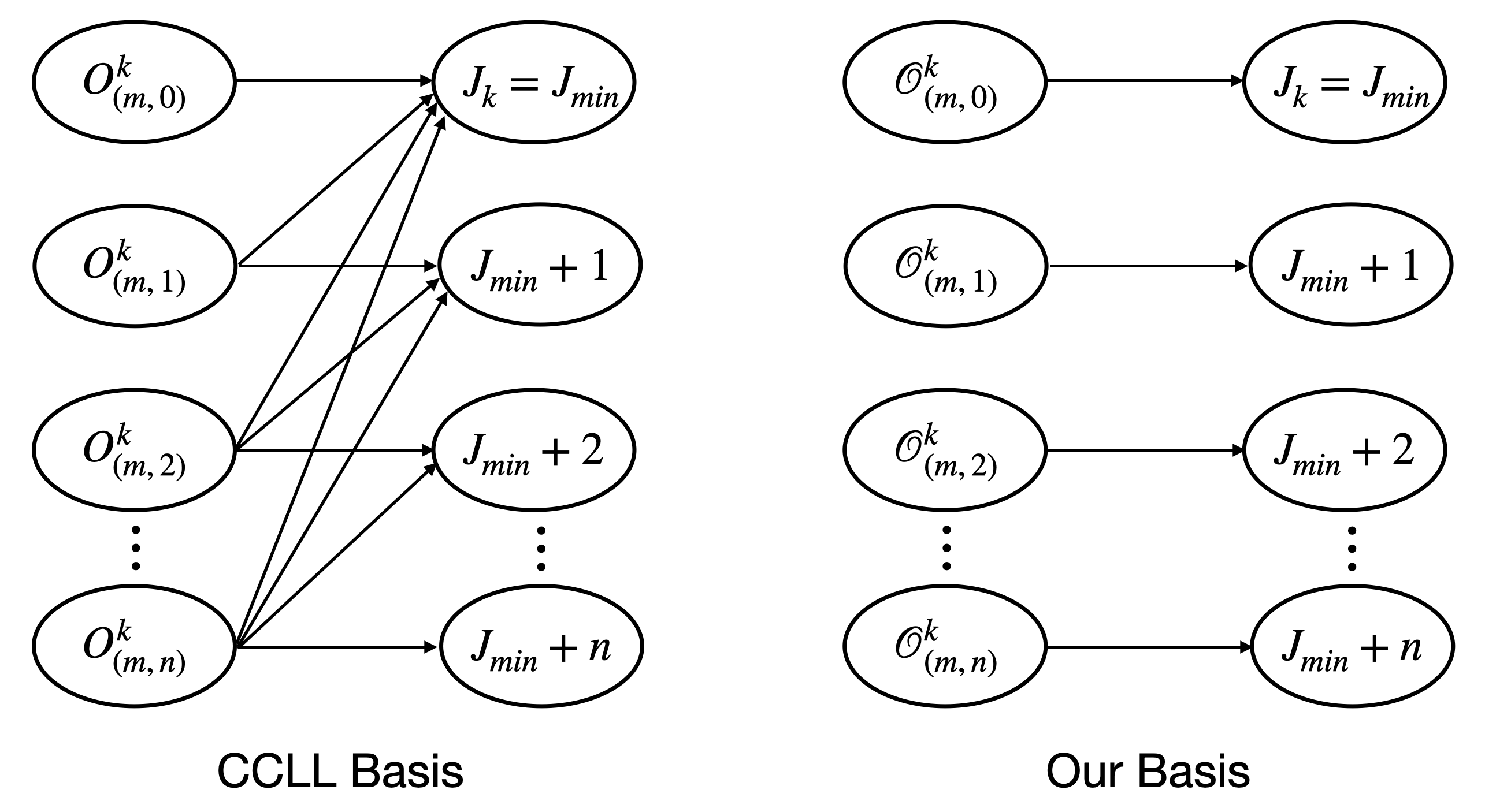}
    \caption{Mapping between   partial waves and EFT contact terms in the CCLL basis of \eq{ccll},  and in our EFT basis defined in \eq{basisdef}. Our basis in  \eq{basisdef} has been designed to simplify this mapping (see text for more details). }
    \label{fig:one}
\end{figure}

We have checked that the principal operators shown in  Table~\ref{tab:ffZh_ops} already contribute to partial waves with a fixed $J$ value.  Thus  the principals  and their $s$-descendants  remain unchanged by this procedure.\footnote{Note that this is not always the case for the principal operators  for other processes presented in Ref.~\cite{Luty2023}. For instance, in the case of the $f\bar{f} \to WW/WZ$ processes  a few of the principal operators contribute to more than a single $J$.  In such cases, one must extract the maximum $J$ component of the principals following the procedure discussed here.} As far as the $(t-u)$-descendants  in \eq{ccll} are concerned, the operator ${ O}^k_{(m,n)}$ is traded for an operator which contributes only to $J= J_{k}+n$ partial waves, $J_{k}$ being the single $J$ to which the principal operator ${\cal O}^k_{(0,0)}$ contributes. This is because each power of $(t-u)$ results in a factor of $\cos \Theta$  so that the amplitude for the operator ${\cal O}^k_{(m,n)}$ is proportional to, 
\bea
(\cos^n \Theta) s^{m+n} {P}_k.
\eea
 The maximum value of $J$ the above amplitude contributes to in the partial wave expansion would be $J_k+n$ because of the $\cos^n \Theta$  factor. Thus each principal operator/$s$-descendant, ${\cal O}^k_{(m,0)}$, will be associated to an infinite tower of fixed $J$ operators, ${\cal O}^k_{(m,n)}$, with $J= J_{k}+n$ and $n$ ranging from 1 to infinity. We would call the operator, ${\cal O}^k_{(m,n)}$,  the $n$-th $J$-descendant of ${\cal O}^k_{(m,0)}$.   Using the  procedure presented in App.~\ref{app:proj} (see in particular, \eq{Eq.App.B11}), we present the first $J$-descendant of the principal operators of Table~\ref{tab:4-pt_principals} in Table~\ref{tab:ffZh_Jdesc_ops}.


By construction,  only three point vertices, principal operators and their $s$-descendants can contribute  to minimum $J$ modes in our basis. To see this, consider for instance the principal operators contributing to  the $f \bar{f} \to Z h$ process in Table~\ref{tab:4-pt_principals}. The  upper (lower) block in Table~\ref{tab:4-pt_principals} shows the principal operators with  $J=1$ ($J=0$). For the $f\bar{f} \to Zh$ process, $J_{min}=0$ for the two helicity channels, $(++;0)$ and $(--;0)$  (see \eq{jmin}) and the rest of the 10 helicity channels have $J_{min}=1$. It is clear from their definition that none of the  $J$-descendants can  contribute to helicity channels for which $J_{min}=0$. Now consider the helicity channels   for which $J_{min}=1$.  For these modes, while the  $J$-descendants of the $J= 1$ principals definitely do not contribute, contributions from  the   $J$-descendants of the $J=0$ principals are possible  as they still have a $J=1$ component. The $J=0$ principal operators, however, have a vanishing contribution to the   $J_{min}=1$ helicity channels (see  Table~\ref{tab:LR_amp_J1}  in App.~\ref{lrrl}) so that their $(t-u)$-descendants---that have the same amplitude as the principals up to a factor of $\cos^n \Theta$---would also  not contribute to these modes. This in turn implies that the $J$-descendants   arising from these $(t-u)$-descendants also do not contribute to these modes.

While the above discussion has been focussed on the $f\bar{f} \to Zh$ process, the above arguments are in fact completely general as long as the four particles involved in the scattering are distinct from each other. We can therefore conclude that for any such process,  minimum $J$ modes receive contributions only from   three-point vertices, principal operators and their $s$-descendants provided the EFT  basis is constructed by following the steps discussed here. 


The mapping of partial waves to EFT operators, in particular to the contact term operators,   is much more straightforward in our basis than in the CCLL basis of \eq{ccll}. This is shown schematically in Fig.~\ref{fig:one} which shows  that, in our basis, the only contact term contributions  the $J=J_{min}$ partial wave coefficients receive are from the $N_p$ principals and their $s$-descendants. For higher values, $J=J_{min}+n$, the partial wave coefficients receive contributions only from  the $N_p$ $J$-descendants of the principal operators, ${\cal O}^k_{(0,n)}$,  and their $s$-descendants,  ${\cal O}^k_{(m,n)}$. In the basis of \eq{ccll}, however, partial wave coefficients of a fixed $J$ receive contributions from an infinite number of $(t-u)$-descendants. This is because the $(t-u)$-descendant amplitude, $(t-u)^n P_k \sim (\cos^n \Theta)  P_k$,    contributes to partial waves with $J=J_k, J_k+1, J_k+2,\dots, J_k+n$. 








\subsection{Mapping minimum $J$ partial waves to EFT Wilson coefficients}
\begin{table}[p]
    \begin{adjustbox}{max width=\textwidth}
    \centering
    \begin{tabular}{|c|c|c|c|c|c|c|}
\hline
   EFT & \multicolumn{6}{ c |}{Helicity Amplitudes}\\
    \cline{2-7}
     WCs & $(+-;\,0)$ & $(+-;\,+)$ &$(+-;\,-)$ & $(-+;\,0)$ & $(-+;\,+)$ & $(-+;\,-)$ \\[0.5ex]
     \hline\rule{0pt}{4ex}
    $c^{3p}_1$ & $Q^Z_{f_R}\dfrac{m_Z}{\E} \alpha_0$ & $Q^Z_{f_R}\dfrac{m_Z}{\E} \alpha_0$  & $Q^Z_{f_R}\dfrac{m_Z}{\E} \alpha_0$  & $Q^Z_{f_L}\dfrac{m_Z}{\E} \alpha_0$  & $Q^Z_{f_L}\dfrac{m_Z}{\E} \alpha_0$  & $Q^Z_{f_L}\dfrac{m_Z}{\E} \alpha_0$  \\[2ex] 
    \hline\rule{0pt}{4ex}
    $c^{3p}_2$ & $-Q^Z_{f_R}\dfrac{m_Z}{\E}\alpha_0$ & $-Q^Z_{f_R}\dfrac{m_Z}{\E}\alpha_0$ & $-Q^Z_{f_R}\dfrac{m_Z}{\E}\alpha_0$ & 0 & 0 & 0 \\[2ex]
    \hline\rule{0pt}{4ex}
    $c^{3p}_3$ & 0 & 0 & 0 & $-Q^Z_{f_L}\dfrac{m_Z}{\E}\alpha_0$ & $-Q^Z_{f_L}\dfrac{m_Z}{\E}\alpha_0$ & $-Q^Z_{f_L}\dfrac{m_Z}{\E}\alpha_0$ \\[2ex]
    \hline\rule{0pt}{4ex}
    $c^{3p}_1\,c^{3p}_2$ & $-Q^Z_{f_R}\dfrac{m_Z}{E}\alpha_0$ & $-Q^Z_{f_R}\dfrac{m_Z}{E}\alpha_0$ & $-Q^Z_{f_R}\dfrac{m_Z}{E}\alpha_0$ & 0 & 0 & 0 \\[2ex]
    \hline\rule{0pt}{4ex}
    $c^{3p}_1\,c^{3p}_3$ & 0 & 0 & 0 & $-Q^Z_{f_L}\dfrac{m_Z}{E}\alpha_0$ & $-Q^Z_{f_L}\dfrac{m_Z}{E}\alpha_0$ & $-Q^Z_{f_L}\dfrac{m_Z}{E}\alpha_0$ \\[2ex]
    \hline
    \hline\rule{0pt}{4ex}
    $c^{3p}_4$ & $\dfrac{\E}{\Lambda}\left(\dfrac{8\varepsilon_0}{\alpha_1}Q^Z_{f_R}\alpha_0\right)$ & $\dfrac{\E}{\Lambda}2Q^Z_{f_R}\alpha_1\alpha_0 $ & $\dfrac{\E}{\Lambda}2Q^Z_{f_R}\alpha_1\alpha_0$ & $\dfrac{\E}{\Lambda}\left(\dfrac{8\varepsilon_0}{\alpha_1}Q^Z_{f_L}\alpha_0\right)$ & $\dfrac{\E}{\Lambda}2Q^Z_{f_L}\alpha_1\alpha_0$ & $\dfrac{\E}{\Lambda}2Q^Z_{f_L}\alpha_1\alpha_0$ \\[2.5ex]
    \hline\rule{0pt}{4ex}
    $c^{3p}_5$ & 0 & $-i\dfrac{\E}{\Lambda}2Q^Z_{f_R}\alpha_2\alpha_0$ & $i\dfrac{\E}{\Lambda}2Q^Z_{f_R}\alpha_2\alpha_0$ & 0 & $-i\dfrac{\E}{\Lambda}2Q^Z_{f_L}\alpha_2\alpha_0$ & $i\dfrac{\E}{\Lambda}2Q^Z_{f_L}\alpha_2\alpha_0$ \\[2.5ex]
    \hline\rule{0pt}{4ex}
         
     $c^1_{(0,0)}$ & $-\dfrac{\E}{\Lambda}$ & $-\dfrac{\E}{\Lambda}$ & $-\dfrac{\E}{\Lambda}$ & 0 & 0 & 0 \\[1.8ex]
     \hline\rule{0pt}{4ex}
     $c^2_{(0,0)}$ & 0 & 0 & 0 &  $-\dfrac{\E}{\Lambda}$ & $-\dfrac{\E}{\Lambda}$ & $-\dfrac{\E}{\Lambda}$ \\[1.8ex]
    \hline\rule{0pt}{4ex}
    $c^{3p}_2c^{3p}_4$ & $-\dfrac{E}{\Lambda}\left(\dfrac{8\varepsilon_0}{\alpha_1}Q^Z_{f_R}\alpha_0\right)$ & $-\dfrac{E}{\Lambda}2Q^Z_{f_R}\alpha_1\alpha_0 $ & $-\dfrac{E}{\Lambda}2Q^Z_{f_R}\alpha_1\alpha_0$ & $0$ & $0$ & $0$ \\[2.5ex]
    \hline\rule{0pt}{4ex}
    $c^{3p}_3c^{3p}_4$ & $0$ & $0$ & $0$ & $-\dfrac{E}{\Lambda}\left(\dfrac{8\varepsilon_0}{\alpha_1}Q^Z_{f_L}\alpha_0\right)$ & $-\dfrac{E}{\Lambda}2Q^Z_{f_L}\alpha_1\alpha_0$ & $-\dfrac{E}{\Lambda}2Q^Z_{f_L}\alpha_1\alpha_0$ \\[2.5ex]
    \hline\rule{0pt}{4ex}
    $c^{3p}_2c^{3p}_5$ & 0 & $i\dfrac{E}{\Lambda}2\alpha_2Q^Z_{f_R}\alpha_0$ & $-i\dfrac{E}{\Lambda}2Q^Z_{f_R}\alpha_2\alpha_0$ & 0 & $0$ & $0$ \\[2.5ex]
    \hline\rule{0pt}{4ex}
    $c^{3p}_3c^{3p}_5$ & 0 & $0$ & $0$ & 0 & $i\dfrac{E}{\Lambda}2Q^Z_{f_L}\alpha_2\alpha_0$ & $-i\dfrac{E}{\Lambda}2Q^Z_{f_L}\alpha_2\alpha_0$ \\[2.5ex]
    \hline
     \hline\rule{0pt}{4ex}
      $c^1_{(1,0)}$ & $-\dfrac{\E^3}{\Lambda^3}$ & $-\dfrac{\E^3}{\Lambda^3}$ & $-\dfrac{\E^3}{\Lambda^3}$ & 0 & 0 & 0 \\[1.8ex]
     \hline\rule{0pt}{4ex}
     $c^2_{(1,0)}$ & 0 & 0 & 0 &  $-\dfrac{\E^3}{\Lambda^3}$ & $-\dfrac{\E^3}{\Lambda^3}$ & $-\dfrac{\E^3}{\Lambda^3}$ \\[1.8ex]
     \hline\rule{0pt}{4ex}
     $c^3_{(0,0)}$ & $-\dfrac{\E^3}{\Lambda^3}\left(\dfrac{2\varepsilon_0}{\alpha_1}\right)$ & $-\dfrac{\E^3}{\Lambda^3}\dfrac{\alpha_1}{2}$ & $-\dfrac{\E^3}{\Lambda^3}\dfrac{\alpha_1}{2}$ & 0 & 0 & 0 \\[1.8ex] 
     \hline\rule{0pt}{4ex}
    $c^4_{(0,0)}$ & 0 & 0 & 0 & $\dfrac{\E^3}{\Lambda^3}\left(\dfrac{2\varepsilon_0}{\alpha_1}\right)$ & $\dfrac{\E^3}{\Lambda^3}\dfrac{\alpha_1}{2}$ & $\dfrac{\E^3}{\Lambda^3}\dfrac{\alpha_1}{2}$ \\[1.8ex]
    \hline\rule{0pt}{4ex}
    $c^5_{(0,0)}$ & 0 & $i\dfrac{\E^3}{\Lambda^3}\dfrac{\alpha_2}{2}$ & $-i\dfrac{\E^3}{\Lambda^3}\dfrac{\alpha_2}{2}$ & 0 & 0 & 0 \\[1.8ex]
    \hline\rule{0pt}{4ex}
    $c^6_{(0,0)}$ & 0 & 0 & 0 & 0 & $i\dfrac{\E^3}{\Lambda^3}\dfrac{\alpha_2}{2}$  & $-i\dfrac{\E^3}{\Lambda^3}\dfrac{\alpha_2}{2}$ \\[1.8ex]
    \hline
    \hline\rule{0pt}{3ex}
    $c^{k}_{(m,0)}$ & \multicolumn{6}{ c |}
    {$(E^2/\Lambda^2)^m\;\vec{V}^k_{(0,0)}$} \\[0.5em]
    \hline
    \end{tabular}
    \end{adjustbox}
    \hspace*{-3cm}
    \caption{EFT contributions---up to all orders in the expansion---  to,  $\hat{a}^{J_{min}}_{\lambda_{f} \lambda_{\bar{f}}; \lambda_Z}$, the minimum $J$ partial wave coefficients rescaled by the normalisation factor in \eq{rescale}. These contributions arise from the three point vertices, principal operators (see Table~\ref{tab:3-point_ops} and  \ref{tab:4-pt_principals}) and their $s$-descendants. We show here only the helicity channels with opposite fermion helicities.  Here, $E=\sqrt{s}$ and the factors, $\alpha_1 = \left[1 + (m_Z^2 - m_h^2)/s\right]$, $\alpha_2 = \sqrt{\left[1 - (m_Z + m_h)^2/s\right]\left[1 - (m_Z - m_h)^2/s\right]}$, $\varepsilon_0={m_Z^2}/{s}$, and $\alpha_0 = 1/(1-\varepsilon_0)$, have been defined such that $\varepsilon_0 \to 0$ and   $\alpha_i \to 1$ in the high energy limit $E \gg m_Z,\,m_h$.}
    \label{tab:full_Amps_Exact}
\end{table}

We have shown that in our EFT basis only  3-point corrections, principals and their $s$-descendants contribute to  the partial wave coefficients, $a^{J_{min}}_{\lambda_f \lambda_{\bar{f}}; \lambda_Z}$,  with $J=J_{min}$. We now   provide the explicit map between the Wilson coefficients  and these partial wave coefficients for the $f\bar{f} \to Zh$ process. We will work in the limit of vanishing fermion masses which  will allow us to split the contributions of the operators to the different helicity channels into three `block-diagonal' tables,  Table~\ref{tab:full_Amps_Exact}, Table~\ref{tab:LR_amp_J0} and Table~\ref{tab:LR_amp_J1}. The first table  contains the contributions  from three point couplings and the  $N_p^{LL,RR}=6$ principal operators---with $LL/RR$  fermion chiralities---to the $N_\lambda^{opp}=6$ helicity amplitudes with $J=1$ and opposite fermion helicities (`$+-$' or `$-+$'). The second table, Table~\ref{tab:LR_amp_J0}, contains the contributions of the two $J=0$ principal operators to the two helicity channels with $J_{min}=0$. Finally, Table~\ref{tab:LR_amp_J1} shows the contributions  of 3-point vertices and the $N_p^{LR,RL}=6$ principal operators---with $LR/RL$ chiralities---to the $N_\lambda^{id}=6$ partial wave coefficients  with $J=1$ and identical fermion helicities (`$++$' and `$--$'). We discuss Table~\ref{tab:full_Amps_Exact} in this section and  Tables~\ref{tab:LR_amp_J0} and~\ref{tab:LR_amp_J1} in App~\ref{lrrl}.

 In Table~\ref{tab:full_Amps_Exact},  we have shown the partial wave coefficients  appearing in \eq{par_wave} after rescaling them by  a normalisation factor, 
 \bea \label{rescale}
\hat{a}^{J}_{\lambda_f\lambda_{\bar{f}};\lambda_Z}&=&\frac{a^{J}_{\lambda_f\lambda_{\bar{f}};\lambda_Z}}{{\cal N}_{\lambda_f \lambda_{\bar{f}};\lambda_Z}}. 
\eea
The normalisation factor  is  given by, ${\cal N}_{\lambda_f \lambda_{\bar{f}};\lambda_Z} =(g/\ctw)^2 \epsilon_{\lambda_Z}$,  where, $\epsilon_{\lambda_Z = \pm} = 1$ and   $\epsilon_{\lambda_Z = 0} = E_z/m_Z$, $E_z$ being the energy of the $Z$-boson in the center of mass frame.  As far as 3-point corrections are concerned, we have shown contributions due to diagrams with one and two EFT insertions in separate rows of Table~\ref{tab:full_Amps_Exact}.  

Note that the contact term operators in our basis--- the principals and descendants---are uniquely defined by their  contribution to the different partial wave coefficients as this completely fixes the amplitude by \eq{par_wave}. Thus if one of the rows in Table~\ref{tab:full_Amps_Exact},  corresponding to a contact term contribution, is a linear combination of other rows, the corresponding operator must be  redundant. We can define such a  row, $i$, of Table~\ref{tab:full_Amps_Exact} as a vector, $\vec{V}_i=\left\{v_1, v_2, \dots \right\}$. In a non-redundant EFT basis such as ours, all these vectors must be linearly independent.

Table~\ref{tab:full_Amps_Exact} has been organised into blocks according to the order in $\E/\Lambda$ at which a particular operator contributes  to the amplitude. Here, $\E=\sqrt{s}$, is  the centre of mass energy of the scattering process. We see that  all 3-point contributions---whether involving   single or double EFT insertions---appear by  ${\cal O}(\E^2/\Lambda^2)$.  As far as the contact terms are concerned, the first contribution arises   in the  $E/\Lambda$ block where two principals contribute.  As we go to higher orders,  the number of contact term contributions eventually becomes $N_p^{LL,RR}=6$ because every principal operator, or one of its descendants, contributes at these orders. In Table~\ref{tab:full_Amps_Exact} this occurs for  any order larger than or equal to $ \E^3/\Lambda^3$.

 



The power of writing the EFT expansion in terms of principals and descendants can be seen from the fact that  the EFT contribution  to the $J=J_{min}$ partial waves up to all  orders can be  determined from Table~\ref{tab:full_Amps_Exact}. This is because although Table~\ref{tab:full_Amps_Exact}   explicitly contains only contributions of operators up to ${\cal O}(E^3/\Lambda^3)$, all other operators contributing to $J=J_{min}$ partial modes  are $s$-descendants. Their contribution is, therefore, given by multiplying powers of $s/\Lambda^2$ to the entries  in the $E^3/\Lambda^3$ block. We show this in the last row of Table~\ref{tab:full_Amps_Exact} where we show the contribution of an $s$-descendant  by multiplying a factor of $(s/\Lambda^2)^n=(\E^2/\Lambda^2)^{n}$ to the vector of the corresponding principal.

\paragraph{High-energy limit} The mapping presented in  Table~\ref{tab:full_Amps_Exact} becomes especially simple in the  high energy limit, $s \gg m_{h,W,Z}^2$.  In  this limit, $\varepsilon_0 \to 0$, and all the factors $\alpha_i \to 1$. Thus the entries of the table simply become  ${\cal O}(1)$ numbers times factors of $(\E/\Lambda)^n$, where  $n$ is an integer power that can be fixed by dimensional analysis. Let us now show that in  the high energy limit the number or principals must always equal the number of helicity channels, $N_\lambda$. This is of course  the case for the $f\bar{f} \to Zh$ process but we now show that this must always be the case for scattering processes with non-identical particles.  

First we will argue that the number of principal operators must be greater than or equal to $N_\lambda$, i.e. $N_p \geq N_\lambda$. The relation in \eq{cpt} allows $N_\lambda$ of the partial wave coefficients, of a given $J$,   to be independent after the requirements of unitarity and CPT invariance are imposed. As the general principles of quantum field theory (QFT) impose no further restrictions on the partial wave coefficients, all these $N_\lambda$  partial wave coefficients must in fact become independent at  a sufficiently high EFT order. As the number of operators contributing at higher orders in $E/\Lambda$ is $N_p$, this implies that we must have $N_p \geq N_\lambda$.\footnote{ This fact    can also be seen in a more direct way by rewriting these operators in the HEFT (see in particular, Ref.~\cite{georgi}). For the purpose of counting operators we can go to the limit of vanishing gauge couplings in HEFT. In this limit HEFT becomes a theory of massless gauge bosons and goldstones and each helicity mode in Table~\ref{tab:full_Amps_Exact} and Table~\ref{tab:LR_amp_J1} can be mapped to a scattering between the massless fields, $f_{L,R}$, $Z^{\mu \nu}_{L,R}= Z^{\mu \nu} \pm i\tilde{Z}^{\mu \nu}$ and $\phi_0$, the goldstone boson eaten by the $Z$ boson in the massive theory. One can thus associate  distinct HEFT operators with each helicity channel in Table~\ref{tab:full_Amps_Exact} and Table~\ref{tab:LR_amp_J1}  in this limit. It can also be seen that hermitian operators formed using these fields will automatically satisfy \eq{cpt}.  This  shows that there must be at least $N_\lambda$ principal operators.} 

Next let us show that in the high energy limit we must have, $N_p \leq N_\lambda$. To this end we consider  the ${\cal O}(E^3/\Lambda^3)$ block in Table~\ref{tab:full_Amps_Exact} that contains contributions from each principal operator or one of its $s$-descendants; there are thus $N_p^{LL,RR}$ contributing operators at this order. Let us make a slight change of basis in the partial wave space and consider the following linear combinations of partial wave amplitudes,
\begin{align}
\{\hat{a}^1_{+-;0}\,,\;(\hat{a}^1_{+-;+} + \hat{a}^1_{+-;-}),\;i(\hat{a}^1_{+-;+} - \hat{a}^1_{+-;-}),\;\hat{a}^1_{-+;0}\,,\;(\hat{a}^1_{-+;+} + \hat{a}^1_{-+;-}),\;i(\hat{a}^1_{-+;+} - \hat{a}^1_{-+;-})\} \nonumber
\end{align}
that are all real by \eq{cpt}. Let us represent the   contribution of the $i$-th the operator to the above partial wave coefficients at ${\cal O}(\E^3/\Lambda^3)$  by the vector, $\vec{\hat{V}}_i=\vec{r}_i (\E^3/\Lambda^3)$. Here $\vec{r}_i$ is an  array of, $N_\lambda^{opp}$, real numbers. As we have already discussed, in a non redundant basis,  the  $\vec{r}_i$ corresponding to the  different operators  must be linearly independent. This implies that  that we must have,  $N_p^{LL,LR}\leq N_\lambda^{opp}$ in a non redundant basis.  By a similar argument one can conclude that, $N_p^{LR, RL}\leq N_\lambda^{id}$, for the principal operators with opposite fermion chiralities, see the $(E^4/\Lambda^4)$ block of  Table~\ref{tab:LR_amp_J1}. Together these two inequalities imply $N_p \leq N_\lambda$.  As we have already argued that, $N_p\geq N_\lambda$,  we can  infer that, 
\begin{equation}
N_p= N_\lambda
\end{equation}
in the high energy limit. While we used the $f\bar{f} \to Zh$ case as an example, these arguments are completely general and can be applied to any process with 4 distinct particles. 

In the high energy limit, therefore, the  principal operators can be thought of as   independent directions in the  $N_\lambda$ dimensional vector space formed by the minimum $J$ partial wave coefficients, $a^{J^{min}}_{\lambda_f\lambda_{\bar{f}};\lambda_Z}$. 

The number of principal operators  is exactly, $N_\lambda$, even at low energies  for the $f\bar{f} \to Zh$ process. Incidentally, this is also the case for many other processes, see Ref.~\cite{Luty2023}. In general, however, the number of principal operators can exceed $N_\lambda$. For instance in the $ff \to WW/WZ$ process $N_p=38$ but $N_\lambda=36$. The above arguments imply that of these 38 principal operators only 36 remain independent in the high energy limit. The other 2 are redundant in this limit unless one is sensitive  to ${\cal O}(m^2_{h,w,Z}/s)$ effects.


\paragraph{EFT predictions at lower orders and helicity directions}
We see  from Table~\ref{tab:full_Amps_Exact}  that in the upper blocks, i.e. at lower orders in $\E/\Lambda$, the number of contributing operators is smaller than,  $N^{opp}_\lambda=6$, the maximum number of independent partial wave coefficients with opposite fermion helicities allowed by \eq{cpt}. Similarly for partial waves with identical fermion helicities, we see from  Table~\ref{tab:LR_amp_J0} and~\ref{tab:LR_amp_J1}, that at lower orders in $\E/\Lambda$ the number of EFT contributions is smaller than the maximum number allowed by \eq{cpt}.  Thus, EFT at lower orders  implies either that some partial wave coefficients should vanish or that  there must be correlations between them  in addition to  \eq{cpt}.\footnote{It is interesting to note something similar that happens in  SMEFT. In SMEFT, in addition to $E/\Lambda$, there is another expansion parameter, $g v/\Lambda$. At lower orders in $g v/\Lambda$    correlations exist between HEFT Wilson coefficients/anomalous couplings.  These correlations are broken order by order in $g v/\Lambda$ ~\cite{Gupta:2014rxa,higgsbasis,Karmakar:2023rdt,Karmakar:2024gla,Chakraborty:2024ciu}. }

For instance in  Table~\ref{tab:full_Amps_Exact}, at the lowest order, there are three  operators  contributing to all the 6  independent partial wave coefficients. Of these only two contributions are linearly independent as a linear combination of $c_2^{3p}$ and $c_3^{3p}$ gives a contribution proportional to $c_1^{3p}$. This implies that, at this order, there must be 4 correlations between the partial wave coefficients. These are given by,
\bea
&& \hat{a}^1_{+-;+} - \hata^1_{+-;0} = 0\;,\quad \hata^1_{+-;-} - \hata^1_{+-;0} = 0 \quad \nonumber\\
&& \hata^1_{-+;+} - \hata^1_{-+;0} = 0\;,\quad \hata^1_{-+;-} - \hata^1_{-+;0} = 0,
\label{corr1}
\eea
where we have expressed these correlations in terms of the scaled partial wave coefficients defined in \eq{rescale}. Note that the above conditions might seem like 8 correlations as the partial wave coefficients are complex. If we impose \eq{cpt}, however, \eq{corr1} contains only 4 new correlations.

As shown  in the second  block of Table~\ref{tab:full_Amps_Exact}, the above linear combinations  receive contributions at  ${\cal O}(\E/\Lambda)$  where all the correlations in \eq{corr1} are broken. At this order the number of independent partial wave coefficients with opposite fermion helicities becomes, $N_\lambda^{opp}=6$, that is the maximum  value allowed by \eq{cpt}.  We can thus  define the following linear combinations of  partial wave coefficients, 
\begin{align}
    \textbf{a}_{1,1} &= (1+\varepsilon_2) \hata^1_{+-;0} + (1+\varepsilon_2) \hata^1_{-+;0} - \frac{\varepsilon_2}{2}(\hata^1_{+-;+} + \hata^1_{+-;-} + \hata^1_{-+;+} + \hata^1_{-+;-})\nn \\
    \textbf{a}_{1,2} &= (1+\varepsilon_2)\hata^1_{+-;0} - (1+\varepsilon_2)\hata^1_{-+;0} - \frac{\varepsilon_2}{2}(\hata^1_{+-;+} + \hata^1_{+-;-} - \hata^1_{-+;+} - \hata^1_{-+;-}) \nn\\
  \delta\textbf{a}_{1,3} &= \frac{1}{\alpha_3}(\hata^1_{+-;+} + \hata^1_{+-;-} - 2\hata^1_{+-;0})\nn \\
    \delta\textbf{a}_{1,4} &= \frac{i}{\alpha_2}(\hata^1_{+-;+} - \hata^1_{+-;-})\nn \\
    \delta\textbf{a}_{1,5} &= -\frac{1}{\alpha_3}\left[Q^Z_{f_L}(\hata^1_{+-;+} + \hata^1_{+-;-} - 2\hata^1_{+-;0}) - Q^Z_{f_R}(\hata^1_{-+;+} + \hata^1_{-+;-} - 2\hata^1_{-+;0})\right] \nn \\
    \delta\textbf{a}_{1,6} &= -\frac{i}{\alpha_2}\left[Q^Z_{f_L}(\hata^1_{+-;+} - \hata^1_{+-;-}) + Q^Z_{f_R}(\hata^1_{-+;-} - \hata^1_{-+;+})\right]
\label{pdirections}
\end{align}
that get non-vanishing contributions at different orders in $\E/\Lambda$. Here,
\begin{eqnarray}
    \varepsilon_1 &=&\dfrac{4\varepsilon_0}{\alpha_1}\nonumber~~~~~~~\alpha_3 =\left(\alpha_1 - \dfrac{4\varepsilon_0}{\alpha_1}\right)~~~~~~~ \varepsilon_2 = \varepsilon_1/\alpha_3,
\end{eqnarray}  
have been again defined such that $\alpha_i \to1$ and $\epsilon_i\to 0$ in the high energy limit. In the above equation  $\delta^p \textbf{a}_{J,k}$ denotes a linear combination of partial wave coefficients that gets its first EFT contribution at ${\cal O}({(\E/\Lambda)}^p)$, where the first subscript refers to the $J$ value of the partial wave coefficients; for $p=0$ we have omitted the prefix $\delta^0$. In the above equations, while  $\textbf{a}_{1,1}$ and $\textbf{a}_{1,2}$ gets contributions already at  ${\cal O}(1)$, the $\delta\textbf{a}_{1,3-6}$  get their lowest order contributions at ${\cal O}(\E/\Lambda)$.  
Note that due to the presence of the  of factors of, $\alpha_i$ and $\epsilon_i$, in their definition, the linear combinations above are energy dependent, i.e. they change depending on the energy bin under consideration. These linear combinations are also not unique, for instance any linear combination of $\delta \mathbf{a}_{1,3}$, $\delta \mathbf{a}_{1,4}$, $\delta \mathbf{a}_{1,5}$ and $\delta\mathbf{a}_{1,6}$ will    also gets its lowest order EFT contribution at ${\cal O}(\E/\Lambda)$. The particular choice of linear combinations above has been made to ensure that they are real (see \eq{cpt}) and that they can be mapped to Wilson coefficients in a straightforward way.

We present this mapping in Table~\ref{tab:pdir}. We see from Table~\ref{tab:pdir}, that the principal directions can be  written to all orders in the EFT expansion as,\footnote{Note that this equation applies only for $J=1$ and  $k=1$-6. In App.~\ref{lrrl} we define helicity directions, $\delta^p\textbf{a}_{1,k}$,  with $k=6$-12 to which \eq{pow} does not apply.}
\bea
\delta^p\textbf{a}_{J,k}= \delta^p {\mathbf{\hat{a}}}_{J,k}^{0}\left(\frac{E}{\Lambda}\right)^p \eta_k +\sum_{m=m_k}^\infty \delta^p {\textbf{a}}_{J,k}^{2m+1-p}\left(\frac{E}{\Lambda}\right)^{2 m+1}
\label{pow}
\eea
where $\eta_k=1$ ($\eta_k=0$ )  for $k=1-4$ ($k=5,6$), $m_{1,2}=0$ and $m_{3,4,5,6}=1$. The $\delta^p {\mathbf{\hat{a}}}_{J,k}^{0}$ and $\delta^p {\textbf{a}}_{J,k}^{m}$ can be read off from Table~\ref{tab:pdir}.  While the  $\delta^p {\textbf{a}}_{J,k}^{m}$ are all independent of energy, the  $\delta^p {\mathbf{\hat{a}}}_{J,k}^{0}$ are not. This is because the first term above contains corrections due to 3-point functions for $k=1-4$.  The energy dependence can be easily factored out by defining, 
 \bea
 \delta^p {\mathbf{\hat{a}}}_{J,k}^{0}&=&\delta^p {\mathbf{{a}}}_{J,k}^{0} \alpha_0\frac{m_Z}{E}~~~~{\rm for~} k=1,2 \nonumber\\
 \delta^p {\mathbf{\hat{a}}}_{J,k}^{0}&=&\delta^p {\mathbf{{a}}}_{J,k}^{0} \alpha_0~~~~~~~~~{\rm for~} k=3,4,
  \eea
  where the  $\delta^p {\mathbf{{a}}}_{J,k}^{0}$ are now  independent of energy. 
\begin{table}[h]
 \begin{adjustbox}{max width=\textwidth}
    \centering
    \begin{tabular}{|c|}
\hline
Principal directions to all orders in EFT\\
\hline
       \hline\rule{0pt}{5ex}
         $\begin{aligned}
             \textbf{a}_{1,1} = \dfrac{m_Z}{E}\left[c^{3p}_1(\Ql+\Qr) - (1+c^{3p}_1)(\Qr \,c^{3p}_2 + \Ql c^{3p}_3)\right]\alpha_0 - \sum_{m=0}^{\infty}\left(\dfrac{E}{\Lambda}\right)^{2m+1}(c^1_{(m,0)}+c^2_{(m,0)})
         \end{aligned}$ \\[0.5cm]
         $\begin{aligned}
             \textbf{a}_{1,2} = \dfrac{m_Z}{E}\left[c^{3p}_1(\Qr - \Ql) - (1 + c^{3p}_1)(\Qr c^{3p}_2 - \Ql c^{3p}_3)\right]\alpha_0 - \sum_{m=0}^{\infty} \left(\dfrac{E}{\Lambda}\right)^{2m+1} (c^1_{(m,0)} - c^2_{(m,0)})
         \end{aligned}$ \\[0.5cm]
         \hline
         \hline\rule{0pt}{5ex}
         $\begin{aligned}
             \delta \textbf{a}_{1,3} =  \dfrac{E}{\Lambda} 4\Qr(1 - c^{3p}_2)\,c^{3p}_4\, \alpha_0 - \sum_{m=1}^{\infty} \left(\dfrac{E}{\Lambda}\right)^{2m+1}c^3_{(m-1,0)} 
         \end{aligned}$ \\[0.7cm]
         $\begin{aligned}
             \delta \textbf{a}_{1,4} =  \dfrac{E}{\Lambda} 4\Qr(1 - c^{3p}_2)\,c^{3p}_5\, \alpha_0 - \sum_{m=1}^{\infty} \left(\dfrac{E}{\Lambda}\right)^{2m+1}c^5_{(m-1,0)} 
         \end{aligned}$ \\[0.7cm]
         $\delta \textbf{a}_{1, 5} = \dfrac{E}{\Lambda}4\Ql\Qr (c^{3p}_2 - c^{3p}_3)c^{3p}_4 \alpha_0 + \displaystyle\sum_{m=1}^{\infty}\left(\dfrac{E}{\Lambda}\right)^{2m+1} \left(Q^Z_{f_L}\,c_{(m-1,0)}^3 + Q^Z_{f_R}\,c_{(m-1,0)}^4\right) $ \\[0.7cm]
         $\delta \textbf{a}_{1, 6} = \dfrac{E}{\Lambda}4\Ql\Qr (c^{3p}_2 - c^{3p}_3)c^{3p}_5 \alpha_0 + \displaystyle\sum_{m=1}^{\infty}\left(\dfrac{E}{\Lambda}\right)^{2m+1} \left(Q^Z_{f_L}\,c_{(m-1,0)}^5 - Q^Z_{f_R}\,c_{(m-1,0)}^6\right) $ \\[0.5cm]
        \hline 
    \end{tabular}
    \end{adjustbox}
    \caption{Expressions, to all orders in the EFT expansion, for the principal directions involving partial wave coefficients with opposite fermion helicities.}
    \label{tab:pdir}
\end{table}

Let us focus on the contributions of the principals and descendants in  Table~\ref{tab:pdir}. It is clear that the  linear combinations, $\delta^p\textbf{a}_k$,  are nothing but  directions in the space of $J=J_{min}$ partial wave coefficients that correspond to  different principal operators---or their  linear combinations---followed by a series of  $s$-descendants. We will therefore call these, `$J=1$  helicity-directions' or simply `principal directions'. In the next subsection we will similarly define helicity directions for partial wave coefficients with higher $J$ values. These helicity directions encode the EFT dimensional analysis into the partial wave expansion. In the next section we will see that differential observables can be mapped to these helicity directions in a straightforward way.

In App~\ref{lrrl} in Table~\ref{tab:LR_RL_hel_dir}  we similarly construct  principal directions for partial wave coefficients with identical fermion helicities.  Considering all the helicity channels together, for the  12 independent $J=1$  partial wave coefficients, 2 linear combinations appear at ${\cal O}(E^0/\Lambda^0)$ in the EFT expansion, 6 appear at ${\cal O}(E/\Lambda)$ and 4 at ${\cal O}(E^2/\Lambda^2)$. As far as the  2 $J=0$ partial wave coefficients are concerned, we obtain 2 linear combinations that receive their first contribution only at ${\cal O}(\E^2/\Lambda^2)$.

\subsection{Higher $J$ partial wave amplitudes} 

Now we follow the procedure outlined in the previous subsection to derive the mapping between EFT Wilson coefficients and partial wave coefficients of higher $J$ values, for helicity amplitudes with opposite fermion helicities. We consider, in particular, the $J=2$ helicity amplitudes. As there are no principal operators with $J>1$, these helicity amplitudes get contributions from the $J$-descendants of the principal operators in Table~\ref{tab:4-pt_principals}; these have been  presented in Table~\ref{tab:ffZh_Jdesc_ops}. We show the contributions of these to the $J=2$ partial wave coefficients in  Table~\ref{tab:full_Amps_Exact_J2}. The table entries  show partial wave coefficients after being rescaled according to \eq{rescale}.
\begin{table}[t]
    \begin{adjustbox}{max width=\textwidth}
    \centering
    \begin{tabular}{|c|c|c|c|c|c|c|}
\hline
   EFT & \multicolumn{6}{ c |}{Helicity Amplitudes}\\
    \cline{2-7}
     Couplings & $(+-;\,0)$ & $(+-;\,+)$ &$(+-;\,-)$ & $(-+;\,0)$ & $(-+;\,+)$ & $(-+;\,-)$ \\[0.5ex]
     \hline\rule{0pt}{4ex}
     $c^1_{(0,1)}$ & $i\dfrac{\E^3}{\Lambda^3}\dfrac{\alpha_2}{\sqrt{6}}$ & $-i\dfrac{\E^3}{\Lambda^3}\dfrac{\alpha_2}{2\sqrt{2}}$ & $-i\dfrac{\E^3}{\Lambda^3}\dfrac{\alpha_2}{2\sqrt{2}}$ & 0 & 0 & 0 \\[2.3ex]
     \hline\rule{0pt}{4ex}
     $c^2_{(0,1)}$ & 0 & 0 & 0 & $-i\dfrac{\E^3}{\Lambda^3}\dfrac{\alpha_2}{\sqrt{6}}$ & $i\dfrac{\E^3}{\Lambda^3}\dfrac{\alpha_2}{2\sqrt{2}}$ & $i\dfrac{\E^3}{\Lambda^3}\dfrac{\alpha_2}{2\sqrt{2}}$ \\[2.3ex]
     \hline
     \hline\rule{0pt}{4ex}
     $c^1_{(1,1)}$ & $i\dfrac{\E^5}{\Lambda^5}\dfrac{\alpha_2}{\sqrt{6}}$ & $-i\dfrac{\E^5}{\Lambda^5}\dfrac{\alpha_2}{2\sqrt{2}}$ & $-i\dfrac{\E^5}{\Lambda^5}\dfrac{\alpha_2}{2\sqrt{2}}$ & 0 & 0 & 0 \\[2.3ex]
     \hline\rule{0pt}{4ex}
     $c^2_{(1,1)}$ & 0 & 0 & 0 & $-i\dfrac{\E^5}{\Lambda^5}\dfrac{\alpha_2}{\sqrt{6}}$ & $i\dfrac{\E^5}{\Lambda^5}\dfrac{\alpha_2}{2\sqrt{2}}$ & $i\dfrac{\E^5}{\Lambda^5}\dfrac{\alpha_2}{2\sqrt{2}}$ \\[2.3ex]
    \hline\rule{0pt}{4ex}
    $c^3_{(0,1)}$ & $i\dfrac{\E^5}{\Lambda^5}\left(\dfrac{2\,\alpha_2\,\varepsilon_0}{\sqrt{6}\,\alpha_1}\right)$ & $-i\dfrac{\E^5}{\Lambda^5}\dfrac{\alpha_1 \alpha_2}{4\sqrt{2}}$ & $-i\dfrac{\E^5}{\Lambda^5}\dfrac{\alpha_1 \alpha_2}{4\sqrt{2}}$ & 0 & 0 & 0 \\[2.3ex] 
     \hline\rule{0pt}{4ex}
    $c^4_{(0,1)}$ & 0 & 0 & 0 & $i\dfrac{\E^5}{\Lambda^5}\left(\dfrac{2\,\alpha_2\,\varepsilon_0}{\sqrt{6}\,\alpha_1}\right)$ & $-i\dfrac{\E^5}{\Lambda^5}\dfrac{\alpha_1 \alpha_2}{4\sqrt{2}}$ & $-i\dfrac{\E^5}{\Lambda^5}\dfrac{\alpha_1 \alpha_2}{4\sqrt{2}}$ \\[2.3ex]
    \hline\rule{0pt}{4ex}
    $c^5_{(0,1)}$ & 0 & $-\dfrac{\E^5}{\Lambda^5}\dfrac{\alpha_2^2}{4\sqrt{2}}$ & $\dfrac{\E^5}{\Lambda^5}\dfrac{\alpha^2_2}{4\sqrt{2}}$ & 0 & 0 & 0 \\[2.3ex]
    \hline\rule{0pt}{4ex}
    $c^6_{(0,1)}$ & 0 & 0 & 0 & 0 & $\dfrac{\E^5}{\Lambda^5}\dfrac{\alpha_2^2}{4\sqrt{2}}$ & $-\dfrac{\E^5}{\Lambda^5}\dfrac{\alpha^2_2}{4\sqrt{2}}$ \\[2.3ex]
    \hline
    \hline\rule{0pt}{3ex}
    $c^k_{(m,\,1)}$ & \multicolumn{6}{ c |}{$(E^2/\Lambda^2)^m\, \vec{V}^k_{(0,\,1)}$} \\[1.4ex]
    \hline
    \end{tabular}
    \end{adjustbox}
    \hspace*{-3cm}
    \caption{EFT contributions---up to all orders in the expansion---to the  partial wave  coefficients, $\hat{a}^{J}_{\lambda_{f} \lambda_{\bar{f}}; \lambda_Z}$ (normalised as in \eq{rescale}) with $J=2$ and opposite fermion helicities. The contributions are from the $J$-descendants of the principal operators (see Table~\ref{tab:ffZh_Jdesc_ops}) and their $s$-descendants.  See the caption of Table~\ref{tab:full_Amps_Exact} for the definition of $\varepsilon_0$ and  $\alpha_{0,1,2}$.}
    \label{tab:full_Amps_Exact_J2}
\end{table}

Note that to compute the contributions to the different higher $J$ partial wave amplitudes, one does not necessarily need to  explicitly construct the $J$-descendants following the procedure in Sec.~\ref{sec:proj} and App.~\ref{sec:proj}.  These contributions can also be found  by using the fact that the  $(t-u)$-descendant amplitude, $i^{n(\textrm{mod} \,2)}(t-u)^n P_k$,  is given by, 
\bea
{\cal M}_{\lambda_f, \lambda_{\bar f };\lambda_Z} =i^{n(\textrm{mod} \,2)} s^n P^k_{\lambda_f, \lambda_{\bar f };\lambda_Z} \left(\frac{\alpha_2}{2}\right)^n\cos^n \Theta~d^{J_k}_{\sigma \lambda} (\Theta)
\label{calm}
\eea
where $P^k_{\lambda_f, \lambda_{\bar f };\lambda_Z}$ is the contribution of the principal amplitude $P_k$ to the partial wave coefficient, ${a}^{J_k}_{\lambda_f, \lambda_{\bar f };\lambda_Z}$. Here $n(\rm{mod}~2)$ is the remainder obtained when $n$ is divided by 2 and  the factor of $i^{n(\textrm{mod} \,2)}$ has been chosen to ensure \eq{Eq.App.A8} is satisfied. To evaluate the contribution to the partial wave amplitude with maximum $J$-value,  $J=J_k+n$, we can   simply use the orthogonality of Wigner functions,
\begin{equation}
   {a}^{J_k+n}_{\lambda_f, \lambda_{\bar f };\lambda_Z}=\frac{2(J_k +n) +1}{2}\int_0 ^{\pi} d\Theta \mbox{ sin}\Theta \; {\cal M}_{\lambda_f, \lambda_{\bar f };\lambda_Z}(\Theta)\;d^{J_k+n}_{\sigma\lambda}(\Theta)
\end{equation}
where ${\cal M}_{\lambda_f, \lambda_{\bar f };\lambda_Z}$ must be substituted from \eq{calm}.
\begin{table}[t]
    \centering
    \hspace*{-1cm}
    \begin{tabular}{|c|}
    \hline
        $J=2$ helicity directions to all orders in EFT\\
       \hline\rule{0pt}{5ex}
         $\begin{aligned}
         \delta^3 \textbf{a}_{2, 1} = \dfrac{1}{\sqrt{6}}\,\sum_{m = 1}^{\infty}\left(\dfrac{E}{\Lambda}\right)^{2m+1}  c^1_{(m-1,1)} 
         \end{aligned}$ \\[0.6cm]
        
       $\begin{aligned}
         \delta^3 \textbf{a}_{2, 2} = \dfrac{1}{\sqrt{6}}\,\sum_{m = 1}^{\infty}\left(\dfrac{E}{\Lambda}\right)^{2m+1}  c^2_{(m-1,1)}
         \end{aligned}$ \\[0.6cm]
        \hline
        \hline\rule{0pt}{5ex}
         $\begin{aligned}
             \delta^5 \textbf{a}_{2, 3} = \dfrac{1}{2\sqrt{2}}\sum_{m=2}^{\infty} \left(\dfrac{E}{\Lambda}\right)^{2m+1} c^3_{(m-2,\,1)}
         \end{aligned}$ \\[0.6cm]

        $\begin{aligned}
            \delta^5 \textbf{a}_{2, 4} = \dfrac{1}{2\sqrt{2}}\sum_{m=2}^{\infty} \left(\dfrac{E}{\Lambda}\right)^{2m+1} c^5_{(m-2,1)}
        \end{aligned}$ \\[0.6cm]

         $\begin{aligned}
             \delta^5 \textbf{a}_{2, 5} = \dfrac{1}{2\sqrt{2}}\sum_{m=2}^{\infty} \left(\dfrac{E}{\Lambda}\right)^{2m+1}  c^4_{(m-2,1)}
         \end{aligned}$ \\[0.6cm]

          $\begin{aligned}
            \delta^5 \textbf{a}_{2, 6} = \dfrac{-1}{2\sqrt{2}}\sum_{m=2}^{\infty} \left(\dfrac{E}{\Lambda}\right)^{2m+1} c^6_{(m-2,1)}
        \end{aligned}$ \\[0.6cm]
        \hline
        
    \end{tabular}
   \caption{Expressions, to all orders in the EFT expansion, for the $J=2$ helicity directions involving partial wave coefficients with opposite  fermion helicities.}
    \label{tab:J2dir}
\end{table}

We see from Table~\ref{tab:full_Amps_Exact_J2} that at the lower orders in $E/\Lambda$, the $J=2$ partial wave coefficients again have correlations in addition to  what is implied by \eq{cpt}. These correlations are broken by EFT effects order by order in $\E/\Lambda$. As in \eq{pdirections} we can define linear combinations of $J=2$ partial wave coefficients, $\delta^p \mathbf{a}_{2,k}$, that receive the lowest order EFT  contributions at ${\cal O} (E^p/\Lambda^p)$. These are given by,
\begin{align}
    \delta^3 \textbf{a}_{2, 1} &= -\frac{i}{\alpha_2}[(1+\varepsilon_2)\hata^2_{+-;0} + \frac{\varepsilon_2}{\sqrt{3}}( \hata^2_{+-;-} + \hata^2_{+-;+})]\nn\\
    \delta^3 \textbf{a}_{2, 2} &= \frac{i}{\alpha_2}[(1+\varepsilon_2)\hata^2_{-+;0} + \frac{\varepsilon_2}{\sqrt{3}}(\hata^2_{-+;-} + \hata^2_{-+;+})] \nn\\
    \delta^5 \textbf{a}_{2, 3} &= \frac{i}{\alpha_2 \alpha_3}(\hata^2_{+-;-} + \hata^2_{+-;+} + \sqrt{3}\, \hata^2_{+-;0}) \nn\\
    \delta^5 \textbf{a}_{2, 4} &= \dfrac{1}{\alpha_2^2}(\hata^2_{+-;-} - \hata^2_{+-;+}) \nn\\
    \delta^5 \textbf{a}_{2, 5} &= \frac{i}{\alpha_2 \alpha_3}(\hata^2_{-+;-} + \hata^2_{-+;+} + \sqrt{3}\, \hata^2_{-+;0}) \nn\\
    \delta^5 \textbf{a}_{2, 6} &= \dfrac{1}{\alpha_2^2}(\hata^2_{-+;-} - \hata^2_{-+;+}).
    \label{heldir2}
\end{align}
where all these linear combinations are real by \eq{cpt}. We call these the $J=2$  helicity directions. Up to ${\cal O} (E^{p-1}/\Lambda^{p-1})$ the linear combination, $\delta^p \mathbf{a}_{2,k}$,  vanishes so  the helicity directions can also be thought of as correlations among partial wave coefficients that exist up to ${\cal O} (E^{p-1}/\Lambda^{p-1})$. Thus \eq{heldir2} presents 2 correlations among $J=2$ partial wave coefficients that last up to ${\cal O} (E^{2}/\Lambda^{2})$ and 4  that last up to  ${\cal O} (E^{4}/\Lambda^{4})$

The $J=2$  helicity directions can also be written as a power series in $E/\Lambda$ as follows,
\bea
\delta^p\textbf{a}_{2,k}= \sum_{j=(p-1)/2}^\infty \delta^p {\textbf{a}}_{2,k}^{2j+1-p}\left(\frac{E}{\Lambda}\right)^{2 j+1}.
\label{pow2}
\eea
All the $\delta^p \mathbf{a}^m_{2,k}$ are energy independent in this case.  Table~\ref{tab:J2dir} provides the expression for the $\delta^p \mathbf{a}_{2,k}$ in terms of  the Wilson coefficients  to all orders in the EFT expansion.





\section{The $f \bar{f} \to Zh$ process in the SMEFT up to dimension-8 level }

\label{sec4} 

To interpret our results within the SMEFT we need to express the Wilson coefficients, $c_{(m,n)}$, in \eq{basisdef} in terms of SMEFT Wilson coefficients. Conceptually the most important thing about the SMEFT interpretation is the fact that in the SMEFT, in addition to $E/\Lambda$, there is another expansion parameter, $\xi=gv/\Lambda$.  Some of the Wilson coefficients, $c_{(m,n)}$,    will be suppressed by powers of  $\xi$. In Table~\ref{tab:smeft}, we show the Wilson coefficients in \eq{basisdef} that are non-zero in the SMEFT at the dimension-8 level. We  also show one possible SMEFT operator that will contribute to each of these Wilson coefficients. 

We see that  while all the 3-point functions and principal operators  in Table~\ref{tab:3-point_ops} and Table~\ref{tab:4-pt_principals} are generated at the dimension-8 level, only the first $s$-descendant and $J$-descendant of  the principal operators with $k=1,2,7,8$ are non-vanishing at this level.  All  other Wilson coefficients in \eq{basisdef} can be set to 0 as they arise at dimension 10 or higher.  Among the non-zero Wilson coefficients in Table~\ref{tab:smeft},  the 3-point vertices and the Wilson coefficients of the principal operators  with $k=1,2,7,8$ arise already at the dimension-6 level; all other  Wilson coefficients arise at the dimension-8 level. 

\begin{table}[p]
\begin{adjustbox}{max width=\textwidth}
\centering
\begin{tabular}{c}
\begin{tabular}{||c|c|c|c|c||}
\hline
Class & WC & SMEFT Operator &$\left(\xi^{n_6}, \xi^{n_8}\right)$ &$d$ \\\hline \rule{0pt}{3.5ex}
\multirow{10}{*}{3-point} & $c^{3p}_1$& 
$(H^\dagger H)(D^{\mu}H^\dagger D_{\mu}H)$& $\left(\xi^2, \xi^4\right)$&\multirow{16}{*}{6}\\\rule{0pt}{4ex}
& $c^{3p}_2$ & $(iH^\dagger \overset{\leftrightarrow}{D}_{\mu}H)(\overline{f}_R\gamma^\mu f_R)$ & $\left(\xi^2, \xi^4\right)$& \\\rule{0pt}{4ex}
& $c^{3p}_3$ & $(iH^\dagger \overset{\leftrightarrow}{D}_{\mu}H)(\overline{F}_L\gamma^\mu F_L)$ & $\left(\xi^2, \xi^4\right)$ &\\\rule{0pt}{4ex}
& $c^{3p}_4$ & $(H^\dagger H)W^a_{\mu\nu}W^{a\mu\nu}$ & $\left(\xi, \xi^3\right)$& \\\rule{0pt}{4ex}
& $c^{3p}_5$ & $(H^\dagger H)W^a_{\mu\nu}\widetilde{W}^{a\mu\nu}$ & $\left(\xi, \xi^3\right)$&\\\rule{0pt}{4ex}
& $c^{3p}_6$ & $W^a_{\mu\nu}\overline{F}_L\sigma^{\mu\nu}\sigma^a\widetilde{H}f_R + \text{h.c.}$ &$\left(\xi, \xi^3\right)$& \\\rule{0pt}{4ex}
& $c^{3p}_7$ & $iW^a_{\mu\nu}\overline{F}_L\sigma^{\mu\nu}\sigma^a\widetilde{H}f_R + \text{h.c.}$ &$\left(\xi, \xi^3\right)$& \\[0.1cm]
\cline{1-4}\rule{0pt}{3.5ex}
\multirow{18}{*}{Principals} & $c^1_{(0,0)}$ & $(iH^\dagger \overset{\leftrightarrow}{D}_{\mu}H)(\overline{f}_R\gamma^\mu f_R)$  &$\left(\xi, \xi^3\right)$&\\\rule{0pt}{4ex}
& $c^2_{(0,0)}$ & $(iH^\dagger \overset{\leftrightarrow}{D}_{\mu}H)(\overline{F}_L\gamma^\mu F_L)$  &$\left(\xi, \xi^3\right)$& \\\rule{0pt}{3.5ex}
& $c^7_{(0,0)}$ & $W^a_{\mu\nu}\overline{F}_L\sigma^{\mu\nu}\sigma^a\widetilde{H}f_R + \text{h.c.}$ &$\left(\xi^0, \xi^2\right)$& \\\rule{0pt}{4ex}
& $c^8_{(0,0)}$ & $iW^a_{\mu\nu}\overline{F}_L\sigma^{\mu\nu}\sigma^a\widetilde{H}f_R + \text{h.c.}$ &$\left(\xi^0, \xi^2\right)$& \\[0.1cm] \cline{2-5}\rule{0pt}{3.5ex}
 & $c^3_{(0,0)}$ & $i\vert H \vert^2\widetilde{B}^{\mu\nu}(\overline{f}_R\gamma_{\mu}\overset{\leftrightarrow}{D}_{\nu}f_R)$ &$\left(0, \xi\right)$& \multirow{24}{*}{$8$}\\\rule{0pt}{4ex}
& $c^4_{(0,0)}$ & $i\vert H \vert^2\widetilde{W}^{a\mu\nu}(\overline{F}_L\gamma_{\mu}\sigma^a \overset{\leftrightarrow}{D}_{\nu}F_L)$ &$\left(0, \xi\right)$& \\\rule{0pt}{4ex}
& $c^5_{(0,0)}$ & $\vert H \vert^2\widetilde{B}^{\mu\nu}D_{\mu}(\overline{f}_R\gamma_{\nu}f_R)$ &$\left(0, \xi\right)$& \\\rule{0pt}{4ex}
& $c^6_{(0,0)}$ & $\vert H \vert^2\widetilde{W}^{a\mu\nu}D_{\mu}(\overline{F}_L\gamma_{\nu}\sigma^aF_L)$ &$\left(0, \xi\right)$& \\\rule{0pt}{4ex}
& $c^9_{(0,0)}$ & $(H^\dagger \overset{\leftrightarrow}{D}_{\mu}H)(\overline{F}_L\overset{\leftrightarrow}{D}{}^{\mu}f_R)\widetilde{H} + \text{h.c.}$ &$\left(0, \xi^2\right)$& \\\rule{0pt}{4ex}
& $c^{10}_{(0,0)}$ & $i(H^\dagger \overset{\leftrightarrow}{D}_{\mu}H)(\overline{F}_L\overset{\leftrightarrow}{D}{}^{\mu}f_R)\widetilde{H} + \text{h.c.}$ &$\left(0, \xi^2\right)$& \\\rule{0pt}{4ex}
& $c^{11}_{(0,0)}$ & $i(H^\dagger \overset{\leftrightarrow}{D}_{\mu}H)D^{\mu}(\overline{F}_Lf_R)\widetilde{H} + \text{h.c.}$ &$\left(0, \xi^2\right)$& \\\rule{0pt}{4ex}
& $c^{12}_{(0,0)}$ & $(H^\dagger \overset{\leftrightarrow}{D}_{\mu}H)D^{\mu}(\overline{F}_Lf_R)\widetilde{H} + \text{h.c.}$ &$\left(0, \xi^2\right)$& \\ \cline{1-4}\rule{0pt}{4ex}
\multirow{6}{*}{$s$-descendants} & $c^1_{(1,0)}$ & $(iH^\dagger \overset{\leftrightarrow}{D}_{\mu}H)\Box(\overline{f}_R\gamma^\mu f_R)$ &$\left(0, \xi\right)$& \\\rule{0pt}{4ex}
& $c^2_{(1,0)}$ & $(iH^\dagger \overset{\leftrightarrow}{D}_{\mu}H)\Box(\overline{F}_L\gamma^\mu F_L)$ &$\left(0, \xi\right)$& \\\rule{0pt}{4ex}
& $c^7_{(1,0)}$ & $W^a_{\mu\nu}\Box(\overline{F}_L\sigma^{\mu\nu}f_R)\sigma^a\widetilde{H} + \text{h.c.}$ &$\left(0, \xi^0\right)$& \\\rule{0pt}{4ex}
& $c^8_{(1,0)}$ & $i{W}^a_{\mu\nu}\Box(\overline{F}_L\sigma^{\mu\nu}f_R)\sigma^a\widetilde{H} + \text{h.c.}$&$\left(0, \xi^0\right)$& \\ \cline{1-4}\rule{0pt}{4ex}
\multirow{6}{*}{$J$-descendants}&  $c^1_{(0,1)}$ & $i(H^\dagger\overset{\leftrightarrow}{D}_{\mu\nu}H)(\overline{f}_R\gamma^{\mu}\overset{\leftrightarrow}{D}{}^{\nu}f_R)$ &$\left(0, \xi\right)$&  \\\rule{0pt}{4ex}
& $c^2_{(0,1)}$ & $i(H^\dagger\overset{\leftrightarrow}{D}_{\mu\nu}H)(\overline{F}_L\gamma^{\mu}\overset{\leftrightarrow}{D}{}^{\nu}F_L)$ &$\left(0, \xi\right)$&  \\\rule{0pt}{5ex}
& $c^7_{(0,1)}$ & i$W^{a,\mu\nu} \left(\left(\overline{F}_L \sigma_{\mu\nu}\, \,\overset{\leftrightarrow}{D}_{\rho}f_R\right) \sigma^a D^{\rho}\widetilde{H}-\dfrac{2}{3}\, \left(\overline{F}_L\sigma_{\alpha\nu}\, \overset{\leftrightarrow}{D^{\alpha}}f_R\right)  \sigma^a D_{\mu}\widetilde{H}\right) + \hc$ &$\left(0, \xi^0\right)$& \\\rule{0pt}{4ex}
& $c^8_{(0,1)}$ & $W^{a,\mu\nu} \left(\left(\overline{F}_L \sigma_{\mu\nu}\, \,\overset{\leftrightarrow}{D}_{\rho}f_R\right) \sigma^a D^{\rho}\widetilde{H}-\dfrac{2}{3}\, \left(\overline{F}_L\sigma_{\alpha\nu}\, \overset{\leftrightarrow}{D^{\alpha}}f_R\right)  \sigma^a D_{\mu}\widetilde{H}\right) + \hc$ &$\left(0, \xi^0\right)$& \\[0.3cm] \hline
 \end{tabular}
\end{tabular}
\end{adjustbox}
\caption{We show the Wilson coefficients in \eq{basisdef} that are non-vanishing at the dimension-8 level in SMEFT. We have also provided in the third column one possible SMEFT operator that can generate each of these Wilson coefficients. Here, $i(H^\dagger\overset{\leftrightarrow}{D}_{\mu\nu}H)=i(H^\dagger D_\mu D_\nu H-h.c.)$,  $F_L$ is the quark or lepton doublet and $f_R=u_R, d_R, e_R$. The operators having, $\tilde{H}$, have been written assuming $f_R=u_R$; for  $f_R=d_R, e_R$ the replacement  $\tilde{H}\to H$ must be made. Other notations and conventions are same as Ref.~\cite{warsaw}. As discussed in the main text, the Wilson coefficients of \eq{basisdef} can be written as $c_i= \xi^{n_6} L_{i,6} + \xi^{n_8} L_{i,8}$ where $L_{i,6}$ ($L_{i,8}$) are linear combinations of dimension-6 (dimension-8) operators (see \eq{smeftmatch}). We present, $\left(\xi^{n_6}, \xi^{n_8}\right)$, for each operator in the fourth column.}
\label{tab:smeft}
\end{table}

All our results in Table~\ref{tab:full_Amps_Exact}--\ref{tab:LR_RL_hel_dir}---including those yet to be discussed in Sec.~\ref{sec5}---can be translated to the SMEFT framework by expressing the Wilson coefficients appearing in Table~\ref{tab:smeft} in terms of SMEFT Wilson coefficients as follows, 
\bea
c_i^{3p}&=& \xi^{p_i} \sum_j \alpha_{ij}\, {\cal C}_j^{6}+ \xi^{(p_i+2)} \sum_j \beta_{ij}\, {\cal C}_j^{8},\nonumber\\
c_{(m,n)}^{k}&=& \xi^{q^k_{mn}} \sum_j \alpha^k_{mnj}\, {\cal C}_j^{6}+\xi^{(q^k_{mn}+2)} \sum_j \beta^k_{mnj}\, {\cal C}_j^{8},
\label{smeftmatch}
\eea
where ${\cal C}^6_j$ (${\cal C}^8_j$) is the Wilson coefficient of a dimension-6 (dimension-8) SMEFT operator; $\alpha_{ij},\, \beta_{ij},\, \alpha^k_{mnj}$ and $\beta^k_{mnj}$ are numerical coefficients; and $p_i,\, q^k_{mn}$ are positive integers. The  suppression factors, $\xi^n= (g v/\Lambda)^n$,  arise due to the presence of additional Higgs doublets in   SMEFT operators that must be set to their vacuum expectation value to obtain the operators in \eq{basisdef}.

The particular linear combinations of dimension-6 and 8 SMEFT operators in \eq{smeftmatch},
\bea 
&&L_{i,6}^{3p}=\sum_j \alpha_{ij} {\cal C}_j^{6},~~~~~~~~~~~~~
L_{i,8}^{3p}=\sum_j \beta_{ij} {\cal C}_j^{8},\nn\\
&&L^k_{m,n,6}=\sum_j \alpha^k_{mnj} {\cal C}_j^{6},~~~~~~~~
L^k_{m,n,8}=\sum_j \beta^k_{mnj} {\cal C}_j^{8}.
\label{smeftdir}
\eea
are of course basis dependent. In a particular SMEFT basis they can be derived by matching \eq{basisdef} to the SMEFT lagrangian in that basis. The powers, $n$, in the suppression factors,  $\xi^n= (g v/\Lambda)^n$, on the other hand, are basis independent. We provide these suppression factors in Table~\ref{tab:smeft} for each of the Wilson coefficients.  A basis independent SMEFT analysis can be performed  by considering only the Wilson coefficients appearing in Table~\ref{tab:smeft}, expressing them in terms of SMEFT Wilson coefficients as in \eq{smeftmatch} and then treating the linear combinations in \eq{smeftdir} as free parameters. These results can then be interpreted in any SMEFT basis by expressing the $L_{i,6}$ and $L_{i,8}$  as in \eq{smeftdir}.

\section{Mapping  differential observables to EFT operators}

\label{sec5}

In the Sec.~\ref{sec3} we showed how the partial wave expansion can be systematically mapped to the EFT expansion to all orders. In this section we use these results to take the final step of mapping experimental observables to the Wilson coefficients.

In order to probe the different helicity  amplitudes for the $f\bar{f} \to Z(\ell\ell) h$ process   it is necessary to access information about the  helicities of the initial fermions and the $Z$-boson. This can be achieved by carefully studying the angular distribution of the leptons arising from the $Z$ boson decay. Writing the partial wave expansion for $f\bar{f} \to Z(\ell\ell) h$---with  the inclusion of the $Z$-decay amplitude---and then squaring it results in the most general angular distribution for the   $f\bar{f} \to Z(\ell\ell) h$ process. In Sec.~\ref{angmom} we define differential observables, the so-called angular moments, that parametrise the most general angular distribution thus obtained.  We then show in Sec.~\ref{mapping2} how a mapping between these observables and the EFT Wilson coefficients can be obtained up to any given order.

  \subsection{Angular moments for  the $f \bar{f} \to Z(\ell\ell) $ process }
  \label{angmom}
\begin{figure}[t]
    \centering
    \includegraphics[width=0.8\linewidth]{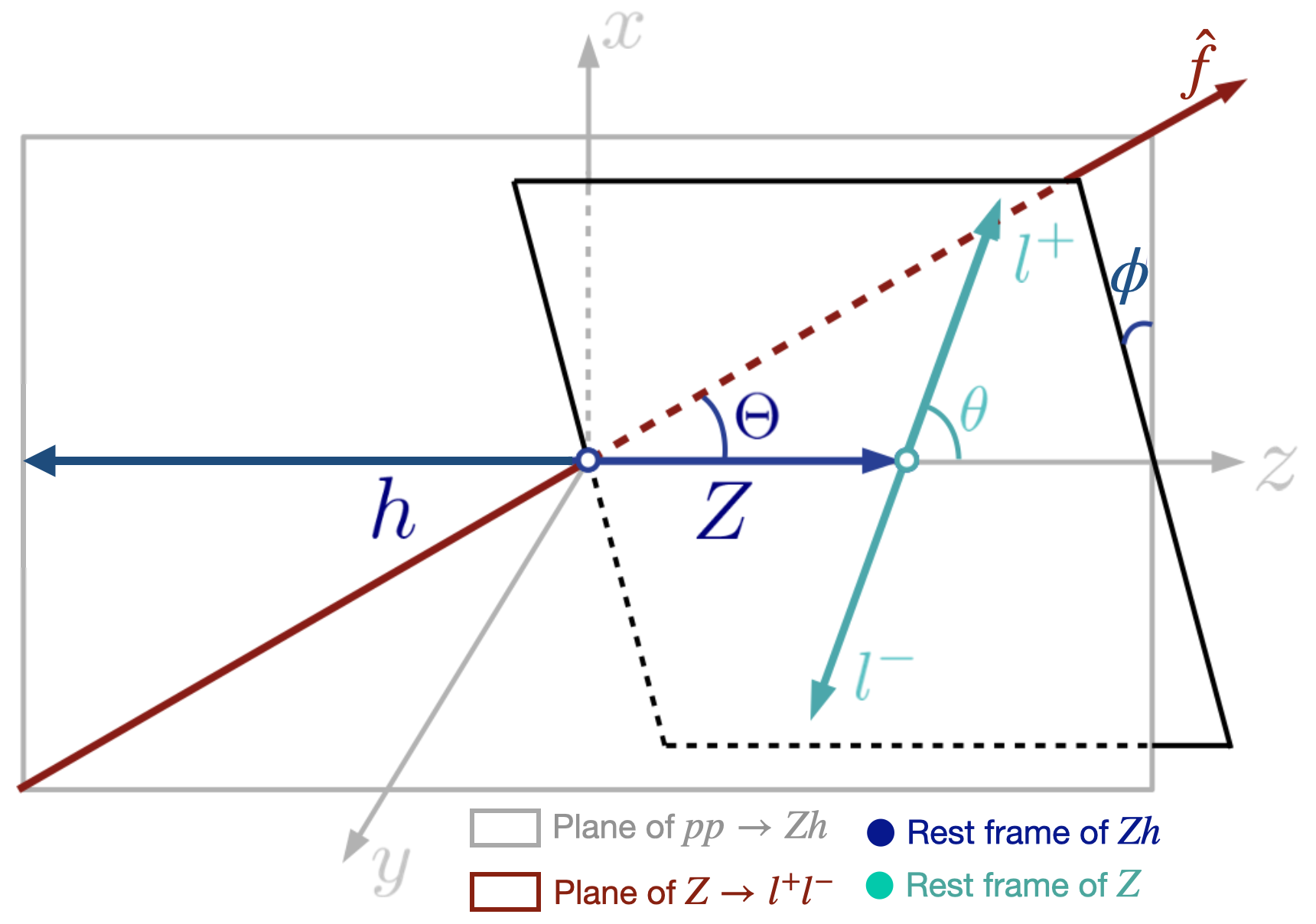}
    \caption{We show the three angles, $\left(\Theta, {\theta}, {\phi}\right)$, in the figure above. The angle, $\Theta$ is the scattering angle of the $f \bar{f} \to Zh$ process. The cartesian axes, $\{x,y,z\}$, are defined  as follows: the $z$ direction is identified as the direction of the $Z$-boson; $y$ is defined to be  normal to the plane of $Z$, i.e. $\hat{y}=\hat{z}\times \hat{f}$, $\hat{f}$ being the direction of the fermion shown by the red arrow;  $x$ is defined such that it completes the right-handed set. The angles, $\theta$ and $\phi$ ($\hat{\theta}$ and $\hat{\phi}$), are the polar and azimuthal angles of the positively charged (positive helicity) lepton in the $Z$ rest frame in the spherical polar coordinate system defined with respect to the  above cartesian   system.}
    \label{fig2}
\end{figure}

 The amplitude for the $f\bar{f} \to Z(\ell\ell) h$ process is given by,
\begin{equation}
    \mathcal{A}_{\lambda_f \lambda_{\bar{f}}}(s, \Theta, \hat{\theta}, \hat{\phi}) = \frac{-ig^Z_l}{\Gamma_Z}\sum_{\lambda_Z= 0, \pm1} \mathcal{M}_{\lambda_f \lambda_{\bar{f}};\lambda_Z}(s, \Theta)\, d^{1}_{\lambda_Z, 1}(\hat{\theta})\, e^{i\lambda_Z\hat{\phi}}
    \label{2to3}
\end{equation}
where $\Gamma_Z$ is the $Z$-decay width and $g^Z_{l}$ is the $Z$ coupling to left or right-handed leptons  given by,  $g^Z_{l}= (g/\ctw) (T^3_{l}+\stw^2)$, with $T^3_{l}=1/2$ ($T^3_{l}=0$) for the former (latter) case. The   Wigner functions arising from the $Z$ decay, $d^{1}_{\lambda, 1}(\hat{\theta})$, are given by,\footnote{Note that we are not considering any EFT effects in the $Z$-decay as these are strongly constrained by LEP data. The EFT effects that modify the couplings, $c^{3p}_{2,3}\frac{g}{\ctw}Q^Z_{f_{R,L}}Z_\mu\bar{l}_{R,L} \gamma^\mu l_{R,L}$, can be readily incorporated by simply modifying,  $g^Z_{l_{R,L}}\to g^Z_{l_{R,L}}-c^{3p}_{2,3}\frac{g}{\ctw}Q^Z_{f_{R,L}}$. The only other EFT correction possible arises from the inclusion of the vertex, $Z_{\mu \nu} \bar{l}_L \sigma^{\mu \nu} l_R$. The fractional contribution of this term to the $Z$-decay width must be within the per-mille level to be within LEP bounds~\cite{lepbounds}. While this vertex  will result in new angular functions in \eq{eq:ang_funcs},  probing 
the corresponding angular moments will be far less sensitive to  EFT effects in the $f \bar{f} \to hZ$ process due to these bounds. We have,  therefore, not included this vertex in  the $Z$ decay amplitude.}
\begin{align}
    d^{1}_{\lambda = 0, 1} = \mbox{sin}\,\hat{\theta} \; \; \; \; \; \mbox{and,} \; \; \; 
    d^{1}_{\lambda = \pm, 1} = \frac{1 + \lambda\,\mbox{cos~}\hat{\theta}}{\sqrt{2}}.
\end{align}
Here  $\mathcal{M}_{\lambda_f \lambda_{\bar{f}};\lambda_Z}$ is the 2 to 2 amplitude defined below \eq{par_wave} and we have used the narrow width approximation.  The angles, $\hat{\theta}$ and $\hat{\phi}$, are the polar and azimuthal angles of the positive helicity lepton in the $Z$-rest frame in the coordinate system defined in Fig.~\ref{fig2}. Notice that as the $Z$-boson is an intermediate state we have coherently summed the amplitude over all its  polarisations

To connect with observations, one needs to compute the squared amplitude and express it in terms of $\theta$ and $\phi$ the polar and azimuthal angle for the positively charged lepton (instead of the corresponding angles for the positive helicity lepton).  Squaring \eq{2to3} and summing over the final state lepton polarisations we obtain,
    \begin{equation}
    \sum_{l_L,l_R} |\mathcal{A}_{\lambda_f \lambda_{\bar{f}}}(s, \Theta, \theta, \phi)|^2 = \alpha_{l_L}|     \mathcal{A}_{\lambda_f \lambda_{\bar{f}}}(s, \Theta, \theta, {\phi})|^2 + \alpha_{l_R}|\mathcal{A}_{\lambda_f \lambda_{\bar{f}}}(s, \Theta, \pi - {\theta}, \pi + {\phi})|^2
    \label{sum}
\end{equation}
where $\alpha_{l_L, l_R} = (g^Z_{l_{L,R}})^2/[(g^Z_{l_L})^2 + (g^Z_{l_R})^2]$. The first term arises due to $Z$ decays to  left-chiral leptons so that the lepton with positive helicity also have a positive charge. The second term shows the contribution due to $Z$ decays to  right-chiral leptons such that the negative helicity lepton carries a positive charge which implies, $(\hat{\theta}, \hat{\phi})=(\pi-{\theta},\, \pi+{\phi})$.

Computing the squared amplitude  in \eq{sum}, using  \eq{2to3}, and averaging over initial fermion spins,  we obtain the fully differentiable cross-section for this process,
\begin{align}
\frac{d \sigma}{d (\cos \Theta)\, d(\cos \theta)\, d\phi\, d s}= K(s)\, \overline{|\mathcal{A}|^2}
&= K(s)\Bigg([\mom^1_{LL}(s,\Theta)\,s^2_{\Theta}\,+ \mom^2_{LL} (s,\Theta)]s^2_{\theta}   + \mom^1_{TT}(s,\Theta)\,c_{\Theta}\,c_{\theta} 
\nonumber\\ 
&+ [\mom^2_{TT}(s,\Theta)(1 + c^2_{\Theta})+ \mom^3_{TT}(s,\Theta)s^2_{\Theta}](1 + c^2_{\theta}) \nonumber\\ 
&+  [\mom^1_{LT}(s,\Theta) + \mom^2_{LT}(s,\Theta)\,c_{\Theta}\,c_{\theta}]\,s_{\Theta}\,s_{\theta}\,c_{\phi} \nonumber\\
&+  [\tilde{\mom}^1_{LT}(s,\Theta) + \tilde{\mom}^2_{LT}(s,\Theta)\,c_{\Theta}\,c_{\theta}] \,s_{\Theta}\,s_{\theta}\,s_{\phi} \nonumber\\
&+ \mom_{TT'}(s,\Theta)\,c_{2\phi}\,s^2_{\Theta}\,s^2_{\theta} + \tilde{\mom}_{TT'}(s,\Theta)\,s_{2\phi}\,s^2_{\Theta}\,s^2_{\theta} \Bigg)
\label{eq:ang_funcs}
\end{align}
where $s_{\alpha} = \mbox{sin~}\alpha$ and $c_{\alpha~} = \mbox{cos~}\alpha$. While the center of mass energy,  $\sqrt{s}$, is known for lepton colliders  it can be experimentally determined in hadron colliders by measuring the invariant mass of the $Zh$ system; this is possible for  decay modes where the Higgs boson is fully reconstructible, such as $h \to \gamma \gamma, b \bar{b}, 4 l$.  The factor, $K(s)$, contains energy dependant factors that  are same for the SM and EFT cross-sections;  for hadron colliders, in particular, this factor will contain the parton distribution functions. As \eq{2to3} is a sum over the amplitude for the three $Z$ polarizations,  the squared amplitude gives nine terms each with a distinct 
$(\theta, \phi)$ dependence. The subscript of the coefficients above shows the particular $Z$-helicity modes that need to be multiplied in order to obtain the corresponding angular function.

Once the dependence on $\Theta$ is taken into account more angular functions arise as shown in \eq{eq:ang_funcs}. In fact to capture the most general  $\Theta$ dependence including all possible values of $J$ in \eq{2to3}, one must further expand the $\mom_i$ above in powers of $\cos \Theta$ as follows,\footnote{Keeping only the $J=0,1$ terms while substituting \eq{par_wave} into \eq{2to3} gives the distribution in \eq{eq:ang_funcs} with the $A_i$  independent of $\Theta$. To see that all the effects due to  the higher values of $J$ in \eq{par_wave} can be incorporated by using the functional form in \eq{theta}, note that the Wigner functions with  $J\geq 2$ can be written as $d^J_{m,m'}= (\sum_p b_p \cos^p \Theta)d^1_{m,m'}$, where the $b_p$ are numerical coefficients.}
\begin{eqnarray}
\mom_i(s,\Theta)=\sum_{n=0}^\infty \mom_{i,n} (s) (\cos \Theta)^{2 n}.
\label{theta}
\end{eqnarray}
The  coefficients  $\mom_{i,n} (s)$ are called angular moments. The fact that only even powers of $\cos \Theta$ appear in \eq{theta} is because, as shown in App.~\ref{app:cpt}, unitarity and CPT invariance imply that the amplitude squared must be invariant under,
 \begin{equation}
    \Theta\;\longrightarrow\;\pi-\Theta,
    \qquad
{\theta}\;\longrightarrow\;\pi-{\theta}.
\end{equation}
We also show in App.~\ref{app:cpt} that the angular moments corresponding to the angular functions that have a $\sin \phi$ or $\sin 2\phi$ probe CP violating EFT contributions.

All the angular functions in \eq{eq:ang_funcs} can be, in principle, probed in a lepton collider. For $pp$ colliders like the LHC, however, an additional subtlety arises. The angles $\Theta$ and $\phi$ are  defined with respect to the direction of the fermion, $f$,  in the $f \bar{f} \to Zh$ process (the definition of the third angle, $\theta$, does not depend on the direction of the fermion). In a $pp$ collider it is impossible to access this information. If  one arbitrarily chooses a direction, $\hat{b}$,  along the beam  to define  $\Theta$ and $\phi$, \eq{eq:ang_funcs} will be correct only for events where the fermion happens to be in this direction, i.e. if $\hat{f}=\hat{b}$. For events where the fermion is in the direction opposite to the chosen direction, i.e. if $\hat{f}=-\hat{b}$, we must make the replacement  $\left(\Theta, \phi \right) \to \left(\pi- \Theta, \pi +\phi \right)$  in \eq{eq:ang_funcs}. If $\hat{b}$ is chosen to be a fixed direction, the angular moments corresponding to functions that are odd under this replacement, namely $\mom^1_{TT,n}, \mom^1_{LT,n}$ and $\tilde{\mom}^1_{LT,n}$, will average to zero. On the other hand, if $\hat{b}$  is chosen to be in the direction of the boost of the $Zh$ system, these moments will not be washed out because of the fact that   direction of this boost is strongly correlated with the direction of the more energetic valence quark.
 
The  angular moments,  $\mom_{i,n} (s)$, are differential observables that can be experimentally measured using multivariate methods. Multivariate methods are essential because   the cross-helicity terms in \eq{eq:ang_funcs}---i.e. the terms arising  from the interference of  amplitudes for  the production of different helicities of the $Z$-boson---vanish upon integration over the leptonic decay angles $\theta$ and $\phi$. This is the case if  one uses   inclusive observables defined in terms of  the $Z$-boson four momentum. For instance it was shown in Ref.~\cite{Banerjeefull} that   the largest effect the Wilson coefficients, $c^{3p}_4$ and $c^{3p}_5$, are respectively on the  angular moments, $\mom^2_{LT,0}$ and $\tilde{\mom}^2_{LT,0}$; the corresponding angular functions vanish upon integration over any of the three angles, $\Theta, \theta$ and $\phi$. To observe these  effects it is thus essential to use multivariate methods that treat  all the three angles differentially. Such studies that go beyond traditional inclusive observables and  aim to probe cross-helicity interference terms in processes with  electroweak gauge bosons  have been dubbed `interference resurrection' studies in the literature (see Ref.~\cite{Hagiwara:1986vm, panico, azatov2, Azatov:2019xxn,atlasir,cmsir} for such studies in diboson channels and Ref.~\cite{Banerjeetensor, Banerjeefull, Banerjeefull2, cmseft} for similar studies in Higgs production and decay channels).


In Ref.~\cite{Banerjeefull},  a phenomenological study  of the $pp \to Z(\ell\ell) h(bb)$ process was performed where the method of moments  was used to extract the angular moments~\cite{James, Dunietz1, Dunietz2} in an unbinned analysis (see also Ref.~\cite{Banerjeetensor}). This method involves deriving a set of weight functions that can be convoluted with the full angular distribution in order to extract the different angular moments (see also Ref.~\cite{Banerjeefull2} where this method was used to probe SMEFT effects in the golden channel, $h \to ZZ^\star \to 4 l$).
The CMS experiment has now adopted an approach~\cite{cmseft} where the full event information is retained and  machine learning techniques are used to efficiently  probe the different terms in the angular distribution shown in \eq{eq:ang_funcs}.  Both Ref.~\cite{Banerjeefull} and~\cite{cmseft}  were restricted to  the terms in \eq{eq:ang_funcs} that arise at the dimension-6 level in the SMEFT and did not consider operators with opposite fermion helicities. The  methods used in these studies can be extended to probe the higher order EFT effects discussed in this work.

\subsection{Mapping between angular moments and EFT Wilson coefficients  up to any given order}
\label{mapping2}

Up to a given order  in $E/\Lambda$, only  partial wave coefficients up to finite value of $J$  enter the expressions for the angular moments. For instance if we truncate the EFT expansion at ${\cal O}(E^3/\Lambda^3)$, only partial wave coefficients with $J\leq 2$ ($J\leq 1$) for opposite (identical)   fermion helicities enter the expressions for the angular moments.  A mapping between angular moments and Wilson coefficients up to this order  can then be obtained by rewriting the partial wave coefficients in \eq{2to3}  in terms of the $J=0,1,2$ helicity directions by inverting eqs.~(\ref{pdirections}),~(\ref{heldir2}),~(\ref{heldirLRJ0}) and~(\ref{heldirLR}). Following the steps discussed in Sec.~\ref{angmom},
we then obtain the mapping between angular moments and helicity directions up to ${\cal O}(E^3/\Lambda^3)$ in Table~\ref{tab:final}.  It is straightforward to convert these expressions for the angular moments in terms of helicity directions to expressions  in terms of Wilson coefficients using Tables~\ref{tab:pdir},~\ref{tab:J2dir} and~\ref{tab:LR_RL_hel_dir}. 

If we want to derive this mapping to orders higher than  $(E^3/\Lambda^3)$, we must include partial wave contributions with higher values of $J$. As the methods discussed in Sec.~\ref{sec3} allow us to derive the mapping between partial wave coefficients and Wilson coefficients for any $J$, it is also  possible to obtain the mapping between the angular moments and the Wilson coefficients to any order in $E/\Lambda$.

Table~\ref{tab:final} has been organised order by order in $E/\Lambda$. We see that up to  ${\cal O}(E^3/\Lambda^3)$ only the moments $A_{i,0}$ are non zero. We have separated the powers of $E$ in Table~\ref{tab:final} as follows, 
\bea
\left(\frac{E}{m_Z}\right)^m \left(\frac{E}{\Lambda}\right)^n
\eea
where only the second  factor arises from the EFT derivatives while the first factor comes from propagators, $Z$-polarisation vectors and the  fermion spinors. While must have $n>0$, $m$ can have any sign.  Also, note that the $\alpha_i$ and $\epsilon_i$ in the table entries  carry energy dependence. The entries in Table~\ref{tab:final} are a quadratic function of the $\delta^p \mathbf{a}^m_{J,k}$ defined in  \eq{pow},~(\ref{pow2}), (\ref{pow3}) (\ref{pow4}) and (\ref{pow5}).  Recall that   $\delta^p \mathbf{a}^m_{J,k}$ is defined such that it is accompanied by a factor   $(E/\Lambda)^{p+m}$. Thus if one of the terms in Table~\ref{tab:final}  is proportional to, 
\bea
 \delta^p \mathbf{a}^r_{J_1,k}\delta^q \mathbf{a}^s_{J_2,l}\left(\frac{E}{m_Z}\right)^m \left(\frac{E}{\Lambda}\right)^n
 \label{countsuper}
\eea
we must have, $n=p+r+q+s$. This fact allows us to know which pairs of, $\delta^p \mathbf{a}^r_{J_1,k}$ and $ \delta^q \mathbf{a}^s_{J_2,l}$,  can appear at a given order in $E/\Lambda$.

 
Measurement of the angular moments at multiple energy bins can be used to systematically determine the helicity directions using Table~\ref{tab:final}. As these helicity directions are linear combinations of Wilson coefficients given by  using Tables~\ref{tab:pdir},~\ref{tab:J2dir} and~\ref{tab:LR_RL_hel_dir},  it is then straightforward to determine/constrain the EFT Wilson coefficients.

The general form of the angular distribution in \eq{eq:ang_funcs} was derived directly from the partial wave expansion. One might ask what new  information does EFT provide?  As we have already discussed before, EFT provides input from dimensional analysis. The amplitude at lower orders is not the most general one and is described by a limited set of parameters equal to the number of Wilson coefficients up to that order. For instance, the $ff \to hZ$ amplitude at tree level in the Standard Model is CP conserving. The CP violating effects are contained in the moment functions in \eq{eq:ang_funcs} having the $s_\phi$ and $s_{2 \phi}$ factors, i.e. the angular moments $\tilde{\mom}^{1}_{LT,n}, \tilde{\mom}^{2}_{LT,n}$ and $\tilde{\mom}_{TT',n}$. These CP violating effects first arise at ${\cal O}(E/\Lambda)$ which corresponds to the dimension-6 level in the SMEFT~\cite{Banerjeetensor, Banerjeefull}. Another example is that the angular moments $\mom^2_{LL,0}$ and $\mom^3_{TT,0}$ arise for the first time at ${\cal O}(E^2/\Lambda^2)$. Finally, all the angular moments $\mom_{i,n}$  in \eq{theta} with $n\geq1$ appear at ${\cal O} (E^6/\Lambda^6)$ or higher.

Apart from these  instances, there is another somewhat less obvious way in which the EFT puts restrictions on the differential distributions at lower orders.  Consider for instance the expressions for the angular moments in Table~\ref{tab:final} at order $(E/\Lambda)^0$. Using only partial wave analysis one may naively assume that  12  $J=1$  partial wave coefficients  (see discussion below \eq{cpt})  would   independently enter the  expressions for the angular moments. Our analysis shows, however, that there are only 2 helicity directions that appear at $(E/\Lambda)^0$  (see Tables~\ref{tab:pdir}, \ref{tab:J2dir} and \ref{tab:LR_RL_hel_dir}). Thus only these  two helicity directions can enter the expressions  for  the angular moments at $(E/\Lambda)^0$. Similarly,    only 8  out of the 12 helicity directions   appear  up to ${\cal O}((E/\Lambda)$ (see   Tables~\ref{tab:pdir},~\ref{tab:J2dir} and~\ref{tab:LR_RL_hel_dir}) and only these linear combination of partial wave coefficients can  enter the  expressions for angular moments at this order. In other words, EFT at lower orders predicts correlations   even among the angular moments. For instance, going from  ${\cal O} (E^2/\Lambda^2)$ to ${\cal O} (E^3/\Lambda^3)$ in Table~\ref{tab:final}, the effect of the new operators  is not to generate new angular moments but to break     correlations between existing  angular moments.

\section{Summary and conclusions}
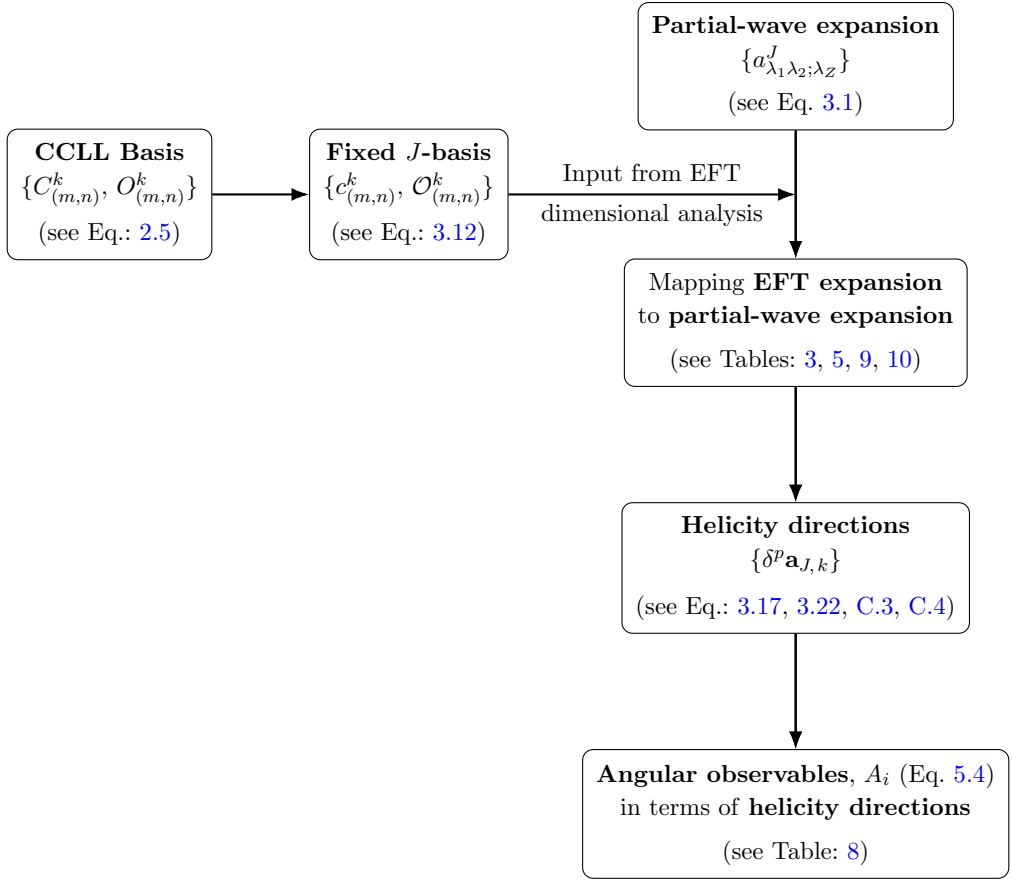
\begin{figure}[t]
\centering
\begin{tikzpicture}[scale=0.85, transform shape,
    box/.style = {
        draw,
        rectangle,
        rounded corners,
        align=center,
        inner sep=6pt
    },
    arrow/.style = {
        ->,
        thick,
        >=Latex
    }
]

\node[box] (ccll) {
    \textbf{CCLL Basis} \\
    $\{C^{k}_{(m,n)},\, O^k_{(m,n)}\}$ \\[0.2cm]
    (see Eq.:~\ref{ccll})
};

\node[box, right=1.5cm of ccll] (fixedJ) {
    \textbf{Fixed $J$-basis} \\
    $\{c^k_{(m,n)},\, \mathcal{O}^k_{(m,n)}\}$ \\[0.2cm]
    (see Eq.:~\ref{basisdef})
};

\node[box, right=2cm of fixedJ, yshift=2cm] (pw) {
    \textbf{Partial-wave expansion} \\
    $\{a^J_{\lambda_1\lambda_2;\lambda_Z}\}$ \\[0.15cm]
    (see Eq.~\ref{par_wave})
};

\node[box, below=2cm of pw] (map) {
    Mapping \textbf{EFT expansion} \\
    to \textbf{partial-wave expansion} \\[0.15cm]
    (see Tables:~\ref{tab:full_Amps_Exact}, \ref{tab:full_Amps_Exact_J2}, \ref{tab:LR_amp_J0}, \ref{tab:LR_amp_J1})
};

\node[box, below=1.8cm of map] (hel) {
    \textbf{Helicity directions} \\
    $\{\delta^p \textbf{a}_{J,\,k}\}$ \\[0.2cm]
    (see Eq.:~\ref{pdirections}, \ref{heldir2}, \ref{heldirLRJ0}, \ref{heldirLR})
};

\node[box, below=1.8cm of hel] (obs) {
    \textbf{Angular observables}, $A_i$  (Eq.~\ref{eq:ang_funcs}) \\
    in terms of \textbf{helicity directions} \\[0.15cm]
    (see Table:~\ref{tab:final})
};

\draw[arrow] (ccll) -- (fixedJ);
\draw[arrow] (pw) -- (map);
\draw[arrow] (map) -- (hel);
\draw[arrow] (hel) -- (obs);
\path (pw) -- (map) coordinate[midway] (pwmapmid);
\draw[arrow] (pw) -- (map);
\draw[arrow] (fixedJ) -- (pwmapmid)
node[midway, above] {Input from EFT}
  node[midway, below] {dimensional analysis};

\draw[arrow] (map) -- (hel);
\draw[arrow] (hel) -- (obs);
\end{tikzpicture}
\caption{A schematic summary of the steps we have taken to arrive at our final results. See the text for more details.}
\label{schematic}
\end{figure}
We have developed methods to map the angular moments for the Higgs-strahlung process---differential observables that parametrise the most general angular distribution of the process, see \eq{eq:ang_funcs} and \eq{theta}---to EFT Wilson coefficients up to any desired order. Some of the main results of this work are in Tables~\ref{tab:full_Amps_Exact},~\ref{tab:full_Amps_Exact_J2},~\ref{tab:LR_amp_J0} and~\ref{tab:LR_amp_J1}, where we explicitly present expressions for  partial wave coefficients with $J=0,1$ and $2$ to all orders in the EFT expansion; our methods can be  similarly utilised to derive such a mapping for any value of $J$. We have then used these results to express  the angular moments in terms of Wilson coefficients up to ${\cal O}(E^3/\Lambda^3)$ in Table~\ref{tab:final}. While we have presented our results up to this order,  our methods allow us to write the angular moments  up to any desired order. It is conceptually straightforward to extend our methods to other 2-to-2 collider processes  such as the Drell-Yan process and diboson production.

We review in Fig.~\ref{schematic} the steps we have taken to arrive at these results. First we write the EFT contributions to all orders using the formalism developed by Ref.~\cite{Luty2023}. In Ref.~\cite{Luty2023} the  amplitude for the $f \bar{f} \to hZ$ process was written as a sum of 12 principal terms each followed by a series of Mandelstam descendants that can be obtained by multiplying factors of, $(s^m (t-u)^n/\Lambda^{2 (m+n)})$, to each principal amplitude.  The operators corresponding to these principal amplitudes have been shown in Table~\ref{tab:3-point_ops} and~\ref{tab:4-pt_principals}.  We call the EFT basis formed by these operators and their descendants as the CCLL basis and show the corresponding lagrangian in \eq{ccll}.

We then proceed to derive a mapping between partial wave coefficients and EFT Wilson coefficients. First, we  modify the EFT basis of  Ref.~\cite{Luty2023}  to one where each operator contributes to partial wave coefficients with a single $J$, see \eq{basisdef}.   This makes  the mapping between  partial wave coefficients and Wilson coefficients  particularly simple (see Fig.~\ref{fig:one}) and allows us to express partial wave coefficients  in terms of the Wilson coefficients to all orders in the EFT, see Tables~\ref{tab:full_Amps_Exact},~\ref{tab:full_Amps_Exact_J2},~\ref{tab:LR_amp_J0} and~\ref{tab:LR_amp_J1}.  This mapping allows us to incorporate EFT dimensional analysis into partial wave analysis and enables us to construct helicity directions, i.e.  linear combinations of partial wave coefficients that get EFT contributions at different orders of $E/\Lambda$, see  Table~\ref{tab:pdir},~\ref{tab:J2dir} and Table~\ref{tab:LR_RL_hel_dir}.  These results can then be used to write the squared amplitude---and thus the differential distributions for the Higgs-strahlung process---to any desired order in $E/\Lambda$. This exercise has been carried out up to ${\cal O} (E^3/\Lambda^3)$ in Table~\ref{tab:final}.

Given the expected increase in experimental sensitivities and the ongoing progress in higher order electroweak and QCD corrections, a systematic approach to  higher order EFT effects is the need of the hour. 
The methods developed in this work allow us to write the differential distributions of collider processes to any desired EFT order. This knowledge is essential for  many different multivariate techniques  such as the method of moments, machine learning techniques, the matrix element method etc.  We thus believe our methods will be indispensable for any   experimental study that aims to probe higher order EFT effects.

\section*{Acknowledgements}
We  would like to thank Joydeep Chakrabortty for  useful discussions. We acknowledge the support from the Department of Atomic Energy (DAE), Government of India, under Project Identification Number RTI 4002.

\refstepcounter{table}
\label{tab:8}

\appendix

\section{Implications of unitarity, CPT and CP invariance on the partial wave coefficients}
\label{app:cpt}
In this appendix, we discuss the constraints on the helicity amplitudes of the $2\!\to\!2$ scattering process $f\overline{f}\to Zh$, as imposed by the general principles of quantum field theory.  The scattering amplitude is defined as
\begin{equation}
    T^{\,f\overline{f}\to Zh}_{\lambda_1\,\lambda_2;\,\lambda_Z}(s,\Theta)
    = \langle p_1,p_2;\lambda_1,\lambda_2 \vert \mathcal{T} 
      \vert p_Z,p_h;\lambda_Z\rangle,
\end{equation}
where $\mathcal{T}$ is the transfer matrix (with $\mathcal{S} = \mathbbm{1} + i\,\mathcal{T}$). Here $p_1$ and $p_2$ denote the four-momenta of the incoming fermion and anti-fermion, with helicities $\lambda_1$ and $\lambda_2$, respectively; $p_Z$ and $p_h$ denote the four-momenta of the outgoing $Z$ boson and Higgs boson, 
and $\lambda_Z$ is the helicity of the outgoing $Z$.  

We begin by examining the implications of unitarity, which expresses the requirement that total probability must be conserved.  In operator form, unitarity of the $\mathcal{S}$-matrix implies
\begin{eqnarray}
    \mathcal{S}\mathcal{S}^{\dagger}&=&\mathbbm{1}\nonumber\\
     \mathcal{T}-\mathcal{T}^{\dagger}&=&i\mathcal{T}\mathcal{T}^{\dagger}\label{Eq.App.A1}
\end{eqnarray}
Next we insert in-state and out-state of our interest in both sides of eq.(\ref{Eq.App.A1}).
\begin{eqnarray}
    \langle p_1,p_2;\lambda_1,\lambda_2\vert \mathcal{T}\vert p_Z, p_h;\lambda_z\rangle - \langle p_1,p_2;\lambda_1,\lambda_2\vert \mathcal{T}^{\dagger}\vert p_Z, p_h;\lambda_z\rangle \nonumber\\
    = i\langle p_1,p_2;\lambda_1,\lambda_2\vert \mathcal{T}\mathcal{T}^{\dagger}\vert p_Z, p_h;\lambda_z\rangle\label{Eq.App.A2}
\end{eqnarray}
Here, we are interested in the helicity amplitudes generated at tree level by the insertion of effective operators arising from dynamics at a scale above the new-physics cutoff $\Lambda$. Consequently, for our purposes, when $m_Z^2 <s \ll \Lambda^{2}$, there is no physical production threshold associated with these heavy degrees of freedom, and the absorptive part of the amplitude on the right-hand side of eq.~(\ref{Eq.App.A2}) vanishes. So, we are left with the following equality,
\begin{eqnarray}
    \langle p_1,p_2;\lambda_1,\lambda_2\vert \mathcal{T}\vert p_Z, p_h;\lambda_z\rangle &=& \left(\langle  p_Z, p_h;\lambda_z\vert \mathcal{T} \vert p_1,p_2;\lambda_1,\lambda_2\rangle\right)^{\ast} \nonumber \\
    T^{f\overline{f}\to Zh}_{\lambda_1\;\lambda_2;\;\lambda_Z}(s,\Theta) &=&  \left(T^{Zh\to f\overline{f}}_{\lambda_Z;\;\lambda_1\;\lambda_2}(s,\Theta)\right)^{\ast}
    \label{Eq.App.A3}
\end{eqnarray}

Now, after applying CPT transformation on the  right-hand side of eq.(\ref{Eq.App.A3}), we find, 
\bea
T^{Zh\to f\overline{f}}_{\lambda_z;\;\lambda_1\;\lambda_2}(s,\Theta) = T^{\overline{f}f\to Zh}_{-\lambda_1\;-\lambda_2;\;-\lambda_Z}(s,\Theta) = T^{f\overline{f}\to Zh}_{-\lambda_2\;-\lambda_1;\;-\lambda_Z}(s,\pi -\Theta).
\eea
This provides the following constraint on the structure of the amplitudes,
\begin{eqnarray}
    T^{f\overline{f}\to Zh}_{\lambda_1\;\lambda_2;\;\lambda_z}(s,\Theta) = \left(T^{f\overline{f}\to Zh}_{-\lambda_2\;-\lambda_1;\;-\lambda_z}(s,\pi-\Theta)\right)^{\ast}. \label{Eq.App.A4}
\end{eqnarray}

In the following discussion, we will drop the superscript $f\overline{f}\!\to Zh$ in the helicity amplitude $T^{\,f\overline{f}\to Zh}_{\lambda_1\,\lambda_2;\,\lambda_Z}(s,\Theta)$, as we will be concerned exclusively with the $f\overline{f}\!\to Zh$ process. Next, we decompose the helicity amplitudes into partial waves with definite total angular momentum $J$ in the following manner, 
\begin{eqnarray}
    {\cal M}_{\lambda_1\,\lambda_2\,;\lambda_Z}(s,\Theta) = 4\pi \sum_{J=J_{\text{min}}} (2J+1)d^J_{\sigma,\lambda_Z}(\Theta)a^J_{\lambda_1\;\lambda_2;\;\lambda_Z}(s).\label{Eq.App.A5}
\end{eqnarray}
Substituting eq.(\ref{Eq.App.A5}) in Eq.(\ref{Eq.App.A4}) and using the identity $d^{J}_{\sigma,-\lambda_Z}(\pi-\Theta)=(-)^{J-\sigma}d^{J}_{\sigma,\lambda_Z}(\Theta)$, finally we reach the following relation on the partial wave amplitudes,
\begin{eqnarray}
    a^J_{\lambda_1\;\lambda_2;\;\lambda_Z}(s) = (-1)^{J-\sigma}\left(a^J_{-\lambda_2\;-\lambda_1;\;-\lambda_Z}(s)\right)^{\ast}.\label{Eq.App.A6}
\end{eqnarray}

Next, we examine the implications of the constraint in eqs.~(\ref{Eq.App.A4},\,\ref{Eq.App.A6}) 
for the parametrization of the $2\to2$ scattering amplitude in terms of principal and descendant contributions.  
We express the amplitude in the form,
\begin{eqnarray}
    {\cal M}_{\lambda_1\,\lambda_2;\,\lambda_Z}(s,\Theta) 
    = \sum_{i=1}^{N_p}
{P}^i_{\lambda_1\,\lambda_2;\,\lambda_Z}(s,\Theta)
    \left(1 + 
    \sum_{\substack{m,n\ge 0 \\ (m,n)\neq (0,0)}} 
    c^{i}_{(m,n)}\, s^{m} (t-u)^{n} \right).
    \label{Eq.App.A7}
\end{eqnarray}
where ${P}^i_{\lambda_1\,\lambda_2;\,\lambda_Z}(s,\Theta)$ denotes the contribution of $i$ th principal operator to the above helicity amplitude.  
Each principal amplitude is multiplied by a series in the Mandelstam variables, representing the tower of descendants associated with that principal.  
As expected, the principal amplitudes alone satisfy the constraints in eq.~(\ref{Eq.App.A4}); therefore, in order for the full amplitude to respect these constraints, the descendant series must satisfy them independently.  
This requirement imposes the following condition on the coefficients $c^{i}_{(m,n)}$ appearing in the descendant expansion:
\begin{eqnarray}
    c^{i}_{(m,n)} = (-1)^{n}\, \left(c^{i}_{(m,n)}\right)^{\ast},
    \label{Eq.App.A8}
\end{eqnarray}
which implies that all coefficients multiplying odd powers of $(t-u)$ must be purely imaginary, whereas those multiplying even powers of $(t-u)$ must be purely real.

Let us   now
comment on the structure of the angular dependence appearing in the
spin-averaged squared amplitude of the Higgs-strahlung process.  The
helicity amplitude for $f \bar{f}\to Z(\ell\ell)\,h$ process can be written as,
\begin{equation}
    \mathcal{A}_{\lambda_1\lambda_2}
    (s,\Theta,\hat{\theta},\hat{\phi})
    =
    \frac{-ig^Z_\ell}{\Gamma_Z}
    \sum_{\lambda_Z=0,\pm1}
   {\cal M}_{\lambda_1\lambda_2;\lambda_Z}(s,\Theta)\,
    d^{1}_{1,\lambda_Z}(\hat{\theta})\,
    e^{i\lambda_Z\hat{\phi}},
    \label{Eq.App.A9}
\end{equation}
Taking the complex conjugate of
eq.~(\ref{Eq.App.A9}), and using 
eq.~(\ref{Eq.App.A4}) together with the identity
$d^{1}_{1,\lambda_Z}(\hat{\theta})
 = d^{1}_{1,-\lambda_Z}(\pi-\hat{\theta})$, one finds that the full
amplitude satisfies the constraint
\begin{equation}
    \left(
    \mathcal{A}_{\lambda_1\lambda_2}
    (s,\Theta,\hat{\theta},\hat{\phi})
    \right)^{\!*}
    =
    -\,
    \mathcal{A}_{-\lambda_2,-\lambda_1}
    (s,\pi-\Theta,\pi-\hat{\theta},\hat{\phi}).
    \label{Eq.App.A10}
\end{equation}
This relation connects each helicity configuration to its opposite
helicity under simultaneous transformations of the production and decay
angles, $(\Theta,\hat{\theta})\to(\pi-\Theta,\pi-\hat{\theta})$. Consequently, the spin-averaged squared amplitude,
\begin{equation}
    \overline{|\mathcal{A}|^2}
    = \sum_{\lambda_1,\lambda_2}
      |\mathcal{A}_{\lambda_1\lambda_2}(s,\Theta,\hat{\theta},\hat{\phi})|^2
    = 
      \overline{|\mathcal{A}_{LR/RL}|^2}
      +\overline{|\mathcal{A}_{RR}|^2}
      +\overline{|\mathcal{A}_{LL}|^2},\label{Eq.App.A11}
\end{equation}
inherits this symmetry. Explicitly, one finds
\begin{align}
    \overline{|\mathcal{A}_{LR/RL}|^2} 
        &= |\mathcal{A}_{++}(s,\Theta,\hat{\theta},\hat{\phi})|^2
         + |\mathcal{A}_{++}(s,\pi-\Theta,\pi-\hat{\theta},\hat{\phi})|^2, \label{Eq.App.A12}
\\[2mm]
    \overline{|\mathcal{A}_{LL}|^2}
        &= -\,\mathcal{A}_{-+}(s,\Theta,\hat{\theta},\hat{\phi})\,
            \mathcal{A}_{-+}(s,\pi-\Theta,\pi-\hat{\theta},\hat{\phi}), \label{Eq.App.A13}
\\[2mm]
    \overline{|\mathcal{A}_{RR}|^2}
        &= -\,\mathcal{A}_{+-}(s,\Theta,\hat{\theta},\hat{\phi})\,
            \mathcal{A}_{+-}(s,\pi-\Theta,\pi-\hat{\theta},\hat{\phi}), \label{Eq.App.A14}
\end{align}
each of which is manifestly even under
\begin{equation}
    \Theta\;\longrightarrow\;\pi-\Theta,
    \qquad
    \hat{\theta}\;\longrightarrow\;\pi-\hat{\theta}.
\end{equation}

 Note that the above symmetry applies to each term in \eq{sum} individually. Writing  this requirement in terms of $\theta$ and $\phi$, the angles corresponding to the positively charged lepton, we obtain,
\begin{equation}
    \Theta\;\longrightarrow\;\pi-\Theta,
    \qquad
{\theta}\;\longrightarrow\;\pi-{\theta},
\end{equation}
for both the terms. All angular structures appearing in the squared and
spin-averaged amplitude of the Higgs–strahlung process must be even
under these transformations. This symmetry strongly constrains the
allowed angular dependences of the differential cross section.

It is useful to note that, under the additional assumption of CP
conservation, the amplitude in eq.~(\ref{Eq.App.A9}) satisfies the
further constraint,
\begin{equation}
    \left(
    \mathcal{A}_{\lambda_1\lambda_2}
    (s,\Theta,\hat{\theta},\hat{\phi})
    \right)^{\!*}
    =
    -\,\mathcal{A}_{\lambda_1\lambda_2}
    (s,\Theta,\hat{\theta},2\pi-\hat{\phi}),
    \label{Eq.App.A15}
\end{equation}
Consequently, each of the three
contributions appearing on the right-hand side of
eq.~(\ref{Eq.App.A11}) inherits this symmetry, and must therefore be
even under the transformation $\hat{\phi}\to 2\pi-\hat{\phi}$. Once again \eq{Eq.App.A15} applies  to each term of \eq{sum} individually.  This allows us to state the above requirement in terms of the angles, $\theta$ and $\phi$, by simply replacing $(\hat{\theta}, \hat{\phi}) \to ({\theta}, {\phi})$ in \eq{Eq.App.A15}.  In the
CP-conserving case, the angular dependence in ${\phi}$ can thus
enter the spin-averaged squared amplitude only through the even
structures $\cos{\phi}$ and $\cos 2{\phi}$.  By contrast, CP
violation manifests itself through the appearance of the odd structures
$\sin{\phi}$ and $\sin 2{\phi}$, which arise from the
interference between the CP-even and CP-odd parts of the amplitude.

\section{Projection Operators}
\label{app:proj}

As discussed  in Sec.~\ref{sec:proj}, any operator contributing to the $f\bar{f} \to Zh$ process can be decomposed into a sum of fixed $J$ operators,
    \bea
O_{i}= \sum_{J=0}^{J_{max} }\alpha_{(i|J)} O_{(i|J)}
\label{above}
\eea
This can be obtained by writing the operator as, 
\bea
O_i=\left[ \bar{f} f\right]_{\mu \nu \rho...} \left[ hZ\right]^{\mu \nu \rho...}
\label{eg}
\eea
where $\left[ hZ\right]^{\mu \nu \rho...}$ is a rank-$n$  tensor constructed from the Higgs and  $Z$-boson fields, whereas  $\left[ \bar{f} f\right]_{\mu \nu \rho...}$ is a rank-$n$ object built from fermion and anti-fermion fields. As discussed in Sec.~\ref{sec:proj}, one can now decompose this operator as in \eq{above} by inserting a complete set of projection operators $\sum_{i,J} ({\cal P}^{i,J})^2$ where the projection operator ${\cal P}^{i,J}$ acting on a rank-$n$ tensor projects it on to a tensor that transforms under the Lorentz group with a fixed $J$ (see \eq{eq:PT}).

We now discuss the construction of these projection operators. It will be convenient  to work in momentum space. For clarity, let us first consider the case 
in which both  $\left[\bar{f} f\right]$ and  $\left[ hZ\right]$ are rank--2 Lorentz tensors. In this setting, it is 
convenient to introduce the following set of tensors,
\begin{eqnarray}
    \theta_{\mu\nu} 
    &=& 
    \eta_{\mu\nu}-\frac{P_{\mu}P_{\nu}}{\mu^2},
    \qquad
    L_{\mu\nu} = \frac{P_{\mu}P_{\nu}}{\mu^2},
    \label{Eq.APP.B2}
\end{eqnarray}
where $P^\mu$ denotes the total four-momentum carried by the two-particle 
incoming (or outgoing) state associated with the corresponding tensor blocks 
$\left[\bar{f} f\right]$ and  $\left[ hZ\right]$. In the center-of-mass frame one has  \(P^\mu = (E_{\rm cm},\,0,0,0)\) and \r{\( \mu \equiv E_{\rm cm}\)}. The operator $\theta_{\mu\nu}$ projects a four-vector onto the subspace transverse to $P_\mu$, while $L_{\mu\nu}$ projects onto the longitudinal direction parallel to $P_\mu$.

We define, $\mathcal{P}^{i,J}$, to be a projection operator that projects an arbitrary rank--2 tensor onto its 
spin-$J$ component under the little group $\mathrm{SO}(3)$. The additional 
index $i$ distinguishes cases where the same spin $J$ appears multiple times in the decomposition. A complete set of Lorentz-covariant projection operators 
$\left(\mathcal{P}^{i,J}\right)^{\mu_1\mu_2}{}_{\nu_1\nu_2}$, which decompose a 
general rank--2 tensor into irreducible representations of the little group 
$\mathrm{SO}(3)$, must satisfy the following completeness and idempotency 
conditions,
\begin{eqnarray}
\sum_{i,J}
\left(\mathcal{P}^{i,J}\right)^{\mu_1\mu_2}{}_{\nu_1\nu_2}
&=&
\delta^{\mu_1}_{\nu_1}\,\delta^{\mu_2}_{\nu_2},
\nonumber\\[6pt]
\left(\mathcal{P}^{i,J}\right)^{\mu_1\mu_2}{}_{\nu_1\nu_2}
\left(\mathcal{P}^{i,J}\right)^{\nu_1\nu_2}{}_{\rho_1\rho_2}
&=&
\left(\mathcal{P}^{i,J}\right)^{\mu_1\mu_2}{}_{\rho_1\rho_2}.
\label{Eq.APP.B3}
\end{eqnarray}

For the present case, the full set of projection operators 
may be written in terms of $\theta^{\mu}{}_{\nu}$ and $L^{\mu}{}_{\nu}$ defined in eq.(\ref{Eq.APP.B2}) in the following manner,
\begin{eqnarray}
\left(\mathcal{P}^{0,0}\right)^{\mu_1\mu_2}{}_{\nu_1\nu_2}
&=&
L^{\mu_1}{}_{\nu_1}\,L^{\mu_2}{}_{\nu_2},
\label{Eq.App.B4}
\\[6pt]
\left(\mathcal{P}^{1,0}\right)^{\mu_1\mu_2}{}_{\nu_1\nu_2}
&=&
\frac{1}{3}\,\theta^{\mu_1\mu_2}\,\theta_{\nu_1\nu_2},
\label{Eq.App.B5}
\\[6pt]
\left(\mathcal{P}^{0,1}\right)^{\mu_1\mu_2}{}_{\nu_1\nu_2}
&=&
\frac{1}{2}\!\left(
L^{\mu_1}{}_{\nu_1}\,\theta^{\mu_2}{}_{\nu_2}
+
\theta^{\mu_1}{}_{\nu_1}\,L^{\mu_2}{}_{\nu_2}
+ (\nu_1 \leftrightarrow \nu_2)
\right),
\label{Eq.App.B6}
\\[6pt]
\left(\mathcal{P}^{1,1}\right)^{\mu_1\mu_2}{}_{\nu_1\nu_2}
&=&
\frac{1}{2}\!\left(
L^{\mu_1}{}_{\nu_1}\,\theta^{\mu_2}{}_{\nu_2}
+
\theta^{\mu_1}{}_{\nu_1}\,L^{\mu_2}{}_{\nu_2}
- (\nu_1 \leftrightarrow \nu_2)
\right),
\label{Eq.App.B7}
\\[6pt]
\left(\mathcal{P}^{2,1}\right)^{\mu_1\mu_2}{}_{\nu_1\nu_2}
&=&
\frac{1}{2}\!\left(
\theta^{\mu_1}{}_{\nu_1}\,\theta^{\mu_2}{}_{\nu_2}
- (\nu_1 \leftrightarrow \nu_2)
\right),
\label{Eq.App.B8}
\\[6pt]
\left(\mathcal{P}^{0,2}\right)^{\mu_1\mu_2}{}_{\nu_1\nu_2}
&=&
\frac{1}{2}\!\left(
\theta^{\mu_1}{}_{\nu_1}\,\theta^{\mu_2}{}_{\nu_2}
+ (\nu_1 \leftrightarrow \nu_2)
\right)
-
\left(\mathcal{P}^{1,0}\right)^{\mu_1\mu_2}{}_{\nu_1\nu_2}.
\label{Eq.App.B9}
\end{eqnarray}
Equations~\eq{Eq.App.B4} and \eq{Eq.App.B5} project a general rank--2 tensor onto two distinct spin--0 components. Equations~\eq{Eq.App.B6}--\eq{Eq.App.B8} correspond to three independent 
spin--1 projections.  
Finally, eq.~\eqref{Eq.App.B9} isolates the spin--2 component, which is the 
symmetric, traceless part of the tensor.

In principle, a tensor of rank $n$ can be decomposed into definite angular-momentum
components in complete analogy with the rank--2 case discussed above. Since our EFT 
basis is constructed by systematically retaining only the highest--$J$ component of 
each descendant operator, we focus here on the projection operator that extracts the 
spin-$n$ component of a rank-$n$ tensor. We define this projection as
\begin{eqnarray}
    \left[\mathcal{P}^{n}\,\Xi\right]^{\mu_1\cdots\mu_n}
    \equiv
    \left(\mathcal{P}^{n}\right)^{\mu_1\cdots\mu_n}{}_{\nu_1\cdots\nu_n}\,
    \Xi^{\nu_1\cdots\nu_n},
    \label{Eq.App.B10}
\end{eqnarray}
where $\mathcal{P}^{n}$ projects a general rank-$n$ tensor $\Xi$ onto its spin-$n$ 
component, corresponding to complete symmetrization of all indices together with the 
removal of all pairwise traces.

To construct $\mathcal{P}^n$ explicitly, we follow Ref.~\cite{ProjOps}, where a compact 
recursion relation for the symmetric traceless projector is derived. After adapting 
this result to our present notation the recursion relation becomes,
\begin{eqnarray}
\left(\mathcal{P}^{n}\right)^{\mu_1\cdots\mu_n}{}_{\nu_1\cdots\nu_n}
&=&
\frac{1}{n}
\sum_{i=1}^{n}
\theta^{\mu_i}{}_{\nu_n}\,
\left(\mathcal{P}^{n-1}\right)^{\mu_1\cdots[\mu_i]\cdots\mu_n}{}_{\nu_1\cdots\nu_{n-1}}
\nonumber
-
\frac{2}{n\,(2(n-2)+d)}\sum_{i<j}^{n}
\theta^{\mu_i\mu_j}\,
\theta_{\nu_n\alpha}\,\\
&&
\times
\left(\mathcal{P}^{n-1}\right)^{\mu_1\cdots[\mu_i]\cdots[\mu_j]\cdots\mu_n\alpha}{}_{\nu_1\cdots\nu_{n-1}},
\label{Eq.App.B11}
\end{eqnarray}
where $\mathcal{P}^{n-1}$ is the spin-$(n-1)$ projector acting on a rank-$(n-1)$ 
tensor, the notation $[\mu_i]$ indicates that the corresponding index is omitted 
from the list, and $d$ is the spatial dimensionality ($d=3$ in our case). Starting from $\left(\mathcal{P}^1\right)^{\mu_1}{}_{\nu_1} = \theta^{\mu_1}{}_{\nu_1}$,
one may iterate the recursion relation~\eq{Eq.App.B11} to obtain the explicit 
form of the spin-$n$ projector for any desired rank $n$.

\section{Helicity amplitudes with identical fermion helicities}
\label{lrrl}

In this appendix, we discuss the mapping of operators having fermions with opposite chiralities --- i.e. the 3-point operators ${\cal O}^{3p}_{6,7}$ and the four point principal operators ${\cal O}^{7-12}_{0,0}$ and their descendants--- to partial wave amplitudes with identical fermion helicities  and $J=0,1$. We work in the limit of vanishing fermion masses. We present these contributions in Table~\ref{tab:LR_amp_J0} and Table~\ref{tab:LR_amp_J1} which show the  rescaled partial wave coefficients of \eq{rescale}.

\begin{table}[t]
    \centering
    \begin{tabular}{|c|c|c|}
    \hline\rule{0pt}{3.5ex}
    EFT WCs  & $(++;\,0)$ & $(--;\,0)$ \\[0.2cm]
    \hline\rule{0pt}{4ex}
    $c_{(0,0)}^{11}$ & $-i\dfrac{E^2}{\Lambda^2}\dfrac{\alpha_2}{\alpha_1} $ & $i\dfrac{E^2}{\Lambda^2}\dfrac{\alpha_2}{\alpha_1}$ \\[2.5ex]
    \hline\rule{0pt}{4ex}
    $c_{(0,0)}^{12}$ & $\dfrac{E^2}{\Lambda^2}\dfrac{\alpha_2}{\alpha_1}$ & $\dfrac{E^2}{\Lambda^2}\dfrac{\alpha_2}{\alpha_1}$ \\[2.5ex]
    \hline
    \end{tabular}
    \caption{Contributions---up to all orders in EFT--- to,  $\hat{a}^{J_{min}}_{\lambda_{f} \lambda_{\bar{f}}; \lambda_Z}$, the minimum $J$ partial wave coefficients rescaled by the normalisation factor in \eq{rescale}. These contributions arise from the three point vertices, principal operators (see Table~\ref{tab:3-point_ops} and  \ref{tab:4-pt_principals}) and their $s$-descendants. We show here only the helicity channels with  opposite fermion helicities and $J_{min}=0$.  See the caption of 
    Table~\ref{tab:full_Amps_Exact} for the definition of $\alpha_1$ and $\alpha_2$.}
    \label{tab:LR_amp_J0}
\end{table}
\begin{table}[p]
    \begin{adjustbox}{max width=\textwidth}
    \centering
    \begin{tabular}{|c|c|c|c|c|c|c|}
\hline
   EFT & \multicolumn{6}{ c |}{Helicity Amplitudes}\\
    \cline{2-7}
     WCs & $(++;\,0)$ & $(++;\,+)$ &$(++;\,-)$ & $(--;\,0)$ & $(--;\,+)$ & $(--;\,-)$ \\[0.5ex]
     \hline\rule{0pt}{4ex}
     
    $ c^{3p}_6$ & $\dfrac{E}{\Lambda}\dfrac{m_Z}{E} \alpha_0$ & $\dfrac{E}{\Lambda}\dfrac{m_Z}{E} \alpha_0$ & $\dfrac{E}{\Lambda}\dfrac{m_Z}{E} \alpha_0$ & $-\dfrac{E}{\Lambda}\dfrac{m_Z}{E} \alpha_0$ & $- \dfrac{E}{\Lambda}\dfrac{m_Z}{E}\alpha_0$ & $-\dfrac{E}{\Lambda}\dfrac{m_Z}{E} \alpha_0$\\[2ex]
    \hline\rule{0pt}{4ex}
     $ c^{3p}_7$ & $i \dfrac{E}{\Lambda}\dfrac{m_Z}{E}\alpha_0$ & $i\dfrac{E}{\Lambda}\dfrac{m_Z}{E} \alpha_0$ & $i\dfrac{E}{\Lambda}\dfrac{m_Z}{E} \alpha_0$ & $i\dfrac{E}{\Lambda}\dfrac{m_Z}{E} \alpha_0$ & $i\dfrac{E}{\Lambda}\dfrac{m_Z}{E} \alpha_0$ & $i\dfrac{E}{\Lambda}\dfrac{m_Z}{E} \alpha_0$\\[2ex]
    \hline\rule{0pt}{4ex}
    $c^{3p}_6\,c^{3p}_1$ & $\dfrac{m_Z}{E}\dfrac{E}{\Lambda}\alpha_0$ & $\dfrac{m_Z}{E}\dfrac{E}{\Lambda}\alpha_0$ & $\dfrac{m_Z}{E}\dfrac{E}{\Lambda}\alpha_0$ & $-\dfrac{m_Z}{E}\dfrac{E}{\Lambda}\alpha_0$ & $-\dfrac{m_Z}{E}\dfrac{E}{\Lambda}\alpha_0$ & $-\dfrac{m_Z}{E}\dfrac{E}{\Lambda}\alpha_0$ \\[2ex]
    \hline\rule{0pt}{4ex}
    $c^{3p}_7\,c^{3p}_1$ & $i\dfrac{m_Z}{E}\dfrac{E}{\Lambda}\alpha_0$ & $i\dfrac{m_Z}{E}\dfrac{E}{\Lambda}\alpha_0$ & $i\dfrac{m_Z}{E}\dfrac{E}{\Lambda}\alpha_0$ & $i\dfrac{m_Z}{E}\dfrac{E}{\Lambda}\alpha_0$ & $i\dfrac{m_Z}{E}\dfrac{E}{\Lambda}\alpha_0$ & $i\dfrac{m_Z}{E}\dfrac{E}{\Lambda}\alpha_0$ \\[2ex]
    \hline
    \hline\rule{0pt}{4.5ex}
    $c_{(0,0)}^7$ & $\dfrac{E^2}{\Lambda^2}\left(\dfrac{2\varepsilon_0}{\alpha_1}\right)$ & $\dfrac{E^2}{\Lambda^2}\dfrac{1}{2}(\alpha_1 -\alpha_2)$ & $\dfrac{E^2}{\Lambda^2}\dfrac{1}{2}(\alpha_1 +\alpha_2)$ & $-\dfrac{E^2}{\Lambda^2}\left(\dfrac{2\varepsilon_0}{\alpha_1}\right)$ & $-\dfrac{E^2}{\Lambda^2}\dfrac{1}{2}(\alpha_1 +\alpha_2)$ & $-\dfrac{E^2}{\Lambda^2}\dfrac{1}{2}(\alpha_1 -\alpha_2)$ \\[2.5ex]
    \hline\rule{0pt}{4.5ex}
    
    $c_{(0,0)}^8$ & $i\dfrac{E^2}{\Lambda^2}\left(\dfrac{2\varepsilon_0}{\alpha_1}\right)$ & $i\dfrac{E^2}{\Lambda^2}\dfrac{1}{2}(\alpha_1 -\alpha_2)$ & $i\dfrac{E^2}{\Lambda^2}\dfrac{1}{2}(\alpha_1 +\alpha_2)$ & $i\dfrac{E^2}{\Lambda^2}\left(\dfrac{2\varepsilon_0}{\alpha_1}\right)$ & $i\dfrac{E^2}{\Lambda^2}\dfrac{1}{2}(\alpha_1 +\alpha_2)$ & $i\dfrac{E^2}{\Lambda^2}\dfrac{1}{2}(\alpha_1 -\alpha_2)$ \\[2.5ex]
    \hline\rule{0pt}{4.5ex}

     $c_{(0,0)}^9$ & $-\dfrac{E^2}{\Lambda^2}$ & $-\dfrac{E^2}{\Lambda^2}$ & $-\dfrac{E^2}{\Lambda^2}$ & $\dfrac{E^2}{\Lambda^2}$ & $\dfrac{E^2}{\Lambda^2}$ & $\dfrac{E^2}{\Lambda^2}$\\[1.8ex]
     \hline\rule{0pt}{4.5ex}
     
     $c_{(0,0)}^{10}$ & $i\dfrac{E^2}{\Lambda^2}$ & $i\dfrac{E^2}{\Lambda^2}$ & $i\dfrac{E^2}{\Lambda^2}$ & $i\dfrac{E^2}{\Lambda^2}$ & $i\dfrac{E^2}{\Lambda^2}$ & $i\dfrac{E^2}{\Lambda^2}$\\[1.8ex]
     \hline\rule{0pt}{4.5ex}
     $c^{3p}_6c^{3p}_4$ & $\dfrac{E^2}{\Lambda^2}\left(\dfrac{8\varepsilon_0}{\alpha_1}\alpha_0\right)$ & $\dfrac{E^2}{\Lambda^2}2\alpha_1\alpha_0 $ & $\dfrac{E^2}{\Lambda^2}2\alpha_1\alpha_0$ & $-\dfrac{E^2}{\Lambda^2}\left(\dfrac{8\varepsilon_0}{\alpha_1}\alpha_0\right)$ & $-\dfrac{E^2}{\Lambda^2}2\alpha_1\alpha_0 $ & $-\dfrac{E^2}{\Lambda^2}2\alpha_1\alpha_0$ \\[2.5ex]
    \hline\rule{0pt}{4.5ex}
    $c^{3p}_7c^{3p}_4$ & $i\dfrac{E^2}{\Lambda^2}\left(\dfrac{8\varepsilon_0}{\alpha_1}\alpha_0\right)$ & $i\dfrac{E^2}{\Lambda^2}2\alpha_1\alpha_0 $ & $i\dfrac{E^2}{\Lambda^2}2\alpha_1\alpha_0$ & $i\dfrac{E^2}{\Lambda^2}\left(\dfrac{8\varepsilon_0}{\alpha_1}\alpha_0\right)$ & $i\dfrac{E^2}{\Lambda^2}2\alpha_1\alpha_0 $ & $i\dfrac{E^2}{\Lambda^2}2\alpha_1\alpha_0$ \\[2.5ex]
    \hline\rule{0pt}{4.5ex}
    $c^{3p}_6c^{3p}_5$ & 0 & $-i\dfrac{E^2}{\Lambda^2}2\alpha_2\alpha_0$ & $i\dfrac{E^2}{\Lambda^2}2\alpha_2\alpha_0$ & 0 & $i\dfrac{E^2}{\Lambda^2}2\alpha_2\alpha_0$ & $-i\dfrac{E^2}{\Lambda^2}2\alpha_2\alpha_0$ \\[2.5ex]
    \hline\rule{0pt}{4.5ex}
    $c^{3p}_7c^{3p}_5$ & 0 & $\dfrac{E^2}{\Lambda^2}2\alpha_2\alpha_0$ & $-\dfrac{E^2}{\Lambda^2}2\alpha_2\alpha_0$ & 0 & $\dfrac{E^2}{\Lambda^2}2\alpha_2\alpha_0$ & $-\dfrac{E^2}{\Lambda^2}2\alpha_2\alpha_0$ \\[2.5ex]
    \hline
       \hline\rule{0pt}{4.5ex}
    $c_{(1,0)}^7$ & $\dfrac{E^4}{\Lambda^4}\left(\dfrac{2\varepsilon_0}{\alpha_1}\right)$ & $\dfrac{E^4}{\Lambda^4}\dfrac{1}{2}(\alpha_1 -\alpha_2)$ & $\dfrac{E^4}{\Lambda^4}\dfrac{1}{2}(\alpha_1 +\alpha_2)$ & $-\dfrac{E^4}{\Lambda^4}\left(\dfrac{2\varepsilon_0}{\alpha_1}\right)$ & $-\dfrac{E^4}{\Lambda^4}\dfrac{1}{2}(\alpha_1 +\alpha_2)$ & $-\dfrac{E^4}{\Lambda^4}\dfrac{1}{2}(\alpha_1 -\alpha_2)$ \\[2.5ex]
    \hline\rule{0pt}{4.5ex}
    
    $c_{(1,0)}^8$ & $i\dfrac{E^4}{\Lambda^4}\left(\dfrac{2\varepsilon_0}{\alpha_1}\right)$ & $i\dfrac{E^4}{\Lambda^4}\dfrac{1}{2}(\alpha_1 -\alpha_2)$ & $i\dfrac{E^4}{\Lambda^4}\dfrac{1}{2}(\alpha_1 +\alpha_2)$ & $i\dfrac{E^4}{\Lambda^4}\left(\dfrac{2\varepsilon_0}{\alpha_1}\right)$ & $i\dfrac{E^4}{\Lambda^4}\dfrac{1}{2}(\alpha_1 +\alpha_2)$ & $i\dfrac{E^4}{\Lambda^4}\dfrac{1}{2}(\alpha_1 -\alpha_2)$ \\[2.5ex]
    \hline\rule{0pt}{4.5ex}

     $c_{(1,0)}^9$ & $-\dfrac{E^4}{\Lambda^4}$ & $-\dfrac{E^4}{\Lambda^4}$ & $-\dfrac{E^4}{\Lambda^4}$ & $\dfrac{E^4}{\Lambda^4}$ & $\dfrac{E^4}{\Lambda^4}$ & $\dfrac{E^4}{\Lambda^4}$\\[1.8ex]
     \hline\rule{0pt}{4.5ex}
     
     $c_{(1,0)}^{10}$ & $i\dfrac{E^4}{\Lambda^4}$ & $i\dfrac{E^4}{\Lambda^4}$ & $i\dfrac{E^4}{\Lambda^4}$ & $i\dfrac{E^4}{\Lambda^4}$ & $i\dfrac{E^4}{\Lambda^4}$ & $i\dfrac{E^4}{\Lambda^4}$\\[1.8ex] 
     \hline\rule{0pt}{4.5ex}
     $c_{(0,1)}^{11}$ & $\dfrac{E^4}{\Lambda^4}\dfrac{\alpha_2}{\alpha_1}$ & 0 & 0 & $-\dfrac{E^4}{\Lambda^4}\dfrac{\alpha_2}{\alpha_1}$ & 0 & 0 \\[2.5ex]
    \hline\rule{0pt}{4.5ex}
    $c_{(0,1)}^{12}$ & $i\dfrac{E^4}{\Lambda^4}\dfrac{\alpha_2}{\alpha_1}$ & 0 & 0 & $i\dfrac{E^4}{\Lambda^4}\dfrac{\alpha_2}{\alpha_1}$ & 0 & 0\\[2.5ex]
     \hline
    \end{tabular}
    \end{adjustbox}
    \hspace*{-3cm}
\label{tab:Table1}
    \caption{EFT contributions---up to all orders in the expansion---to,  $\hat{a}^{J_{min}}_{\lambda_{f} \lambda_{\bar{f}}; \lambda_Z}$, the minimum $J$ partial wave coefficients rescaled by the normalisation factor in \eq{rescale}. These contributions arise from the three point vertices, principal operators (see Table~\ref{tab:3-point_ops} and  \ref{tab:4-pt_principals}) and their $s$-descendants/$J$-descendants. We show here only the helicity channels with  opposite fermion helicities and $J=1$. See the caption of Table~\ref{tab:full_Amps_Exact} for the definition of $\varepsilon_0$ and $\alpha_{0,1,2}$.}
    \label{tab:LR_amp_J1}
\end{table}

First consider the two partial wave coefficients with $J=0$ corresponding to the helicity channels $(++;0)$ and $(--;0)$. These helicity channels get the lowest order  contribution from the two independent operators $\mathcal{O}^{11}_{0,0}$ and $\mathcal{O}^{12}_{0,0}$ at ${\cal O} (E^2/\Lambda^2)$ as shown in Table~\ref{tab:LR_amp_J0}. 

As far as the six  $J=1$  partial wave coefficients are concerned we see that there are    only two EFT contributions at ${\cal O} (E/\Lambda)$ resulting in the following correlations between them,
\begin{align}
    \hata^1_{++;+} - \hata^1_{++;0} &= 0, \quad \hata^1_{++;-} - \hata^1_{++;0} = 0 \\
    \hata^1_{--;+} - \hata^1_{--;0} &= 0, \quad \hata^1_{--;-} - \hata^1_{--;0} = 0.
    \label{corrLR}
\end{align}
The above equations contain 4 additional correlations after imposing \eq{cpt}. 


\begin{table}[t]
\begin{adjustbox}{max width=\textwidth}
    \centering
    \begin{tabular}{|c|}
        \hline
        
Principal directions to all orders in EFT\\
\hline
\hline
         $J = 0$ \\
\hline\rule{0pt}{5ex}
     $\begin{aligned}
         \delta^2 \textbf{a}_{0,1} = \sum_{m=1}^{\infty} c^{11}_{(m-1,0)}\left(\frac{E}{\Lambda}\right)^{2m}
     \end{aligned}$ \\[0.6cm]
      $\begin{aligned}
         \delta^2 \textbf{a}_{0,2} = \sum_{m=1}^{\infty} c^{12}_{(m-1,0)}\left(\frac{E}{\Lambda}\right)^{2m}
     \end{aligned}$ \\[0.55cm]
     \hline
     \hline
     $J = 1$ \\
     \hline\rule{0pt}{4.5ex}
         $\delta \textbf{a}_{1,7} = \dfrac{E}{\Lambda}\dfrac{m_Z}{E} c^{3p}_6(1 + c^{3p}_1) \alpha_0 + \dfrac{E^2}{\Lambda^2}\left[2(c^{3p}_4 c^{3p}_6 + c^{3p}_5 c^{3p}_7)\,\alpha_0 \alpha_1 - c^9_{(0,0)}\right] - \displaystyle\sum_{m=2}^{\infty}\left(\dfrac{E}{\Lambda}\right)^{2m} c^9_{(m-1,\,0)}  $ \\[0.7cm]
         $\delta \textbf{a}_{1,8} = \dfrac{E}{\Lambda}\dfrac{m_Z}{E} c^{3p}_7(1 + c^{3p}_1) \alpha_0 + \dfrac{E^2}{\Lambda^2}\left[2(c^{3p}_4 c^{3p}_7 - c^{3p}_5 c^{3p}_6)\,\alpha_0 \alpha_1 + c^{10}_{(0,0)}\right] + \displaystyle\sum_{m=2}^{\infty}\left(\dfrac{E}{\Lambda}\right)^{2m} c^{10}_{(m-1,\,0)}  $ \\[0.5cm]
         \hline
         \hline\rule{0pt}{4.5ex}
         $\delta^2 \textbf{a}_{1,9} = -\dfrac{E^2}{\Lambda^2} \left(4\,c^{3p}_5 c^{3p}_7 \alpha_0 - c^7_{(0,0)}\right) + \displaystyle\sum_{m=2} \left(\dfrac{E}{\Lambda}\right)^{2m} c^7_{(m-1,\,0)}$ \\[0.7cm]
         $\delta^2 \textbf{a}_{1,10} = \dfrac{E^2}{\Lambda^2} \left(4\,c^{3p}_5 c^{3p}_6 \alpha_0 + c^8_{(0,0)}\right) + \displaystyle\sum_{m=2} \left(\dfrac{E}{\Lambda}\right)^{2m} c^8_{(m-1,\,0)}$ \\[0.7cm]
         $\delta^2 \textbf{a}_{1,11} = \dfrac{E^2}{\Lambda^2} 8(c^{3p}_4 c^{3p}_6 + c^{3p}_5 c^{3p}_7)(\alpha_1 - \varepsilon_1) \dfrac{\alpha_1}{\alpha_2}\alpha_0 -4 \displaystyle\sum_{m=2}^{\infty}\left(\dfrac{E}{\Lambda}\right)^{2m}c^{11}_{(m-2,\,0)} $ \\[0.7cm]
         $\delta^2 \textbf{a}_{1,12} = \dfrac{E^2}{\Lambda^2} 8(c^{3p}_5 c^{3p}_6 - c^{3p}_4 c^{3p}_7)(\alpha_1 - \varepsilon_1) \dfrac{\alpha_1}{\alpha_2}\alpha_0 +4 \displaystyle\sum_{m=2}^{\infty}\left(\dfrac{E}{\Lambda}\right)^{2m}c^{12}_{(m-2,\,0)} $ \\[0.7cm]
         \hline
    \end{tabular}
    \end{adjustbox}
    \caption{Expressions, to all orders in the EFT expansion, for the  principal    directions involving partial wave coefficients with identical fermion helicities.}
    \label{tab:LR_RL_hel_dir}
\end{table}

At the next order, ${\cal O} (E^2/\Lambda^2)$,  there are 4 new EFT that completely break all the  correlations in \eq{corrLR}. Once again we can construct helicity directions for both $J=0$ and $J=1$ partial wave coefficients. The two $J=0$ helicity directions are defined as, 
\begin{align}
\delta^2 \textbf{a}_{0,1} = i\frac{\alpha_1}{2\alpha_2}(\hata^0_{++;0} - \hata^0_{--;0}) \nonumber\\
    \delta^2 \textbf{a}_{0,2} = \frac{\alpha_1}{2\alpha_2}(\hata^0_{++;0} + \hata^0_{--;0}). 
    \label{heldirLRJ0}
\end{align}
The 6 $J=1$ directions are defined as follows,  
\begin{align}
   {\delta} \textbf{a}_{1,7} &= \frac{1}{4\alpha_2}\left[\alpha_4(\hata^1_{++;+} - \hata^1_{--;-}) - \varepsilon_3(\hata^1_{++;-} - \hata^1_{--;+})\right] \nn\\
    {\delta}\textbf{a}_{1,8} &= -\frac{i}{4\alpha_2}\left[\alpha_4(\hata^1_{++;+} + \hata^1_{--;-}) - \varepsilon_3(\hata^1_{++;-} + \hata^1_{--;+})\right] \nn\\
     \delta^2\textbf{a}_{1,9} &= -\dfrac{1}{2\alpha_2}(\hata^1_{++;+} - \hata^1_{++;-} + \hata^1_{--;+} - \hata^1_{--;-}) \nn\\ 
     \delta^2\textbf{a}_{1,10} &= \dfrac{i}{2\alpha_2}(\hata^1_{++;+} - \hata^1_{++;-} - \hata^1_{--;+} + \hata^1_{--;-})\nn \\ 
     \delta^2 \textbf{a}_{1,11} &= \frac{\alpha_1}{\alpha_2^2}\left[\alpha_5(\hata^1_{++;+} - \hata^1_{--;-}) + \varepsilon_4(\hata^1_{++;-} - \hata^1_{--;+}) - 2\alpha_2(\hata^1_{++;0} - \hata^1_{--;0}) \right] \nn\\
     \delta^2 \textbf{a}_{1,12} &= i\frac{\alpha_1}{\alpha_2^2}\left[\alpha_5(\hata^1_{++;+} + \hata^1_{--;-}) + \varepsilon_4(\hata^1_{++;-} + \hata^1_{--;+}) - 2\alpha_2(\hata^1_{++;0} + \hata^1_{--;0}) \right],
     \label{heldirLR}
\end{align}
where all the above linear combinations are real by \eq{cpt}. Here,  
\begin{align}
    \varepsilon_3 &= \alpha_1 - \alpha_2 \,,\;\;\quad\;\; \varepsilon_4 = \varepsilon_1 - (\alpha_1 - \alpha_2)\\
    \alpha_4 &= \alpha_1 + \alpha_2 \,,\;\;\quad\;\; \alpha_5 = \alpha_1 + \alpha_2 - \varepsilon_1
\end{align}
have been again defined such that $\alpha_i \to 1$ and $\varepsilon_i \to 0$ in the high energy limit.

We can again  write these principal directions to all orders in the EFT expansion parameter as shown in Table~\ref{tab:LR_RL_hel_dir}.  For  the $J = 0$ principal directions we can write, 
\begin{align}
    &\delta^q \textbf{a}_{0, k} =  \sum_{m=1} \delta^q \textbf{a}^{2m - q}_{0,k} \left(\frac{E}{\Lambda}\right)^{2m},
    \label{pow3}
\end{align} 
where the $\delta^q \textbf{a}^{m}_{0,k}$ can be read off from Table~\ref{tab:LR_RL_hel_dir}. For $J = 1$  and  $k = 7,\,8$,  we can write, 
\begin{align}
    &\delta \textbf{a}_{1, k} = \delta \hat{\textbf{a}}_{1,k}^0 \left(\frac{E}{\Lambda}\right) + \delta\hat{\textbf{a}}_{1,k}^{1} \left(\dfrac{E}{\Lambda}\right)^{2}  + \sum_{m=2} \delta \textbf{a}^{2m-1}_{1,k} \left(\frac{E}{\Lambda}\right)^{2m} 
    \label{pow4}
\end{align} 
where the $\delta \hat{\textbf{a}}_{1,k}^0, \delta \hat{\textbf{a}}_{1,k}^1$ are energy dependent. The energy dependence can be extracted by writing, $\delta \hat{\textbf{a}}_{1,k}^0 = \delta \textbf{a}_{1,k}^0\,\alpha_0\,$, $\delta \hat{\textbf{a}}_{1,k}^1 = \delta \textbf{a}_{1,k}^{1,\,3p}\,\alpha_0 \alpha_1 + \delta \textbf{a}_{1,k}^{1}$, such that  $\delta \textbf{a}_{1,k}^0, \delta \textbf{a}_{1,k}^1$ and $\delta \textbf{a}_{1,k}^{1,3p}$ are independent of energy. The  $\delta \textbf{a}_{1,k}^{1,3p}$ and $\delta \textbf{a}_{1,k}^{m}$ are again given by linear combinations of Wilson coefficients shown in  Table~\ref{tab:LR_RL_hel_dir}. Finally, for $J = 1$ and $k = 9$-12, we can write,
\begin{align}
    &\delta^2 \textbf{a}_{1, k} =  \delta^2\hat{\textbf{a}}_{1,k}^{0} \left(\dfrac{E}{\Lambda}\right)^{2}  + \sum_{m=2} \delta^2 \textbf{a}^{2m-2}_{J,k} \left(\frac{E}{\Lambda}\right)^{2m} 
    \label{pow5}
\end{align} 
where $\delta^2\hat{\textbf{a}}_{1,k}^{0} = \delta^2 \textbf{a}_{1, k}^{0,\,3p} \alpha_0 + \delta^2 \textbf{a}_{1, k}^0$ for $k = 9, 10$, and $\delta^2\hat{\textbf{a}}_{1,k}^{0} = \delta^2 \textbf{a}_{1, k}^{0,\,3p} \alpha_0(\alpha_1 - \varepsilon_1)\dfrac{\alpha_1}{\alpha_2}$ for $k = 11, 12$.
The $\delta^2 \textbf{a}_{1, k}^{0,\,3p}$ and $\delta^2 \textbf{a}_{1, k}^{m}$ are independent of energy and the   expressions for them in terms of Wilson coefficients can be obtained from   Table \ref{tab:LR_RL_hel_dir}.

\pagebreak
\begin{table}[h!]
\begin{adjustbox}{max width=\textwidth}
    \centering
    \begin{tabular}{|c|c|}
    \hline
       $\mom_i$  &  Expressions in terms of helicity directions\\
       \hline\rule{0pt}{4ex}
        \multirow{13}{*}{$\mom^1_{LL}$} & $
       \begin{aligned}
       G_{-+0}^2 + G_{+-0}^2 + \dfrac{\alpha_1}{2} \alpha_0\bigg(\smPL(\CaI^0 + \CaII^0) + \smML(\CaI^0 - \CaII^0)\bigg) + \dfrac{\alpha_1^2}{8}\alpha_0^2\left(\left(\CaI^{0}\right)^2 + \left(\CaII^{0}\right)^2\right) 
       \end{aligned}$\\[0.4cm]
       
       \cline{2-2}\rule{0pt}{8ex}
       
          &  $\begin{aligned}
             +\frac{\alpha_1}{2}&\bigg[\smPL (\CaI^1 + \CaII^1 + \rr \alpha_3\,\CaIII^0) + \smML\bigg(\CaI^1 - \CaII^1 + \dfrac{\rr\alpha_3}{\Qr}(\CaV^0+ \Ql\CaIII^0)\bigg) \\
             &+\frac{\alpha_1 \alpha_0}{4}\bigg(2(\CaI^0 \CaI^1 + \CaII^0\,\CaII^1) + \rr\alpha_3\bigg\{(\CaI^0 + \CaII^0)\CaIII^0 + \frac{1}{\Qr}(\CaI^0 - \CaII^0)(\CaV^0 + \Ql \CaIII^0)\bigg\}\bigg)\bigg]\frac{E}{m_Z} \left(\dfrac{E}{\Lambda}\right) 
             \end{aligned}$\\[0.9cm]
            \cline{2-2}\rule{0pt}{4.5ex}
            
             & $-\dfrac{\alpha_1^2\alpha_0^2}{2} \left(\dfrac{\alpha_5 +\varepsilon_4}{\alpha_4 - \varepsilon_3}\right)^2 [(\BaIsX)^2 + (\BaIIsX)^2]\dfrac{E^2}{\Lambda^2} $ \\[0.35cm]
          
          \cline{2-2}
          
          \rule{0pt}{10.5ex}
          & $\begin{aligned}
        &-\dfrac{\alpha_1^2}{4} \dfrac{(\alpha_5 +\varepsilon_4)}{(\alpha_4 - \varepsilon_3)^2}\alpha_0 \bigg[4(\alpha_5 + \varepsilon_4)\bigg(\BaIsX (\BaIsI) + \BaIIsX (\BaIIsI)\bigg) \\
        &+ 2(\alpha_5 \varepsilon_3 + \alpha_4 \varepsilon_4)\bigg(\BaIsX(\BaIIIsX) + \BaIIsX(\BaIVsX)\bigg) \\
        &+ (\varepsilon_3 - \alpha_4)\BaVsX\,\BaIsX + (\alpha_4 - \varepsilon_3)\BaVIsX\,\BaIIsX\bigg]\dfrac{m_Z}{E}\left(\dfrac{E^2}{m_Z^2}\right) \dfrac{E^3}{\Lambda^3}
    \end{aligned}$
          \\[1.4cm]
           \hline
           \hline\rule{0pt}{4.5ex}
    \multirow{7}{*}{$\mom^2_{LL}$} 
    & $\alpha_1^2\alpha_0^2 \left(\dfrac{\alpha_5 +\varepsilon_4}{\alpha_4 - \varepsilon_3}\right)^2 [(\BaIsX)^2 + (\BaIIsX)^2]\dfrac{E^2}{\Lambda^2}$ \\[0.4cm]
    
    \cline{2-2} 
    
    \rule{0pt}{10.5ex}
    & $\begin{aligned}
        &+\dfrac{\alpha_1^2}{2} \dfrac{(\alpha_5 +\varepsilon_4)}{(\alpha_4 - \varepsilon_3)^2}\alpha_0 \bigg[4(\alpha_5 + \varepsilon_4)\bigg(\BaIsX (\BaIsI) + \BaIIsX (\BaIIsI)\bigg) \\
        &+ 2(\alpha_5 \varepsilon_3 + \alpha_4 \varepsilon_4)\bigg(\BaIsX(\BaIIIsX) + \BaIIsX(\BaIVsX)\bigg) \\
        &+ (\varepsilon_3 - \alpha_4)\BaVsX\,\BaIsX + (\alpha_4 - \varepsilon_3)\BaVIsX\,\BaIIsX\bigg]\dfrac{m_Z}{E}\left(\dfrac{E^2}{m_Z^2}\right) \dfrac{E^3}{\Lambda^3}
    \end{aligned}$ \\[1.4cm]
    \hline
    \hline\rule{0pt}{4ex}
    \multirow{15}{*}{$\mom^1_{TT}$} & $2\epsilon_{LR} (\smPT + \smMT + \alpha_0\,\CaI^0)(\smPT - \smMT + \alpha_0\,\CaII^0)\dfrac{m_Z^2}{E^2}$ \\[0.4cm]
    
    \cline{2-2}\rule{0pt}{8ex}
    
    & $\begin{aligned}
        &+2\epsilon_{LR}\bigg[\smPT(\CaI^1 + \CaII^1 + \alpha_3(1+\rr)\CaIII^0) - \smMT\bigg(\CaI^1 - \CaII^1 + \frac{1}{\Qr}\alpha_3(1 + \rr)(\CaV^0 + \Ql\,\CaIII^0)\bigg) \\
        &+\alpha_0(\CaI^0 \CaII^1 +\CaII^0 \CaI^1) + \dfrac{\alpha_0}{2\Qr}\alpha_3(1+\rr)\{\Qr \CaIII^0 (\CaI^0 + \CaII^0) - (\CaI^0 - \CaII^0)(\CaV^0 + \Ql\,\CaIII^0)\}\bigg]\dfrac{m_Z}{E}\dfrac{E}{\Lambda}
    \end{aligned}$\\[0.9cm]
    
    \cline{2-2} \rule{0pt}{8ex}
    & $\begin{aligned}
       &+\epsilon_{LR}\dfrac{1}{2(\Qr)^2} \bigg[2\Qr\,\CaI^1\bigg\{2\Qr\,\CaII^1 - \alpha_3(1 + \rr)(\CaV^0 - \Qr(1 - \qr)\CaIII^0)\bigg\} - \alpha_2^2(\CaVI^0 + \Ql \CaIV^0)^2 \\
        &+\alpha_3(1+\rr)(\CaV^0 + \Qr(1+\qr)\CaIII^0)\{2\Qr\,\CaII^1 - \alpha_3(1+\rr)(\CaV^0 - \Qr(1-\qr)\CaIII^0)\} + \alpha_2^2 (\Qr)^2(\CaIV^0)^2\bigg]\dfrac{E^2}{\Lambda^2}
    \end{aligned}$ \\[0.9cm]
    \cline{2-2}
    \rule{0pt}{8ex}
    & $\begin{aligned}
        &+2\epsilon_{LR}  \bigg[\smPT(\CaI^3 + \CaII^3 + \alpha_3(1+\rr)\CaIII^2) - \smMT\bigg(\CaI^3 - \CaII^3 + \frac{\alpha_3}{\Qr}(1+\rr)(\CaV^2 + \Ql\CaIII^2)\bigg)     \\
        &+ \alpha_0(\CaI^0\, \CaII^3 + \CaII^0\, \CaI^3)  + \dfrac{\alpha_0}{2\Qr}(1 + \rr)\alpha_3\bigg(\Qr(\CaI^0 + \CaII^0)\CaIII^2 - (\CaI^0 - \CaII^0)(\CaV^2 + \Ql\,\CaIII^2)\bigg)\bigg]\dfrac{m_Z}{E}\dfrac{E^{3}}{\Lambda^3}
    \end{aligned}$ \\[1cm]
    \hline
   
    \end{tabular}
    \end{adjustbox}
    \label{tab:moments1}
\end{table}

\pagebreak
\clearpage
\begin{table}[h]
\begin{adjustbox}{max width=\textwidth}
    \centering
    \begin{tabular}{|c|c|}
    \hline
     $\mom_i$  &  Expressions in terms of helicity directions\\
     \hline\rule{0pt}{5ex}
    \multirow{15}{*}{$\mom^2_{TT}$} & $\dfrac{1}{2}(\smPT^2 + \smMT^2)\dfrac{m_Z^2}{E^2} + \dfrac{1}{4}\bigg[2\smPT\alpha_0(\CaI^0 + \CaII^0) + 2\smMT\alpha_0(\CaI^0 - \CaII^0) + \alpha_0^2(\CaI^0)^2 + \alpha_0^2(\CaII^0)^2\bigg]\dfrac{m_Z^2}{E^2}$ \\[0.4cm] 
    \cline{2-2}
    \rule{0pt}{8ex}
    & $\begin{aligned}
    &+\dfrac{1}{2}\bigg[\smPT(\CaI^1 + \CaII^1 + \alpha_3(1+\rr)\CaIII^0) + \smMT\bigg(\CaI^1 - \CaII^1 + \frac{1}{\Qr}\alpha_3(1+\rr)(\CaV^0 + \Ql\,\CaIII^0)\bigg) \\
    &+ \alpha_0(\CaI^0 \,\CaI^1+ \CaII^0\,\CaII^1) + \frac{\alpha_0}{2\Qr}\alpha_3(1+\rr)\{\Qr(\CaI^0 + \CaII^0)\CaIII^0 + (\CaI^0 - \CaII^0)(\CaV^0 + \Ql\,\CaIII^0)\} \bigg]\dfrac{m_Z}{E}\frac{E}{\Lambda}
    \end{aligned} $ \\[0.9cm]
    \cline{2-2}
    \rule{0pt}{7.5ex}
    & $\begin{aligned}
        &+ \dfrac{1}{8(\Qr)^2}\bigg[\alpha_2^2(\CaVI^0 + \Ql\,\CaIV^0)^2 + 2\Qr\alpha_3 (1 + \rr)\{\Qr(\CaI^1 + \CaII^1)\CaIII^0 + (\CaI^1 - \CaII^1)(\CaV^0 + \Ql\,\CaIII^0)\}\\
        &+ (\Qr)^2\left(2(\CaI^1)^2 + 2(\CaII^1)^2+ \alpha_2^2(\CaIV^0)^2\right) + \alpha_3^2 (1 + \rr)^2 \{(\Qr)^2 (\CaIII^0)^2 + (\CaV^0 + \Ql\,\CaIII^0)^2\}\bigg]\dfrac{E^{2}}{\Lambda^2}
    \end{aligned}$  \\[0.9cm]
    \cline{2-2}
    \rule{0pt}{11ex}
    & $\begin{aligned}
          &+\dfrac{1}{2}\bigg[\smPT(\CaI^3 + \CaII^3 + \alpha_3(1+\rr)\CaIII^2) + \smMT\bigg(\CaI^3 - \CaII^3 + \frac{1}{\Qr}\alpha_3(1+\rr)(\CaV^2 + \Ql\,\CaIII^2)\bigg)\\
          &+ \dfrac{\alpha_0}{2\Qr}\alpha_3(1 + \rr) \bigg(\Qr(\CaI^0 +\CaII^0)\CaIII^2 + (\CaI^0 - \CaII^0)(\CaV^2 + \Ql\,\CaIII^2)\bigg) \\
          &+\alpha_0(\CaI^0\, \CaI^3 + \CaII^0\, \CaII^3)\bigg] \dfrac{m_Z}{E}\dfrac{E^3}{\Lambda^3} 
         \end{aligned}$ \\[1.5cm]
    \hline
     \hline\rule{0pt}{4ex}
     
    \multirow{4.5}{*}{$\mom^3_{TT}$} 
    & $ \dfrac{8\alpha_2^2}{(\alpha_4 - \varepsilon_3)^2}\alpha_0^2 [(\BaIsX)^2 + (\BaIIsX)^2]\dfrac{m_Z^2}{E^2}\dfrac{E^2}{\Lambda^2}$ \\[0.4cm]
    \cline{2-2} 
    \rule{0pt}{7.5ex}
    & $\begin{aligned}
        &+\dfrac{4\alpha_2^2}{(\alpha_4 - \varepsilon_3)^2} \alpha_0\bigg[4\{\BaIsX(\BaIsI) + \BaIIsX(\BaIIsI)\} \\
        &+(\alpha_4 + \varepsilon_3)\{\BaIsX(\BaIIIsX) + \BaIIsX(\BaIVsX)\}\bigg]\dfrac{m_Z}{E}\dfrac{E^3}{ \Lambda^3}\\
    \end{aligned}$ \\[0.85cm]
    \hline
    \hline\rule{0pt}{7.5ex}
         \multirow{18}{*}{$\mom^1_{LT}$}   & $
    \begin{aligned}
    &-2\epsilon_{LR}(G_{+-0}G_{+-} - G_{-+0}G_{-+})\dfrac{m_Z}{E}
    - \epsilon_{LR}\,\alpha_0\bigg[G_{+-0} (\CaI^0 + \CaII^0) - G_{-+0} (\CaI^0 - \CaII^0)\bigg]\dfrac{m_Z}{E}  \\ 
    &-\epsilon_{LR}\dfrac{\alpha_1}{2}\alpha_0\bigg[ \smPT(\CaI^0 + \CaII^0) - \smMT(\CaI^0 - \CaII^0) + 2\alpha_0\,\CaI^0 \CaII^0\bigg] \dfrac{m_Z}{E}
    \end{aligned}$\\[0.9cm]
    \cline{2-2}
    \rule{0pt}{11.5ex}
   & $\begin{aligned}
    &-\epsilon_{LR} \bigg[\smPL(\CaI^1 + \CaII^1+\alpha_3(1+\rr)\CaIII^0) - \smML\bigg(\CaI^1 - \CaII^1+\frac{\alpha_3}{\Qr}(1+\rr)(\CaV^0 + \Ql\,\CaIII^0)\bigg) \\
    &+\dfrac{\alpha_1}{2}\smPT(\CaI^1 + \CaII^1+\alpha_3\,\rr\,\CaIII^0) - \dfrac{\alpha_1}{2}\smMT\bigg(\CaI^1 - \CaII^1+ \frac{\alpha_3}{\Qr}\,\rr(\CaV^0 + \Ql\,\CaIII^0)\bigg) \\
    &+ \dfrac{\alpha_1}{4}\alpha_0\bigg(4(\CaI^1\,\CaII^0 + \CaI^0\, \CaII^1) - \frac{\alpha_3}{\Qr}(1+2\rr)\{\CaI^0(\CaV^0 - (\Qr-\Ql)\CaIII^0) - (\CaV^0 + (\Qr+\Ql)\CaIII^0)\CaII^0\}\bigg)\bigg]\dfrac{E}{\Lambda}
    \end{aligned}$ \\[1.5cm]
    \cline{2-2}
    \rule{0pt}{7.5ex}
    & $\begin{aligned}
        -\epsilon_{LR} \dfrac{\alpha_1}{4(\Qr)^2}\bigg[&\alpha_3(\CaV^0 + (\Qr + \Ql)\CaIII^0)\bigg(\Qr(1+2\rr)\CaII^1 - \alpha_3\,\rr(1 + \rr)(\CaV^0 - (\Qr-\Ql)\CaIII^0)\bigg) \\
        &+ \Qr\,\CaI^1\bigg(4\Qr\,\CaII^1 - \alpha_3(1 + 2\rr)(\CaV^0 - (\Qr-\Ql)\CaIII^0)\bigg)\bigg]\dfrac{E}{m_Z}\dfrac{E^2}{\Lambda^2}
    \end{aligned}$ \\[0.9cm]
    \cline{2-2}
    \rule{0pt}{11.5ex}
   & $\begin{aligned}
    -\epsilon_{LR}&\bigg[\smPL(\CaI^3 + \CaII^3 + \alpha_3(1+\rr)\CaIII^2) - \smML\bigg(\CaI^3 - \CaII^3 + \frac{\alpha_3}{\Qr}(1+\rr)(\CaV^2 + \Ql\,\CaIII^2)\bigg) \\
        &+\frac{\alpha_1}{2} \smPT(\CaI^3 + \CaII^3 + \alpha_3\,\rr\,\CaIII^2) - \frac{\alpha_1}{2} \smMT\bigg(\CaI^3 - \CaII^3 + \frac{\alpha_3}{\Qr}\rr(\CaV^2 + \Ql\,\CaIII^2)\bigg) \\
        &+ \dfrac{\alpha_1\alpha_0}{4}\bigg(4(\CaI^3\,\CaII^0 + \CaII^3\,\CaI^0) + \alpha_3(1+2\rr)\bigg\{\CaIII^2(\CaI^0 +\CaII^0) - \frac{1}{\Qr}(\CaI^0-\CaII^0)(\CaV^2+\Ql\,\CaIII^2)\bigg\}\bigg)\bigg] \dfrac{E^3}{\Lambda^3}
    \end{aligned}$\\[1.5cm] 
\hline

\end{tabular}
    \end{adjustbox}
    \label{tab:moments2}
\end{table}
\clearpage

\begin{table}[h!]
    \begin{adjustbox}{max width=\textwidth}
    \centering
    \begin{tabular}{|c|c|}
      \hline
     $\mom_i$  &  Expressions in terms of helicity directions\\
    \hline\rule{0pt}{7ex}
    \multirow{25}{*}{$\mom^2_{LT}$} & $\begin{aligned}
        &-2(\smPL\,\smPT + \smML\,\smMT)\dfrac{m_Z}{E} -\alpha_0\bigg[\smPL(\CaI^0 + \CaII^0) + \smML(\CaI^0 - \CaII^0) \\
        &+ \dfrac{\alpha_1}{2}\bigg(\alpha_0(\CaI^0)^2 + \alpha_0(\CaII^0)^2 + \smPT(\CaI^0 + \CaII^0) + \smMT(\CaI^0 - \CaII^0)\bigg)\bigg]\dfrac{m_Z}{E}
    \end{aligned}
    $ \\[0.9cm]
    \cline{2-2}
    \rule{0pt}{11ex}
    & $\begin{aligned}
    &-\bigg[\smPL(\CaI^1 + \CaII^1 + \alpha_3(1+\rr)\CaIII^0) + \smML\bigg(\CaI^1 - \CaII^1 + \frac{\alpha_3}{\Qr}(1+\rr)(\CaV^0 + \Ql\,\CaIII^0)\bigg) \\
    &+ \dfrac{\alpha_1}{2}\smPT(\CaI^1 + \CaII^1 + \rr\alpha_3\,\CaIII^0) + \dfrac{\alpha_1}{2}\smMT\bigg(\CaI^1 - \CaII^1 + \frac{\alpha_3}{\Qr}\rr(\CaV^0 + \Ql\,\CaIII^0)\bigg) \\
    &+ \dfrac{\alpha_1}{4}\alpha_0\bigg(4(\CaI^0\,\CaI^1 + \CaII^0\,\CaII^1) + \frac{\alpha_3}{\Qr}(1 + 2\rr)\{\CaV^0(\CaI^0 - \CaII^0) + \Qr(\CaI^0(1 + \qr) + (1-\qr)\CaII^0)\CaIII^0\}\bigg)\bigg]\dfrac{E}{\Lambda} \\
    \end{aligned}$ \\[1.5cm]
    \cline{2-2}
    \rule{0pt}{11ex}
    & $\begin{aligned}
        -\frac{\alpha_1}{4(\Qr)^2}\bigg[ &2(\Qr)^2\left((\CaI^1)^2+(\CaII^1)^2\right) + \alpha_3^2 \,\rr(1+\rr)\left\{(\CaV^0)^2 +2\Ql\,\CaIII^0\,\CaV^0 + ((\Qr)^2 + (\Ql)^2)(\CaIII^0)^2\right\} \\
        &+ \alpha_3(1+2\rr)\Qr\bigg(\CaV^0(\CaI^1 - \CaII^1) +\Qr\{\CaI^1(1 + \qr) +(1-\qr)\CaII^1\}\CaIII^0\bigg)\bigg]\frac{E}{m_Z}\frac{E^2}{\Lambda^2} \\
       &- 4\alpha_1\alpha_2\alpha_0^2 \dfrac{(\alpha_5 + \varepsilon_4)}{(\alpha_4 - \varepsilon_3)^2}[(\BaIsX)^2 + (\BaIIsX)^2] \left(\dfrac{E}{m_Z}\dfrac{m_Z^2}{E^2}\right)\dfrac{E^2}{\Lambda^2}
    \end{aligned}$ \\[1.4cm]
    \cline{2-2}
    \rule{0pt}{23.5ex}
    & $\begin{aligned}
       &-\bigg[\smPL(\CaI^3 + \CaII^3 + \alpha_3(1+\rr)\CaIII^2 ) + \smML\bigg(\CaI^3 - \CaII^3 + \frac{\alpha_3}{\Qr}(1+\rr)(\CaV^2 + \Ql\,\CaIII^2  )\bigg)\\
       &+ \dfrac{\alpha_1}{2}\smPT(\CaI^3 + \CaII^3 + \alpha_3\rr\,\CaIII^2 ) + \dfrac{\alpha_1}{2}\smMT \bigg(\CaI^3 - \CaII^3 + \frac{\alpha_3}{\Qr}\rr(\CaV^2 + \Ql\,\CaIII^2  )\bigg) \\
       &+ \dfrac{\alpha_1}{4\Qr}\alpha_0 \bigg(\alpha_3(1+2\rr)\{\Qr\CaIII^2(\CaI^0 + \CaII^0) +(\CaI^0 - \CaII^0)(\CaV^2+\Ql\,\CaIII^2)\}\\
       &+4\Qr(\CaI^0\CaI^3 + \CaII^0\CaII^3)\bigg)\bigg]\dfrac{E^{3}}{\Lambda^3} \\
    &-\dfrac{\alpha_1\alpha_2}{(\alpha_4 - \varepsilon_3)^2} \alpha_0\bigg[(\varepsilon_3 - \alpha_4)\BaVsX\,\BaIsX + (\alpha_4 - \varepsilon_3)\BaVIsX\,\BaIIsX\\[0.15cm]
      &+ \{\varepsilon_3(3\alpha_5 + \varepsilon_4)+\alpha_4(\alpha_5 + 3\varepsilon_4)\}\{\BaIsX(\BaIIIsX) + \BaIIsX (\BaIVsX)\} \\[0.15cm]
      &+8(\alpha_5 + \varepsilon_4)\{\BaIsX(\BaIsI) + \BaIIsX(\BaIIsI)\} \bigg]\left(\frac{m_Z}{E}\frac{E}{m_Z}\right)\frac{E^{3}}{\Lambda^3}
    \end{aligned}$\\[3.5cm]
    \hline   
     \hline\rule{0pt}{7.8ex}
     
       \multirow{10}{*}{$\tilde{\mom}^1_{LT}$} 
    & $\begin{aligned}
    &-\epsilon_{LR}\,\frac{\alpha_2}{\Qr}\bigg[ \Qr(\smPL - \qr\,\smML)\CaIV^0 - \smML\,\CaVI^0 \\&-\dfrac{\alpha_1\alpha_0}{4}\bigg(\CaI^0(\CaVI^0  -(\Qr - \Ql)\CaIV^0)-(\CaVI^0 +(\Qr + \Ql)\CaIV^0)\, \CaII^0\bigg)\bigg]\dfrac{E}{\Lambda} 
    \end{aligned}$ \\[0.95cm]
    \cline{2-2} 
    \rule{0pt}{7.5ex}
    & $\begin{aligned}
        &- \epsilon_{LR} \dfrac{\alpha_1\alpha_2}{4(\Qr)^2}\bigg[\Qr\,\CaI^1(\CaIV^0(\Qr-\Ql) - \CaVI^0) + \Qr(\CaIV^0(\Qr+\Ql) + \CaVI^0)\CaII^1 \\
        &-\rr\alpha_3\bigg(\CaV^0(\CaVI^0 + \Ql\,\CaIV) +(\Ql\,\CaVI^0 - \left((\Qr)^2-(\Ql)^2\right)\CaIV^0)\CaIII^0\bigg)\bigg]\dfrac{E}{m_Z}\dfrac{E^2}{\Lambda^2}
    \end{aligned}$ \\[0.9cm]
    \cline{2-2}
    \rule{0pt}{9ex}
    & $\begin{aligned}
        &+\epsilon_{LR}\dfrac{\alpha_2}{2} \bigg[\smPL (\sqrt{6} \,\KaI^0 - 2\CaIV^2) + \frac{1}{\Qr}\smML (\sqrt{6} \,\Qr\,\KaII^0 + 2\Ql\,\CaIV^2 + 2\CaVI^2) \\
        &+ \dfrac{\alpha_1}{4\Qr}\alpha_0\bigg(\sqrt{6}\Qr\left(\CaI^0(\KaI^0 + \KaII^0) + \CaII^0(\KaI^0 - \KaII^0)\right)+ 2(\CaI^0 - \CaII^0)(\CaVI^2 + \Ql\,\CaIV^2) - 2\Qr (\CaI^0 +\CaII^0)\CaIV^2\bigg)\bigg]\dfrac{E^3}{\Lambda^3}
    \end{aligned}$ \\[1cm]
    \cline{2-2}
    \hline
    \hline \rule{0pt}{5ex}
    
    \multirow{11}{*}{$\tilde{\mom}^2_{LT}$}  
    & $\begin{aligned}
        &-\frac{\alpha_2}{\Qr} \bigg[\Qr(\smPL + \qr\,\smML)\CaIV^0 + \alpha_2\,\smML \,\CaVI^0 + \dfrac{\alpha_1}{4}\alpha_0\bigg\{\Qr(\CaI^0 + \CaII^0)\CaIV^0 +(\CaI^0-\CaII^0)(\CaVI^0 + \Ql\,\CaIV^0)\bigg\}\bigg] \,\dfrac{E}{\Lambda}
        \end{aligned}$  \\[0.45cm]
    \cline{2-2}
    \rule{0pt}{8ex}
    & $\begin{aligned}
        &- \dfrac{\alpha_1 \alpha_2}{4(\Qr)^2} \bigg[\Qr\,\CaI^1(\CaVI^0 + \Qr(1+\qr)\CaIV^0) - \Qr\,\CaII^1(\CaVI^0 - \Qr(1-\qr)\CaIV^0) \\
        &+ \rr\alpha_3\bigg(\CaV^0(\CaVI^0 +\Ql\,\CaIV^0) + \CaIII^0\left(\Ql\,\CaVI^0 +((\Qr)^2 + (\Ql)^2)\CaIV^0\right)\bigg)\bigg] \dfrac{E}{m_Z}\dfrac{E^2}{\Lambda^2}
    \end{aligned}$ \\[0.95cm]
    \cline{2-2} \rule{0pt}{11.5ex}
    & $\begin{aligned}
        &+\dfrac{\alpha_2}{2}\bigg[\smPL(\sqrt{6}\,\KaI^0 - 2\CaIV^2) - \frac{1}{\Qr}\smML(\sqrt{6}\Qr\,\KaII^0 + 2\Ql\,\CaIV^2 + 2\CaVI^2) \\
        &-\dfrac{\alpha_1\alpha_0}{4\Qr}\bigg(2(\CaI^0-\CaII^0)(\CaVI^2+\Ql\,\CaIV^2) +2\Qr(\CaI^0+\CaII^0)\,\CaIV^2 -\sqrt{6}\Qr\{\CaI^0(\KaI^0 - \KaII^0) + \CaII^0(\KaI^0 + \KaII^0)\}\bigg)\bigg]\dfrac{E^{3}}{\Lambda^3} \\
        &+ \alpha_0\alpha_1\alpha_2\dfrac{(\alpha_5 + \varepsilon_4)}{(\alpha_4 - \varepsilon_3)}\bigg[\BaIIsX(\BaIIIsX) - \BaIsX( \BaIVsX)\bigg]\dfrac{E^{3}}{\Lambda^3}
    \end{aligned}$\\[1.5cm]
    
    \hline
    
    \end{tabular}
    \end{adjustbox}
    \label{tab:moments3}
\end{table}
\clearpage
\setcounter{table}{7}
\begin{table}[tt]
\begin{adjustbox}{max width=\textwidth}
    \centering
    \begin{tabular}{|c|c|}
    \hline
     $\mom_i$  & Expressions in terms of helicity directions\\
         \hline \rule{0pt}{4ex}
     \multirow{19}{*}{$\mom_{TT'}$} & $\dfrac{1}{2}\bigg[(\smPT^2 + \smMT^2) +\smPT\alpha_0(\CaI^0 + \CaII^0) + \smMT\alpha_0(\CaI^0 - \CaII^0) + \dfrac{\alpha_0^2}{2}
    \bigg((\CaI^0)^2 + (\CaII^0)^2\bigg)\bigg]\dfrac{m_Z^2}{E^2} $ \\[0.4cm] 
    \cline{2-2} \rule{0pt}{8ex}
    & $\begin{aligned}
    &+\dfrac{1}{2}\bigg[\smPT(\CaI^1 + \CaII^1 + \alpha_3(1+\rr)\CaIII^0) + \smMT\bigg(\CaI^1 - \CaII^1 + \frac{\alpha_3}{\Qr}(1+\rr)(\CaV^0 + \Ql\,\CaIII^0)\bigg) \\
    &+ \alpha_0(\CaI^0\, \CaI^1+ \CaII^0\,\CaII^1) + \frac{\alpha_0}{2}\alpha_3(1+\rr)\bigg(\CaIII^0(\CaI^0 + \CaII^0) + \dfrac{1}{\Qr}(\CaI^0 - \CaII^0)(\CaV^0 + \Ql\,\CaIII^0)\bigg)\bigg]\dfrac{m_Z}{E}\frac{E}{\Lambda}
    \end{aligned} $ \\[0.85cm]
    \cline{2-2}
    \rule{0pt}{11ex}
    & $\begin{aligned}
         &+ \dfrac{1}{8(\Qr)^2}\bigg[2\Qr\,\alpha_3(1 + \rr)\bigg(\Qr(\CaI^1 +\CaII^1) \CaIII^0 + (\CaI^1 -\CaII^1)(\CaV^0 + \Ql\,\CaIII^0)\bigg)- \alpha_2^2(\CaVI^0 + \Ql\,\CaIV^0)^2 \\
         &+ \alpha_3^2(1+\rr)^2 \bigg((\Qr)^2(\CaIII^0)^2 + (\CaV^0 + \Ql\,\CaIII^0)^2\bigg) + (\Qr)^2\bigg(2\left\{(\CaI^1)^2+(\CaII^1)^2\right\} - \alpha_2^2 (\CaIV^0)^2\bigg) \bigg]\dfrac{E^2}{\Lambda^2} \\
          &- \dfrac{4\alpha_0^2\alpha_2^2}{(\alpha_4 - \varepsilon_3)^2} [(\BaIsX)^2 + (\BaIIsX)^2]\dfrac{m_Z^2}{E^2} \dfrac{E^2}{\Lambda^2}
    \end{aligned}$ \\[1.5cm]
    \cline{2-2}
     \rule{0pt}{14.5ex}
    & $\begin{aligned}
          &+\dfrac{1}{2}\bigg[\smPT(\CaI^3 + \CaII^3 + \alpha_3(1+\rr)\CaIII^2) + \smMT\bigg(\CaI^3 - \CaII^3 + \frac{\alpha_3}{\Qr}(1+\rr)(\Ql\,\CaIII^2 + \CaV^2)\bigg) \\
          &+ \dfrac{\alpha_0}{2\Qr}\alpha_3(1 + \rr) \bigg(\Qr(\CaI^0 +\CaII^0)\CaIII^2 + (\CaI^0 - \CaII^0)(\CaV^2 + \Ql\,\CaIII^2)\bigg) +\alpha_0(\CaI^0\, \CaI^3 + \CaII^0\, \CaII^3)\bigg] \dfrac{m_Z}{E}\dfrac{E^{3}}{\Lambda^3} \\
          &-\dfrac{2\alpha_2^2 \alpha_0}{(\alpha_4 - \varepsilon_3)^2}\bigg[4\{\BaIsX (\BaIsI) + \BaIIsX (\BaIIsI)\} \\
          &+ (\alpha_4 + \varepsilon_3)\{\BaIsX(\BaIIIsX) + \BaIIsX(\BaIVsX)\}\bigg]\dfrac{m_Z}{E}\dfrac{E^{3}}{\Lambda^3}
         \end{aligned}$ \\[2cm]
    \hline
    \hline \rule{0pt}{8ex}
    
 \multirow{11}{*}{$\tilde{\mom}_{TT'}$}
    & $\begin{aligned}
    &+\dfrac{\alpha_2}{2\Qr}\bigg[\Qr(\smPT + \qr\smMT)\CaIV^0 + \smMT\,\CaVI^0 \\&+ \frac{\alpha_0}{2}\bigg\{\Qr(\CaI^0 + \CaII^0)\CaIV^0 + (\CaI^0 - \CaII^0)(\CaVI^0 + \Ql\,\CaIV^0)\bigg\}\bigg] \dfrac{m_Z}{E}\dfrac{E}{\Lambda}
    \end{aligned}
    $ \\[0.9cm]
    \cline{2-2}
    \rule{0pt}{7.5ex}
    & $\begin{aligned}
        &+ \dfrac{\alpha_2}{4(\Qr)^2}\bigg[\Qr\, \CaI^1(\CaVI^0 +\Qr(1+\qr)\CaIV^0) - \Qr\,\CaII^1(\CaVI^0 - \Qr(1-\qr)\CaIV^0) \\
        &+\alpha_3(1 + \rr)\bigg\{\CaV^0(\CaVI^0 + \Ql\,\CaIV^0) + \CaIII^0(\Ql\CaVI^0 + (\Qr)^2(1+\qr^2)\CaIV^0)\bigg\}\bigg]\dfrac{E^2}{\Lambda^2}
    \end{aligned}$ \\[0.9cm]
    \cline{2-2} \rule{0pt}{11.5ex}
    & $\begin{aligned}
   & +\dfrac{\alpha_2}{4}\bigg[\smPT(2\CaIV^2 - \sqrt{6}\,\KaI^0) + \frac{1}{\Qr}\smMT(2\Ql\,\CaIV^2 + \sqrt{6}\Qr\,\KaII^0 + 2\CaVI^2) + \dfrac{\alpha_0}{2}\bigg(2\CaIV^2(\CaI^0 + \CaII^0)\\
   &- \sqrt{6}\{\CaI^0(\KaI^0 - \KaII^0) + \CaII^0(\KaI^0 + \KaII^0)\}\bigg)+ \frac{1}{\Qr}\alpha_0(\CaI^0 - \CaII^0)(\CaVI^2 + \Ql\,\CaIV^2)\bigg]\dfrac{m_Z}{E}\dfrac{E^{3}}{\Lambda^3} \\
   &+ \dfrac{4\alpha_0\alpha_2^2}{(\alpha_4 - \varepsilon_3)}\left[\BaIIsX (\BaIIIsX) - \BaIsX(\BaIVsX)\right]\dfrac{m_Z}{E}\dfrac{E^3}{\Lambda^3}
    \end{aligned}$ \\[1.6cm]
    \hline
    \end{tabular}
    \end{adjustbox}
    \caption{Expressions for the angular moments  (see \eq{eq:ang_funcs} and~(\ref{theta})) in terms of the helicity directions  up to ${\cal O} (E^3/\Lambda^3)$ in the EFT expansion. Note that each of the table entries above must be multiplied by a factor of $\mathcal{G}^2$, where $\mathcal{G} = g^2 \sqrt{(g^Z_{l_L})^2 + (g^Z_{l_R})^2}/(c_{\theta_W}^2\,\Gamma_Z)$, to obtain the expression for the angular moments normalised as in \eq{theta}. We write these expressions in terms of the $\delta^p\mathbf{a}_{J,k}^m$  in \eq{pow}~(\ref{pow2}),~(\ref{pow3}),~(\ref{pow4}) and~(\ref{pow5});  these can be written as simple linear combinations of Wilson coefficients using Table~\ref{tab:pdir},~\ref{tab:J2dir},~\ref{tab:LR_RL_hel_dir}. Here, $\epsilon_{LR} = \alpha_{l_L} - \alpha_{l_R}$, $\qr = \Ql/\Qr$, $G_{+-0} =  Q^Z_{f_R} \frac{\alpha_1}{2}\alpha_0$, $G_{-+0} = Q^Z_{f_L} \frac{\alpha_1}{2}\alpha_0$, $G_{+-} = Q^Z_{f_R}\alpha_0$ and $G_{-+} = Q^Z_{f_L}\alpha_0$. }
    \label{tab:final}
\end{table}

\bibliographystyle{JHEP}
\bibliography{References}
\end{document}